%% file: Shower-duplicity-PII.tex
\documentclass[preprint,12pt]{elsarticle}
\usepackage{etoolbox}
\usepackage{hyperref}

\journal{--}

\usepackage{color, graphicx} 
\usepackage{txfonts}

\usepackage{natbib} 
\usepackage{url}
\usepackage{enumitem}
\usepackage{ulem}
\usepackage[switch, modulo]{lineno}

\defcitealias{2024A&A...682A.159J}{Paper~I} 
\newcommand{\bo}[1]{\bf{#1}}

\title{Identifying misclassified meteor showers in the IAU MDC database. Reclassification Proposal.}
\author{T.J. Jopek$^{a}$, L. Neslu\v{s}an$^{b}$, S. \v{D}uri\v{s}ov\'{a}$^{c}$, R. Rudawska$^{d, e}$, M. Hajdukov\'{a}$^{c}$}
\affiliation[1]{Astronomical Observatory Institute, Faculty of Physics and Astronomy, A.M. University, Poznan, Poland}
\affiliation[2]{Astronomical Institute, Slovak Academy of Sciences, Tatranska Lomnica, Slovakia}
\affiliation[3]{Astronomical Institute, Slovak Academy of Sciences, Bratislava, Slovakia}
\affiliation[4]{Starion Group, Leiden, The Netherlands}
\affiliation[5]{ESA ESTEC, Noordwijk, The Netherlands}

\date{\textbf{Submitted to Astronomy \& Astrophysics}}

\makeatletter
\let\newtitle\@title
\let\newauthor\@author
\let\newaffiliation\@date
\let\newdate\@date
\makeatother

\begin{document}
\maketitle
\pdfstringdefDisableCommands{\let\textsuperscript\relax}

\section*{\centering Abstract}

The study of meteoroid streams reveals the full complexity of these structures. 
At present, we have no objective method of deciding whether the parameters of the observed meteoroid stream represent a further solution to an already known object or whether they relate to a new discovery. As result, the Meteor Data Center (MDC) database of the International Astronomical Union (IAU) contains duplicates and false duplicates of the meteor showers.

It is desirable to detect questionable cases and, if possible, correct their status, thereby contributing to the improvement of the content of the IAU MDC database. The correct content of the MDC database is important in its applications, for example, in assessing the threat from meteoroids to Earth's artificial satellites.

Two approaches were used, in the first the internal compatibility of geocentric and heliocentric parameters representing a given flux was verified. In the second, a comparison of two or more solutions of the same stream was made in as much detail as possible.
Fifty-six streams were verified, for which clear suspicion of misclassification was established in our earlier work.

For 43 streams, the misclassification was confirmed, with a proposal to change their status in the MDC to autonomous. In the remaining 13 cases, it was proposed to leave their status unchanged. 

Although we do not consider it 'definitive', our study clearly shows that repeated misclassification of new meteoroid flux solutions has occurred in the past. Correction of these cases will significantly improve the content of the MDC database. As an additional product, based on the approach proposed in this and our earlier work, relevant procedures have been proposed, which, available on the MDC database website, will make it possible to compare new meteoroid data with the contents of the database, thereby avoiding errors in their classification.
\section{Introduction}
\label{Intro}
As in any natural science, in meteor astronomy, databases are a fundamental starting point in many detailed studies. Therefore, their high quality, the absence of errors and artifacts are essential for the results of meteor data analyses to come to correct conclusions.

The International Astronomical Union's (IAU) Meteor Data Center (MDC) shower database is widely used by the meteor astronomers' community. It is used in numerous scientific studies with various outcomes, such as identifying the parent bodies of meteoroid streams and analyzing their dynamical evolution \citep[e.g.][]{Vaubaillon_etal2019}, predicting meteor shower outbursts \citep[e.g.][]{Egal_etal2023}, or developing models of meteoroid environments near Earth \citep[e.g.][]{Moorhead_etal2017, 2025AdSpR..75.1145M}. These models are essential for protecting satellites and space detectors and are used by spacecraft designers and mission planners\footnote{A Meteoroid Handbook for Aerospace Engineers and Managers: \url{https://ntrs.nasa.gov/citations/20200000049}}.

The MDC maintains an official catalogue where new meteor shower discoveries are submitted to receive a unique designation. However, a single meteor shower may be observed by various teams and/or using different techniques, as well as in different years during subsequent encounters of the Earth with a given meteoroid stream. Each submission (a set of mean shower parameters and a set of individual orbits) of the same meteor shower is called a "shower solution", and showers with multiple solutions in the MDC are referred to as multisolution showers (MSS). Showers observed only once and/or submitted by a single team are referred to as single-solution showers (SSS).

The classification of the observed shower—whether it is a new, uncatalogued shower or a new solution of an already catalogued shower—depends entirely on the subjective approach of the authors submitting their observations to the MDC. This sometimes leads to contamination of the MDC with incorrectly assigned shower solutions: (1) If a new observation of previously known meteor shower is incorrectly submitted as the discovery of a new shower, a duplicate shower will appear in the MDC. This means that such a solution, instead of forming an MSS, will create a new SSS with a different designation in the MDC. (2) The opposite problem occurs if an observed shower is incorrectly submitted as a solution to a cataloged shower it does not belong to. In this case, we can talk about a false duplicate solution, as correct solutions of an MSS shower should be recognized as duplicates. In this case, incorrectly assigning solutions to an MSS shower distorts the shower's characteristics. We used the same terminology as in our previous paper \citep{2024A&A...682A.159J} (hereafter referred to as Paper I), which serves as the initial basis for this study.

Regarding problem (1), in Paper I, we identified seven SSS showers as duplicates of previously cataloged showers (listed in Table 7 of Paper I). All of these have already been removed from the MDC Working List. Shower 1196/ZCM, submitted to the MDC in 2022, was also recognized as a duplicate by the authors \citep{Odeh_etal2023}. They did not include its parameters in their final publication, neither as a new solution for its duplicate counterpart shower 246/AMO; therefore, it could not be reassigned to that shower in the MDC.\footnote{The criteria for moving showers between MDC lists are outlined in \citep{2023A&A...671A.155H} and available on the MDC website: \url{https://ceresiaumdc.ta3.sk/shower_status_criteria} 
}
The other six showers found to be duplicates in Paper I (1151/NPA, 1110/CEP, 1108/IHR, 1157/FCD, 1161/THT, 1107/JID) were submitted to the MDC in 2022 and likely also recognized as duplicates, since they did not appear in the final publication of this submission \citep{2024book...Jen}. However, under their respective duplicate counterparts, only the 1151/NPA shower solution was found (under shower 575/SAU), with parameters and mean orbit that exactly matched the submission; therefore, this move was also made in the MDC. In the other cases, the corresponding parameters did not align with the submission, so these solutions will be among the removed showers unless they are resubmitted to the MDC.

As for problem (2), we revealed in Paper I that the IAU MDC meteor shower database contains a considerable number of false duplicates (solutions misclassified as members of the MMS). Showers contaminated with false solutions exhibit biased characteristics that influence studies that use them. Their presence reduces the quality of the database, so it is desirable to verify, remove or possibly correct misclassified  shower solutions. We address these issues in this study, with the aim of restoring the internal consistency of each affected shower. We propose specific solutions for each of the showers requiring reclassification. In doing so, we continue our  efforts to improve the content of the MDC database. 
%
\section{Input data used}
\label{MDC-data}
We used the same meteor shower data set as in Paper I (including 920 meteor showers, 252 of them MSS, as of December 2022), as well as results obtained from it. 
Our starting point is the list of false duplicates taken from Paper I. We provide this list in Table~\ref{tab:false-duplicates}. 
Its content was obtained using two methods: one involving the use of cluster analysis and orbital similarity function (CAM method), and a second one called "maximum-sigmas" based on a direct comparison of selected geocentric and heliocentric shower parameters (MSM method). Details of these methods are available in Paper I.

In Table~\ref{tab:false-duplicates}, for each shower potentially containing false duplicates, the meteor shower codes\footnote{The old and new meteor shower designation rules are described in \citep{2023NewAR..9601671J}.}, and two values of orbital similarity $DH_{Min}$ and $DH_{MI}$ are given. They were calculated using the hybrid D-function  \citep[see][] {1993Icar..106..603J, 1997A&A...320..631J};  the $DH_{Min}$ is the smallest threshold values of orbital similarity by which all members (all duplicates?) of the given MSS can be identified using the CAM method; $DH_{MI}$ gives the maximum acceptable threshold value to reliably identify (by applied cluster analysis) a group of the size given in column $N_S$. The threshold values of $DH_{MI}$ correspond to the probability of 1\% of randomly identifying a group of size $N_S$, (given in each row of Table~\ref{tab:false-duplicates}) in all analysed meteoroid streams orbits. As we see, the size of $N_{S}$$=$$2$ occurs most frequently, so the corresponding thresholds do not depend on the cluster analysis used. 

For $N_{S}$$=$$2$ the threshold $DH_{MI}$ values depend only on the size and type of orbital sample being tested (short, long orbital period streams). Using formulas (21) and (22) \citep[given in Table~9 in the][]{2017P&SS..143...43J}, the plausible threshold values of orbital similarity for pairs of orbits taken from a sample of similar size as used in our study are $0.022$-$0.025$. Therefore, based on a comparison of the $DH_{Min}$ and $DH_{MI}$ values, we can claim that all meteor showers in Table~\ref{tab:false-duplicates} contain false duplicates, especially those with $N_{S}$$=$$2$.    

However, the results given in Table~\ref{tab:false-duplicates} are of different status.
For example, for the first three streams: 3/SIA, 127/MCA and 321/TCB --- the differences between $DH_{Min}$ and $DH_{MI}$ are small, and we can attribute their causes to the limitations of the methods used in Paper I. We therefore  regard these MSS as correctly classified in the MDC, and have omitted them from further discussion. 

In the remainder of this article, we will address the causes of the all other possible misclassifications and propose appropriate corrections to the MDC database.

\begin{table}
\begin{center}
\scriptsize
\caption{$56$ MSS's containing potential false duplicates.
The individual columns stand: the MDC code designation of the shower; $N_{S}$ is the number of solutions found in the MDC for this group; $DH_{Min}$ is the smallest orbital similarity threshold needed to identify all $N_S$ members of the MSS; $DH_{MI}$ is the maximum acceptable threshold value to reliably identify a group of the size $N_{S}$. In the last column, an asterisk marks those MSS among which an internal inconsistency of geocentric and heliocentric parameters was found. Particularly questionable MSS cases are noted in bold in column $DH_{Min}$. 
}
\begin{tabular}{r l c c c c}
\multicolumn{1}{c}{ }&\multicolumn{1}{c}{ Shower Code}& \multicolumn{1}{c}{$N_{S}$} & \multicolumn{1}{c}{$DH_{Min}$} &  \multicolumn{1}{c}{$DH_{MI}$} &  \multicolumn{1}{c}{  }\\
\hline
\hline
  1  & 0003/SIA &    2 &  0.065 &   0.039 &  \\ 
  2  & 0127/MCA &    2 &  0.041 &   0.039 & \\
  3  & 0321/TCB &    2 &  0.089 &   0.059 &  \\ 
  \hline
  4  & 0011/EVI &    3 &  \bo{0.322} &  0.079 &  * \\ 
  5  & 0107/DCH &    3 &  \bo{0.358} &  0.136 & \\ 
  6  & 0121/NHY &    3 &  {0.269} &  0.079 & * \\ 
  7  & 0152/NOC &    4 &  \bo{0.644} &  0.182 & *\\ 
  8  & 0183/PAU &    3 &  {0.294} &  0.136 & \\ 
  9  & 0188/XRI &    3 &  \bo{0.534} &  0.079 & \\ 
 10  & 0219/SAR &    4 &  \bo{0.373} &  0.099 & \\ 
\hline
  3  & 0025/NOA &    2 &  {0.210} &  0.039 & \\ 
  4  & 0032/DLM &    2 &  \bo{0.321} &  0.059 &  * \\ 
  5  & 0040/ZCY &    2 &  {0.259} &  0.059 & \\ 
  6  & 0076/KAQ &    2 &  {0.142} &  0.039 & * \\ 
  7  & 0088/ODR &    2 &  \bo{0.362} &  0.059 & * \\ 
  8  & 0093/VEL &    2 &  \bo{0.412} &  0.059 & \\ 
  9  & 0100/XSA &    2 &  {0.209} &  0.039 & \\ 
 10  & 0105/OCN &    2 &  {0.134} &  0.059 & *\\ 
 11  & 0106/API &    2 &  \bo{0.329} &  0.059 & \\ 
 13  & 0108/BTU &    2 &  \bo{0.344} &  0.059 & \\ 
 14  & 0113/SDL &    2 &  {0.172} &  0.039 & *\\ 
 15  & 0118/GNO &    2 &  \bo{0.798} &  0.059 & *\\ 
 17  & 0124/SVI &    2 &  {0.226} &  0.039 & \\
 19  & 0128/MKA &    2 &  \bo{0.315} &  0.039 & *\\ 
 20  & 0133/PUM &    2 &  {0.139} &  0.039 & \\ 
 21  & 0150/SOP &    2 &  \bo{0.343} &  0.039 & \\ 
 22  & 0151/EAU &    2 &  {0.155} &  0.059 & \\ 
 24  & 0154/DEA &    2 &  {0.178} &  0.039 &  \\ 
 25  & 0167/NSS &    2 &  \bo{0.282} &  0.039 &  \\ 
 26  & 0170/JBO &    2 &  {0.220} &  0.039 & *\\ 
 27  & 0179/SCA &    2 &  {0.234} &  0.039 & *\\ 
 29  & 0186/EUM &    2 &  {0.194} &  0.039 & \\ 
 31  & 0189/DMC &    2 &  \bo{0.454} &  0.039 & \\ 
 32  & 0197/AUD &    2 &  \bo{0.316} &  0.039 & \\ 
 33  & 0199/ADC &    2 &  {0.107} &  0.039 & \\ 
 34  & 0202/ZCA &    2 &  \bo{0.434} &  0.039 & *\\ 
 36  & 0220/NDR &    2 &  {0.144} &  0.039 & \\ 
 37  & 0233/OCC &    2 &  {0.160} &  0.039 & \\ 
 38  & 0253/CMI &    2 &  \bo{0.347} &  0.039 & \\ 
 39  & 0254/PHO &    2 &  {0.199} &  0.039 & *\\ 
 41  & 0324/EPR &    2 &  {0.250} &  0.059 &  \\ 
 42  & 0326/EPG &    2 &  {0.175} &  0.059 &  \\ 
 43  & 0327/BEQ &    2 &  \bo{0.377} &  0.059 & \\ 
 44  & 0334/DAD &    2 &  0.110 &   0.059 &  \\ 
 45  & 0347/BPG &    2 &  {0.230} &  0.059 &  \\ 
 46  & 0372/PPS &    2 &  \bo{0.350} &  0.059 & \\ 
 47  & 0386/OBC &    2 &  {0.266} &  0.059 & \\ 
 
\hline
\hline
\end{tabular}
\label{tab:false-duplicates}
\end{center}
\normalsize
\end{table}

\addtocounter{table}{-1}
\begin{table}
\begin{center}
\scriptsize
\caption{ Continuation.}
\begin{tabular}{r l c c c c}
\multicolumn{1}{c}{ }&\multicolumn{1}{c}{ Shower Code}& \multicolumn{1}{c}{$N_{S}$} & \multicolumn{1}{c}{$DH_{Min}$} &  \multicolumn{1}{c}{$DH_{MI}$} &  \multicolumn{1}{c}{  }\\
\hline
\hline
 48  & 0392/NID &    2 &  \bo{0.317} &  0.059 & *\\ 
 49  & 0490/DGE &    2 &  {0.149} &  0.039 & \\ 
 50  & 0507/UAN &    2 &  \bo{0.326} &  0.059 & \\ 
 51  & 0512/RPU &    2 &  \bo{0.370} &  0.059 & \\ 
 52  & 0555/OCP &    2 &  {0.282} &  0.059 & \\ 
 53  & 0574/GMA &    2 &  \bo{0.480} &  0.059 & \\ 
 54  & 0644/JLL &    2 &  {0.256} &  0.039 & *\\ 
 55  & 0709/LCM &    2 &  {0.142} &  0.039 & \\ 
 56  & 1048/JAS &    2 &  {0.150} &  0.059 & \\ 
\hline
\end{tabular}
\end{center}
\normalsize
\end{table}

%
\section{Methodology}
\label{sec:metodologia}

Searching for duplicates of an already known shower has some analogy with linking observational data of a 'new' asteroid with that of an already known object. Unfortunately, because of the low precision of the meteor observations and, above all, the lack of access to the orbital data of the members of the stream, this kind of approach could not be used here. 

Therefore, in Paper I, in order to verify the status of the members of MSSs among MDC data, a methodology analogous to the searching for streams among observed meteoroid's orbits was used. Despite the limited methodological rigor, this approach proved to be successful; solutions of almost all major streams as well as many minor streams were confirmed as duplicates; and, on the other hand, the correctness of $56$ classifications was questioned; see Table~\ref{tab:false-duplicates}. 

However, the decision of whether a newly found solution of a meteor shower is the first solution of newly discovered, autonomous shower, which has not been known, or it is another solution belonging to an already known shower, is a complex problem, as differences in solutions may, to some extent, also reflect the actual properties of the stream. In this section, we discuss all possible causes of differences in MSS solutions.

The first issue to consider is the existence of largely structured showers and shower complexes.

In more detail, some meteoroid streams have a filamentary structure, and their meteoroids hit Earth, causing showers with very similar mean characteristics.
Then, it is possible either to regard these similar showers as the autonomous entities or as the structures of the same showers.
Or, some mean parameters (or a single mean parameter) of a group of showers can be consecutive, but extended within a large interval, which is hardly acceptable as the interval of values of single shower. In such a case, we usually speak about the shower complex. Thus, the classification also depends on the convention adopted by the authors, as no document has defined this to date.

Our methodology used in the following regards each shower of a shower complex as the autonomous shower. In addition, if a shower filament can be discerned, it is also regarded an autonomous shower.
(If we always knew the uncertainty of parameter determination, $\sigma$, the filament could be discerned by a difference that exceeds 1 $\sigma$.)

Another possible cause of false duplicates in MDC is incorrect meteoroid data and thus inconsistencies in their geocentric and heliocentric parameters. 
In \citep{2023A&A...671A.155H}, the consistency of the data in the MDC was established with their counterparts in the source publications. 
The results shown in Table~\ref{tab:false-duplicates} were found using only data that have passed this verification process.
Therefore, as the next step, we decided to verify whether the cause of false duplicates in MDC is the lack of internal consistency between averaged geocentric parameters and averaged orbital elements of a shower.
For this purpose, we used a software package published by \citep{2024CoSka..54a..57N}. 
We considered the stream to be biased with incompatibility of the data, if the differences between the catalog and the recalculated values of even one parameter were: $\Delta \alpha_g \geq 5.0^{\circ}$, $\Delta \delta_g \geq 2.5^{\circ}$, $\Delta V_g \geq 1.5\,$km$\,$s$^{-1}$, for geocentric parameters, and $\Delta q \geq 0.05\,$au, $\Delta e \geq 0.05$  $\Delta \omega \geq 5.0^{\circ}$, $\Delta \Omega \geq 5.0^{\circ}$ and $\Delta i \geq 2.5^{\circ}$, for the orbital elements. 
Obtained results are discussed in Section \ref{sec:discussion}. All MSSs for which some parameters were found to be internally inconsistent are marked with asterisks in the last column of Table~\ref{tab:false-duplicates}.

Other reasons why two sets of averaged parameters for the same meteoroid stream may differ significantly may be as follows: 
\begin{itemize}
\item manual and automatic astrometry of observed meteor images \citep{2009JIMO...37...21K},
\item the algorithm used to calculate the orbital elements \citep[see e.g.][]{1957SCoA....1..183W,1961deor.book.....D, 1987BAICz..38..222C, 1990BAICz..41..391B,1995A&AS..112..173B,2010epsc.conf..888J,2015P&SS..117..223D},
\item the difference in the epochs of osculation of the orbital elements used to calculate the averaged stream orbits, \citep{2007MNRAS.375..595W}, 
\item the way individual meteoroid data are averaged, \citep[see][]{1999md98.conf..199V,1999SoSyR..33..302V, 2006MNRAS.371.1367J,2010pim8.conf...91J}. 
\end{itemize}
Meteors are observed with different techniques, reduced in different ways, using different astrometric methods.
The orbits of the observed meteoroids are calculated in different ways, taking different approximations of Earth's ephemeris and also different time scales.
Meteoroid streams are almost always identified among orbits corresponding to different epochs of osculation. Usually these differences are not large, but sometimes they approach 50 years, for example, in the case of the solutions of stream 183/PAU from Table~\ref{tab:MSS-1}. Using the work of \citep{2007MNRAS.375..595W} it can be estimated that for two averaged orbits of the same stream, corresponding to distant epochs of osculation, and additionally taking into account other factors mentioned above, the values of the hybrid DH-function can be greater than $0.1$. That is clearly more than the acceptable threshold values  of $0.039$-$0.059$ for the pairs listed in Table~\ref{tab:false-duplicates}.

In order to estimate the effect of these factors on the D values between the averaged orbits of the streams more precisely, it is necessary to have access to the parameters of the members of a given stream. Unfortunately, the requirement to include the parameters of the stream members in the MDC database together with the averaged parameters of the stream was introduced relatively recently \citep{2020P&SS..18204821J}. Hence, with a few exceptions, this access is not available for the MSS listed in Table~\ref{tab:false-duplicates}. 
For this reason, in the present study, when analysing the misclassified duplicates in the MDC, we use only data from Table~\ref{tab:false-duplicates}, Table~\ref{tab:MSS-1} and Table~\ref{tab:DH}. This means that we could use the positions of the shower radians on the sky maps, the averaged orbits, the corresponding DH-values, and the contribution of the individual components to the DH-function values. Furthermore, we were able to take into account the links of MSSs with their parent comets, if confirmed or not, as given in \citep{2024MNRAS.535.3661D}.
%
\section{Discussion of false-duplicates}
\label{sec:discussion}
In this chapter we analyse each MSS in more detail. In the first Subsection we deal with groups with 3 or 4 members. In the next Subsection we discuss those MSS's for which we had two solutions available in our sample.
In the Appendix in Table~\ref{tab:MSS-1}, the geocentric and heliocentric data of these MSS are listed.
Also in the Appendix in Table~\ref{tab:DH}, for all pairs of each MSS (see the second column of that table), the DH-values and their four components from the orbital elements of the compared pairs are given. 

Basically, based on the high values $DH_{Min}$ in Table~\ref{tab:DH}, and in view of the similarity threshold values for pairs of orbits that were adopted in Paper I, ($DH_{MI}=0.039$ for orbits with $i\leq 10^{\circ}$, $DH_{MI}=0.059$ for $i>10^{\circ}$) it can be argued that among the MSS from Table~\ref{tab:MSS-1} there is not a single pair correctly classified. 
In addition to verification using the D-function, all pairs of solutions in Table~\ref{tab:DH} were verified using the MSM (maximum sigma method, see Paper I). In all cases, this verification turned out to be consistent with the result obtained with the D-function.     
So, on that basis, we could end the discussion.

However, the existence of false duplicates in the MDC is the result of a long-standing subjective approach used to decide whether a new meteoroid data set represents another solution to an already existing stream. Since this decision is complex (as described in the previous section), we still do not have an officially accepted rule on this issue to date. The first MSS were created in the book by \cite{2006mspc.book.....J}. The author compiled shower solutions from various publications by different research teams using an approach that is, unfortunately, not clearly enough explained in the publication. Showers created in this way were sometimes named after one of their solutions given in the original paper or renamed.
The author kindly provided the MDC with the 230 meteor showers published in this book at the time of the creation of the shower database in 2007, see \citep{2011msss.conf....7J}. Later, new showers and their solutions were added to the list, in the MDC, according to the classification made by the authors in the publications, which was sometimes not appropriate. Until the unique criteria are defined, the classification problem will persist.
 
Hence, also in view of the reasons mentioned in Section~\ref{sec:metodologia} for which the average orbits of the same stream may differ, we decided to continue this study. Our approach allows us to eliminate clearly incorrect solutions in the MDC, and contributes to the understanding of how to ascertain whether a new solution could be a duplicate of an existing meteoroid stream, crucial for the discovery of new meteor showers.

During the research described in Paper I we limited ourselves to the streams available at MDC until December 2022, with complete orbital data. Currently, we are discussing in detail the individual cases identified in Paper I as false duplicates. To this end, we also included in our considerations those data that arrived at MDC database after December 2022, as well as those available in the literature but never entered into the database. 
\subsection{MSS with three or four members}
\label{unquestionable1}
\noindent
{\bf 11/EVI, eta-Virginids.}  
For the three solutions of this MSS we have $DH_{Min}$$=$$0.322$ and $DH_{M_I}$$=$$0.079$, strongly suggesting erroneous values for some of the stream parameters or incorrect classification.

The four meteors of the first solution 11/EVI/00 were identified by \citep{1971SCoA...12...14L} among the $2400$ orbits of meteoroids photographed in the 1950s. In Table~1a and~1b of \citep{1971SCoA...12...14L}, the shower is called Northern Virginids.
The parameters of the second solution 11/EVI/01, according to \citep{2006mspc.book.....J} (p. 699, Table~7), were calculated by averaging 7 meteoroids. However, we have not been able to determine when and by which technique these 7 objects were observed, or the size sample in which this stream was identified.

The third solution 11/EVI/06 was recently given by \citep{2022JIMO...50...38S}, who sampled $298689$ orbits of automatic TV meteor observation network (``SonotaCo consortium'') and identified 158 members of this shower. 

In Table~\ref{tab:MSS-1}, the averaged orbital elements of this MSS differ significantly in $q$, and the spatial orientation of the orbit, see Table~\ref{tab:DH}. However, for the second solution the internal consistency test failed. Similarly to \cite{2016JIMO...44..151K} (see Table~7 in his paper) we were not able to recalculate the orbital elements using the corresponding geocentric parameters of this shower. Therefore, we recommend moving solution 11/EVI/01 to the list of removed shower's data. 

The geocentric and heliocentric parameters of the other two solutions of this stream are internally consistent. However, the similarity of their orbits is  highly questionable at $DH_{Min}= 0.326$, which is significantly too high to support the hypothesis that these two solutions involve the same stream. 
\begin{figure}
    \centering
    \includegraphics[width=0.34\textwidth,angle=-90]{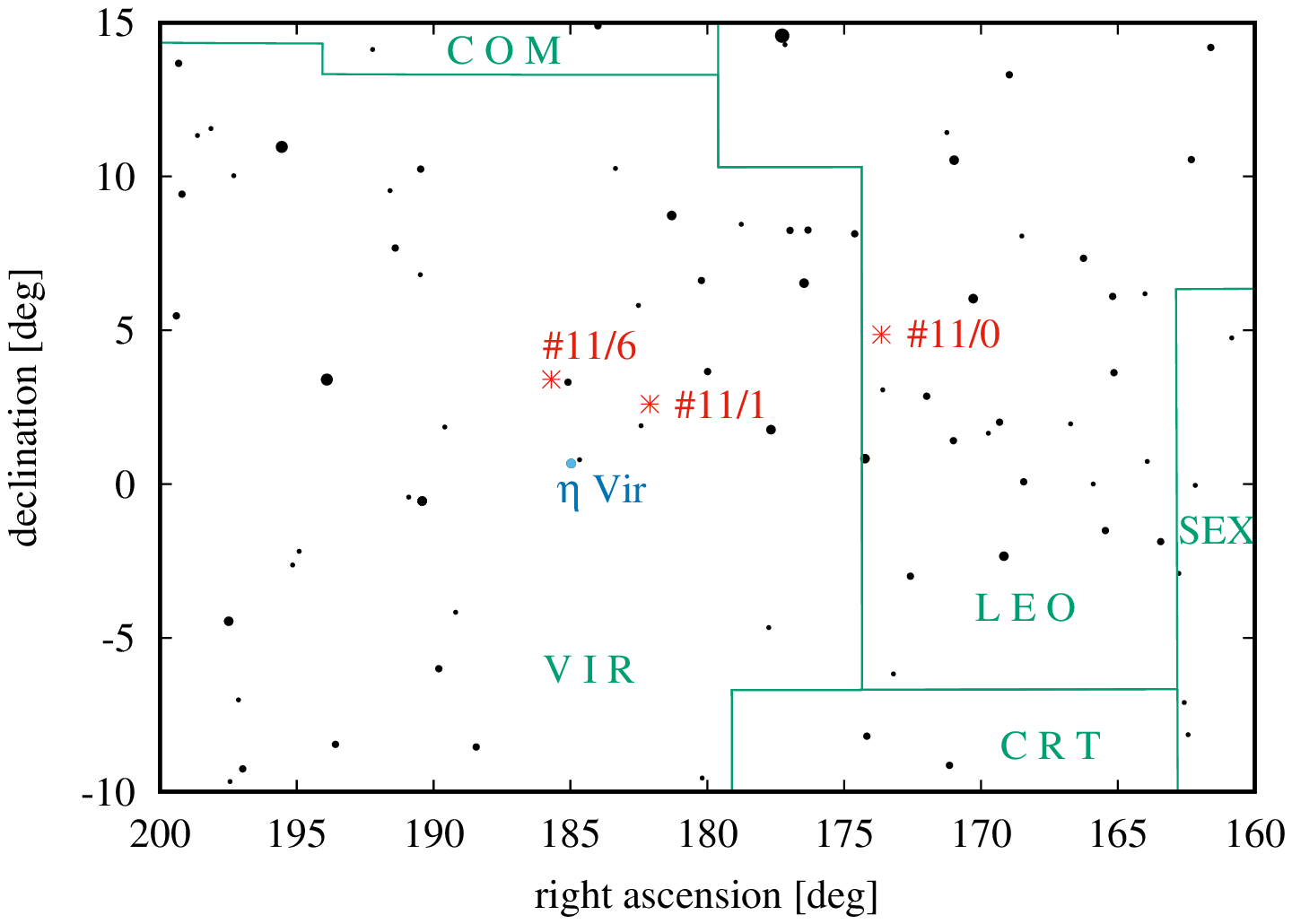}
    \caption{The positions of radiants (depicted by red asterisks) of the three solutions of 11/EVI shower. The star $\eta$-Virginis is shown with a cyan circle, the positions of the other stars with the black circles.
    }
    \label{fig:11radiants}
\end{figure}
In the domain of geocentric parameters, these two solutions have clearly different values of geocentric velocity, $36.0\,$km$\,$s$^{-1}$ and $27.3\,$km$\,$s$^{-1}$, respectively. In addition, their radiants fall into two different constellations Leo and Virgo, see Figure~\ref{fig:11radiants}. Due to these reasons, solution 11/EVI/00 should be considered a false duplicate of the eta-Virginids shower. Hence, in the MDC database, this shower should be represented by a single solution 11/EVI/06. We propose that solution 11/EVI/00 be moved to the MDC Working list, as a standalone shower with the original name Northern Virginids as proposed in  \cite{1971SCoA...12...14L}, and be assigned new IAU MDC codes. 

The current status of the shower 11/EVI is established. However, after the reclassification of its two incorrect solutions, it no longer fulfils the criteria for this status. If the IAU MDC does not receive a new submission for the shower, it should be considered for reassignment to the Working list. All procedures for changing the status of a shower, or for reclassifying solutions or changing their names, must be approved by the IAU Working Group on Meteor Shower Nomenclature (WG) \footnote{\url{https://www.iau.org/science/scientific_bodies/working_groups/276/}}.
\\

{\bf 107/DCH, delta-Chamaeleontids.} 
The stream is represented by three solutions;  $DH_{min}=0.358$ and $DH_{MI}=0.136$ differ significantly. The geocentric and heliocentric parameters of these solutions are internally compatible. Solutions 107/00 and 107/01 are taken from \citep{1975AuJPh..28..591G}, where, in Table 1, the authors assigned them two distinct designations 2.15 and 2.14, respectively. Their classification as a single shower and the name of the shower, delta-Chamaeleontids, were later proposed by \cite{2006mspc.book.....J}.
\citep{1999md98.conf..307J} partly confirmed the identification of solution 107/01, using a different approach and more numerous orbital sample than Gartrell and Elford. In his paper, Jopek gave the stream the name Chamaeleontids. In the MDC database, this solution is designated as 107/DCH/02.
All solutions of this MSS come from the radar meteor data sample, Gartrell and Elford searched 1667 orbits, Jopek used the same orbits and additionally included 2101 ones. All orbits were determined by the Adelaide radio meteor system in Australia.  Solution 107/00 was obtained from only 4 orbits, the others, 107/01 and 107/02, from 47 and 33 orbits, respectively. 

The geocentric parameters and orbits of these solutions differ markedly. 
In particular, the values of the geocentric velocity, eccentricity, and inclination of the orbits. 

\begin{figure}
    \centering
    \includegraphics[width=0.34\textwidth, angle=-90]{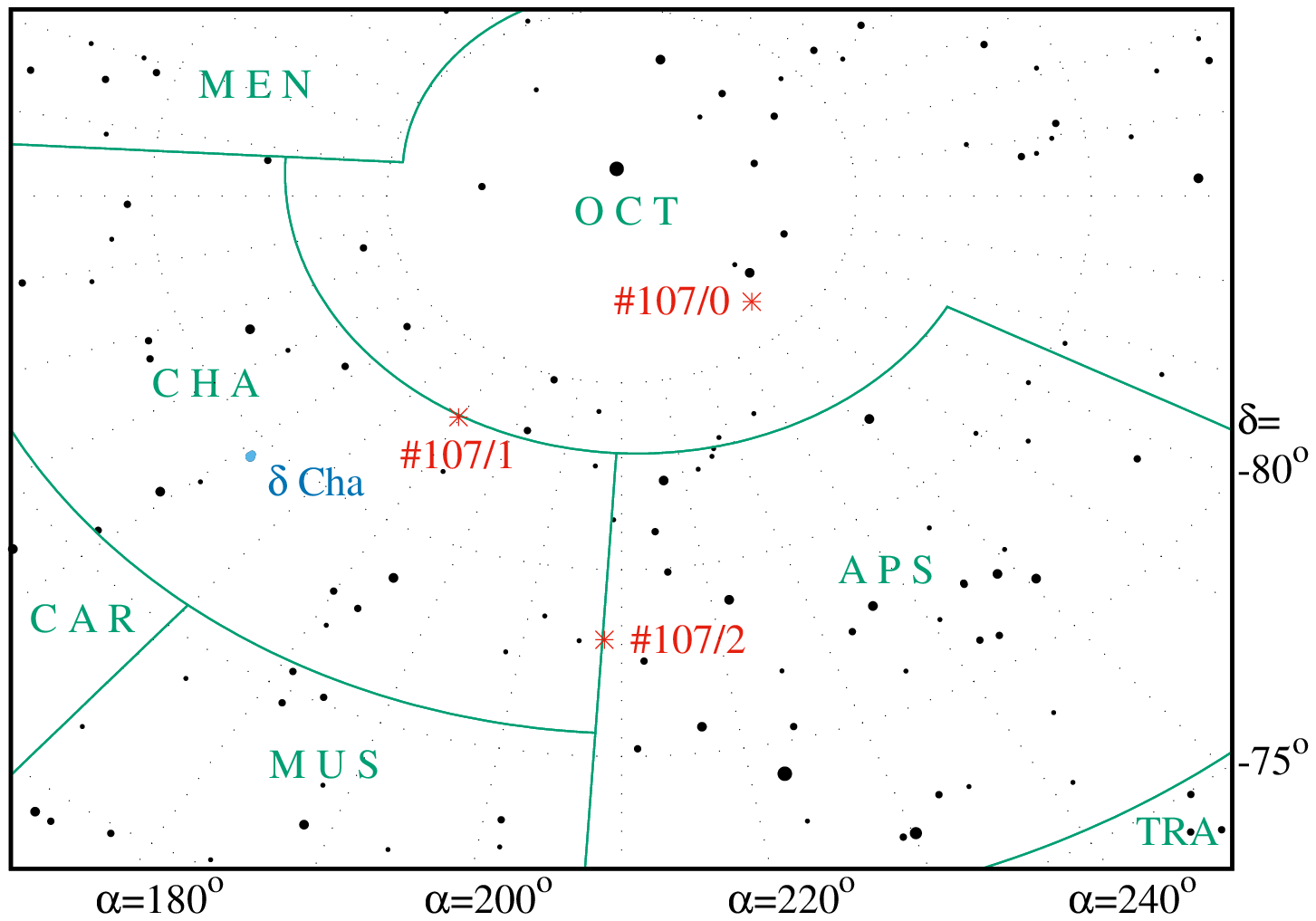} 
    \caption{The positions of radiants (depicted by red asterisks) of three solutions of shower delta-Chamaeleontids, 107/DCH. The position of double star $\delta$-Chamaeleontis is shown with a cyan circle, the positions of the other stars with the black circles.}
    \label{fig:107radiants}
\end{figure}
The positions of their radians differ significantly, as can be seen in Figure~\ref{fig:107radiants}. The radiant 107/00 lies in the constellation Octans, far from the constellation Chamaeleon. The other radians are located on the outskirts of the constellations Chamaeleon and Apus (radiant 107/01), and Chamaeleon and Octans (radiant 107/02).  
The arithmetic mean value of the orbital similarity taken from all pairs of this MSS <$DH$>=$0.388$. However, the orbital similarity of solutions 107/00 and 107/01 is $DH=0.526$ and for solutions 107/00 and 107/02 we have $DH=0.358$. For solutions 107/01 and 107/02, the orbital similarity takes the value $DH=0.279$, see Table~\ref{tab:DH}. For this reason, we can agree with \citep{1975AuJPh..28..591G}, 
who in their study, concluded that solutions 107/00 and 107/01 represent two different streams.
Hence, we recommend that solution 107/00 be a standalone shower. Without the 107/00 solution, the similarity of the other two solutions as measured by the DH-function has improved and is $DH_{Min}$=$0.279$, though it remains significantly higher than the maximum $DH_{MI}=0.059$ threshold allowed for this pair. It should be noted that solutions 107/01 and 107/02 were obtained from two partly overlapping orbital samples, so they cannot be considered fully independent solutions.  

We suggest a further study of this MSS, in which the variability of the orbits of the members of this group will be recognised by means of computer simulation. Perhaps we are dealing with a case of a toroidal stream whose orbits undergo rapid dispersion, which would explain the high value of $DH=0.526$ determined for solutions 107/01 and 107/02 of this stream. Until then, all three solutions should represent three showers on the MDC Working list, with 107/00 and 107/02 receiving new names and codes approved by the WG, while 107/01 retains the current shower name, delta-Chamaeleontids (see Figure~\ref{fig:107radiants}).\\

{\bf 121/NHY  nu-Hydrids.} 
For the three members of this MSS, $DH_{Min}=0.269$ and $DH_{M_I}=0.079$ differ significantly.  As noted in \citep{2016JIMO...44..151K}, there is an internal inconsistency in the geo-helio data of solution 121/NHY/00 given by \citep{1989JIMO...17..242T}. According to Koseki, the reason for the discrepancy is a typographical error in the value of $RA=158^{\circ}$ in \citep{1989JIMO...17..242T}. In order to remove the data inconsistency of this solution, Koseki proposes $RA=168.6^{o}$. An analogous conclusion follows from our calculations, but since for right ascension \citep{1989JIMO...17..242T} gives only three significant digits in her paper, the correction of the typographical error should be $RA=168^{\circ}$ in the B1950 reference frame.

\begin{figure}
    \centering
    \includegraphics[width=0.34\textwidth, angle=-90]{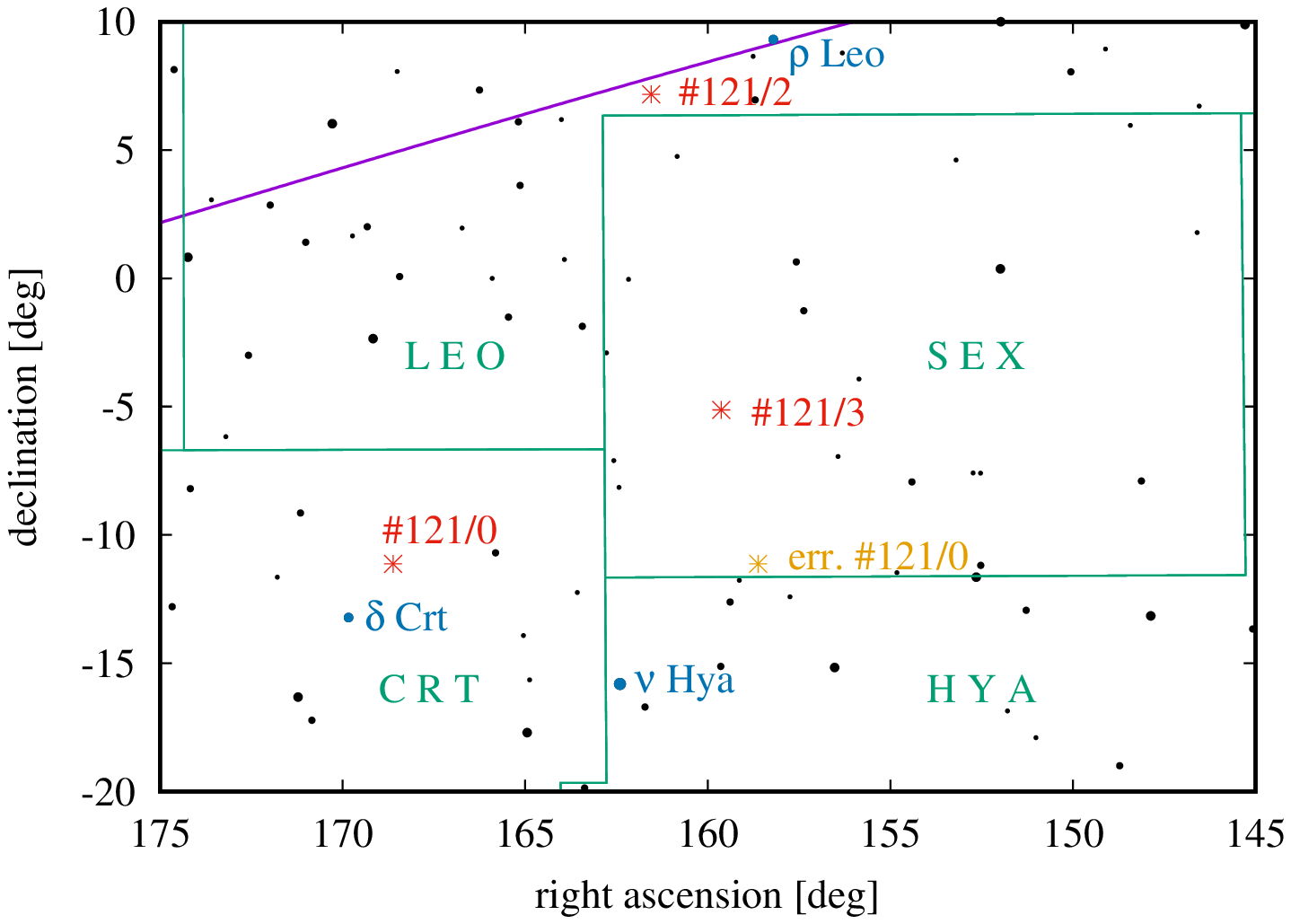}
    \caption{The positions of the radiants of three solutions of shower 121/NHY nu-Hydrids (red asterisks). The originally published, erroneous position of the solution 121/0 is shown with the orange asterisk. The purple line shows the ecliptic. Stars $\nu$-Hydrae, $\rho$-Leonis, and $\delta$-Crateris are plotted with the cyan circles; the other stars with the black circles.}
    \label{fig:121radiants}
\end{figure}

When it comes to the name of this shower, we are facing another problem. The parameters of the 121/NHY/02 solution are taken from \citep{1976Icar...27..265S}, where the author named the identified meteoroid stream, rho-Leonids (Table VI, p. 275) according to its radiant position in the constellation Leo near the star $\rho$ Leonis, see Figure~\ref{fig:121radiants}.  In \citep{1975AuJPh..28..591G}, the solution 121/NHY/03 is not named; instead, the authors proposed the code 3.02, only. 
\citep{1989JIMO...17..242T} proposed the name nu-Hydrids for the solution 121/NHY/00, but she most likely did so based on the coordinate $RA=158^{\circ}$ for this shower, which has now been found to be erroneous, causing an internal inconsistency in the data she gave for this shower. The corrected radiant, with a value of $RA=168^{\circ}$ (B1950), lies in the constellation Crater near the star $\delta$ Crateris, see Figure~\ref{fig:121radiants}.
Consequently, none of the three shower solutions 121/NHY, has a radiant located in the constellation Hydra. Each is located in a different constellation: Crater, Leo and Sextant. 

Returning to the similarity assessment of the orbits, a comparison of all pairs of orbits for 121/NHY showed that their DH values range from $0.201$ to $0.289$, see Table~\ref{tab:DH}. They are too different from the threshold value of $D_{MI}=0.059$ allowed for these pairs,  but, on the other hand, they do not differ enough to determine which of the 121/NHY solutions can be considered a false duplicate. Rather, it would be more appropriate to state that none of the solutions given in Table~\ref{tab:MSS-1} are members of the 121/NHY shower, but that they are three autonomous showers. Our recommendation is to keep these solutions on the MDC 
Working list as standalone streams. We propose that solution 121/NHY/02 retain, if possible, a variation of its original name as given in \citep{1976Icar...27..265S}, e.g. Southern rho-Leonids, while the other two solutions be assigned new names and codes. Shower 121/NHY will be simultaneously moved to the MDC List of removed showers to ensure traceability and identification of all solutions. Alternatively, to minimize changes, one of the solutions could retain the current shower name despite the inconsistency with the old nomenclature rules; the decision will be made by the WG.\\
%

{\bf 152/NOC Northern Daytime omega-Cetids.}  
For this shower the value of $DH_{Min}$=$0.644$ is significantly higher than the orbital similarity threshold of $DH_{MI}$=$0.182$, acceptable to a group of four members. 

\begin{figure}
    \centering
    \includegraphics[width=0.34\textwidth, angle=-90]{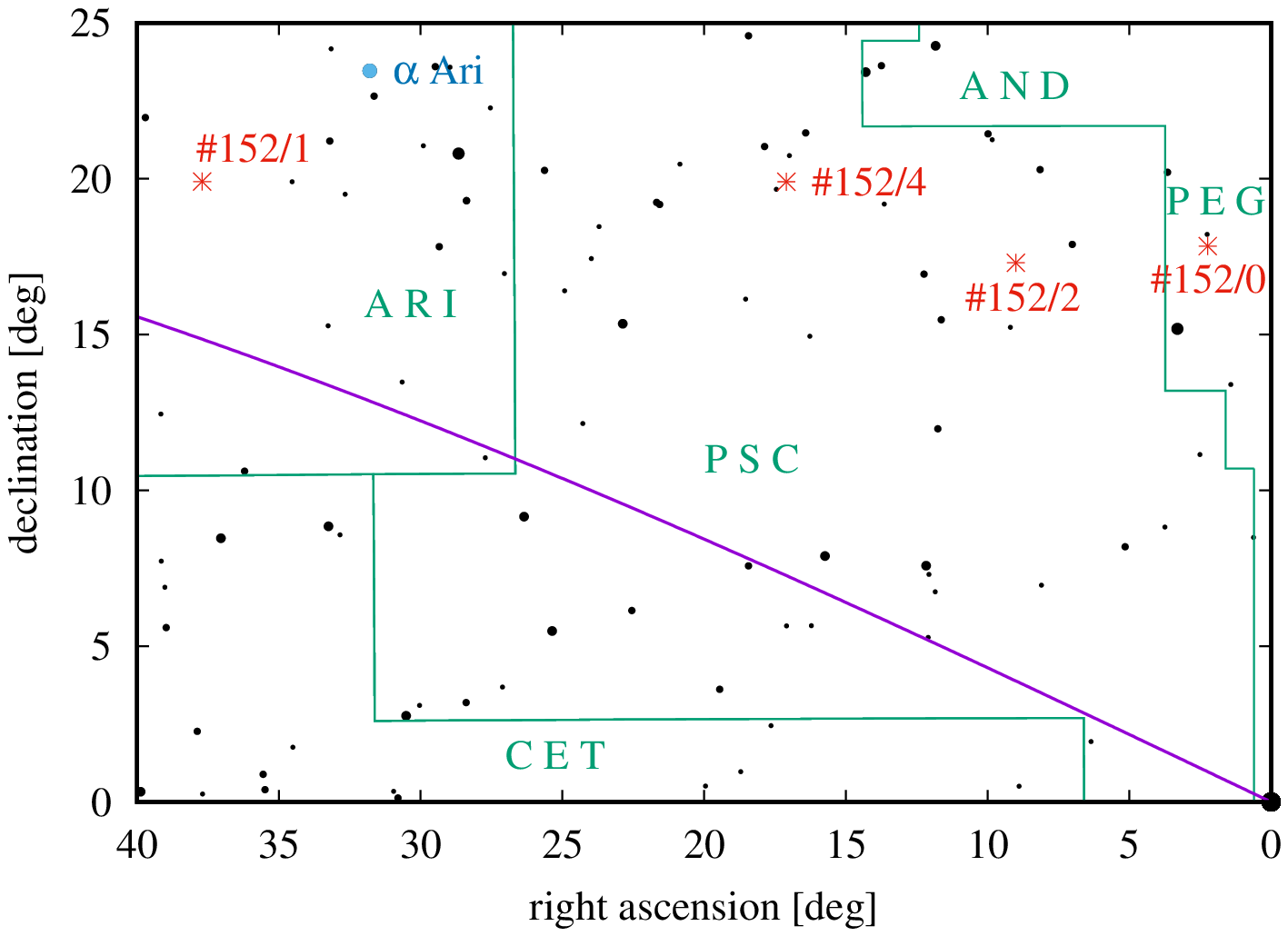}
    \caption{
    The positions of the radiants of four solutions of shower 152/NOC Northern Daytime omega-Cetids (red asterisks). One can see that no solution is located in the constellation Cetus, however. The stars are depicted with the black circles. The purple curve indicates the ecliptic.}
    \label{fig:152radiants}
\end{figure}

The first three solutions were identified among radio meteors, the fourth among video-observed meteors. Solution 152/NOC/00 was 
determined by \citep{1976Icar...27..265S}, who named it gamma-Pegasids. 
Solution 152/NOC/01 by \citep{1964AuJPh..17..205N}, who in Table~4 of his paper designated it as "Gr. 61.5.3". This shower consists of ten members. Nilsson's identification turned out to be correct, as it was confirmed by a completely different cluster analysis method by \citep{1999md98.conf..307J}. In Table~1 of this work, the stream is named alpha-Arietids. 
Solutions 152/NOC/02 and 152/NOC/04 were identified by \citep{2008Icar..195..317B} and \citep{2022JIMO...50...38S}, respectively. The radiants of all four solutions are located in three constellations: Aries, Pisces and Pegasus, see Figure~\ref{fig:152radiants}.
It is not specified in \citep{2006mspc.book.....J} why the shower was given a name related to the constellation Cetus.

In Table~\ref{tab:MSS-1} and Table~\ref{tab:false-duplicates}, compared to the others, the second solution, 152/NOC/01 has different values of orbital angular elements. In particular, it has a much smaller orbit inclination value $i=10.2^{\circ}$. 
The cluster analysis performed without 152/NOC/01 solution showed that the three remaining solutions can be identified as MSS with $DH_{Min}=0.193$ which is close to the acceptable value of $DH_{MI}=0.136$ for three members of the shower. 
As a result of this analysis we can consider the second solution given in Table~\ref{tab:MSS-1} as a false duplicate of the 152/NOC stream. We therefore recommend that 
152/NOC/01 be reclassified in the MDC database as a standalone shower with a new name and IAU MDC codes assigned.

However, one issue remains, for two solutions, 152/NOC/02 and 152/NOC/04, we found an internal inconsistency in the data. The data for these streams comes from the source works by \citep{2008Icar..195..317B} and \citep{2022JIMO...50...38S}, respectively. However, in both cases the inconsistency of the data is close to the assumed critical values of the differences between the parameters calculated by us and those given in the MDC database, (see Section \ref{sec:metodologia}).
Which, in our opinion, justifies leaving these solutions as duplicates of the 152/NOC meteor shower. This shower has already passed the status of an established shower and the elimination of solution 152/NOC/01, based on 16 orbits, will not affect this status.\\ 

{\bf 183/PAU Piscis Austrinids. }   
For the three solutions to this shower, $DH_{Min}=0.294$ while, the acceptable threshold value, $DH_{MI}=0.136$. The first two solutions were obtained from radio observations and the third with video cameras. 
The 183/PAU/00 solution is from \citep{1967SCoA...11..183K}, the 183/PAU/01 was given by \citep{2008Icar..195..317B}, the third 183/PAU/04 was given by \citep{2016Icar..266..331J}. The geo-helio data of these solutions are internally consistent, but the values of the parameters of the first one are clearly different from the others, especially the values of the inclination of the orbits and the geocentric velocities, see Table~\ref{tab:MSS-1}.

When ignores the orbit of 183/00, the orbital similarity of the other two solutions is $DH_{Min}=0.241$, see Table~\ref{tab:DH}, 
which for the remaining pair of solutions still far exceeds the allowable value of $DH_{MI}=0.056$. 
However, in the MDC database there is another solution of this shower provided by \citep{Shiba2023}, that was not used in Paper I, because the shower data have been submitted to the MDC after December 2022, the date to which we were limited in selecting the research sample in Paper I. 
Taking into account the solution 183/06 in MDC and at the same time removing 183/0, for these three solutions, we obtained $DH_{MI}$ = 0.191 with a cut-off value of $DH_{MI}$ = 0.146. And given the comments mentioned in Section 3, these solutions can be considered to represent the same meteoroid stream, especially since we are dealing here with the different way each author averaged the elements of the orbits. Brown calculated the elements of the orbits based on averaged values of geocentric parameters; Jenniskens gave medians; and Shiba counted the arithmetic means of the elements of the stream members' orbits. 

\begin{figure}
    \centering
    \includegraphics[width=0.34\textwidth, angle=-90]{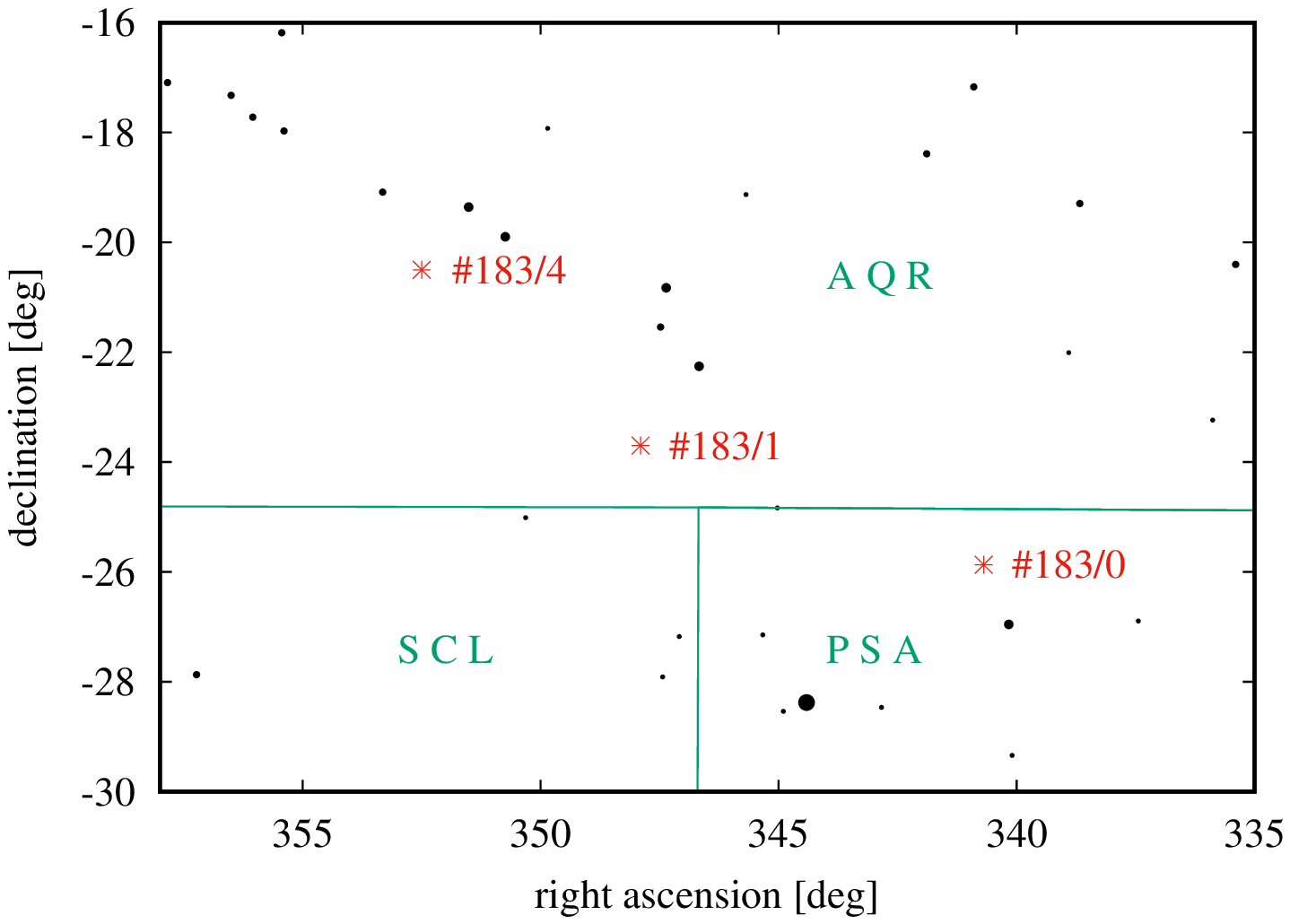}
    \caption{
    The positions of the radiants of three solutions of shower 183/PAU Piscis Austrinids (red asterisks). Only the solution AdNo=0 is located in the constellation Piscis Austrinus. Two other solutions, 183/PAU/01 and 183/PAU/04, are located in constellation Aquarius. The stars are depicted with the black circles.}
    \label{fig:183radiants}
\end{figure}

In view of the above, despite the fact that only the radiant of solution 183/00 is located in the constellation Piscis Austrinus, the others are in the constellation Aquarius, see Figure~\ref{fig:183radiants},  we propose that solutions 183/01, 183/04 remain unchanged in the MDC. On the other hand, we propose that stream 183/00 be given autonomous status with a new name and codes. The shower 183/PAU remains established since removing one of the four current solutions does not affect its status.\\
%

\begin{figure}
    \centering
    \includegraphics[width=0.34\textwidth, angle=-90]{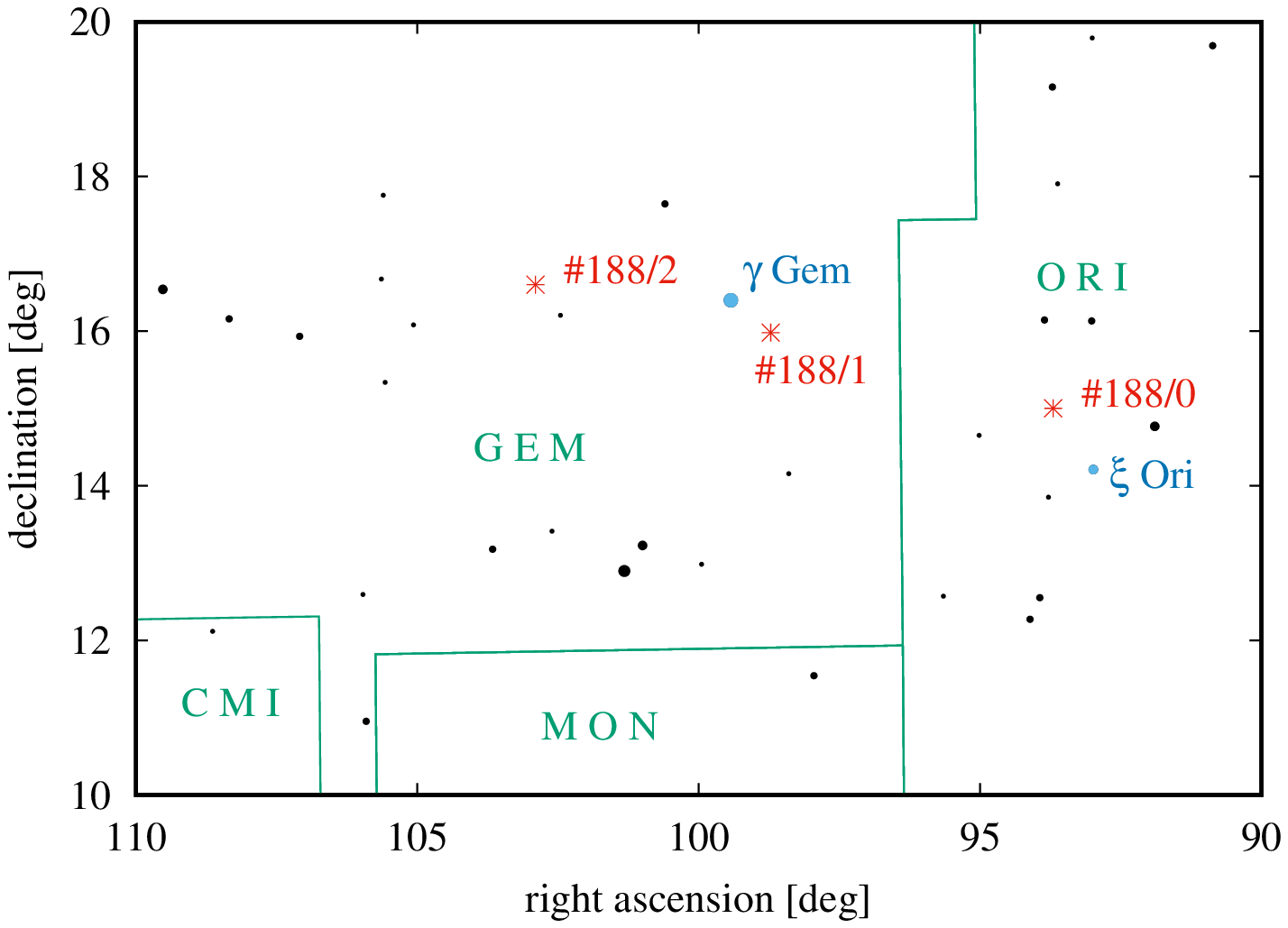}
    \caption{The positions of the radiants of three solutions of shower 188/XRI nu-Orionids (red asterisks). Stars $\xi$-Orionis and $\gamma$-Geminorum are plotted with the cyan circles; the other stars with the black circles.}
    \label{fig:188radiants}
\end{figure}

{\bf 188/XRI Daytime xi-Orionids.} 
The values $DH_{Min} = 0.534$ and $D_{MI}=0.079$ clearly show that there is a misclassification among the solutions of this MSS. All solutions were obtained from radio observations. \cite{1964AuJPh..17..205N} did not provide the name of the identified stream, among $2101$ orbits he identified only 3 members of his stream. 
Similarly for 188/XRI/01, the name of the shower is not provided in  \citep{1967SCoA...11..183K}; the name of the Daytime xi-Orionids is given in \citep{2008Icar..195..317B}. The positions of the radians for these solutions are shown in Figure~\ref{fig:188radiants}. Only the radiant given by Nillson lies in the constellation Orion, the other two lie in the constellation Gemini.

In Tables~\ref{tab:MSS-1} and \ref{tab:DH}, we see that the parameters of the 188/XRI/01 solution differ from the others. This solution was recently suggested to correspond to another SSS shower, 1211/SFG, from the Working list \citep{2024book...Jen}, though this association was not confirmed by the methods we used. 

Excluding solution 188/01 makes $DH_{Min}=0.268$ and $DH_{MI}=0.036$, which still means that we are not dealing with duplicates here. In the absence of access to the source data that the authors of these solutions used to calculate the averaged parameter values, further consideration cannot take place. However, within the framework of the methodology used, all 188/XRI solutions should be treated as stand-alone showers.
Since 188/XRI is on the List of established showers, we propose to change its status and move all three solutions back to the Working list. Solution 188/00 will retain the codes and name Daytime xi-Orionids, while 188/01 and 188/02 will be assigned new names and codes.\\

\begin{figure}
    \centering
    \includegraphics[width=0.34\textwidth, angle=-90]{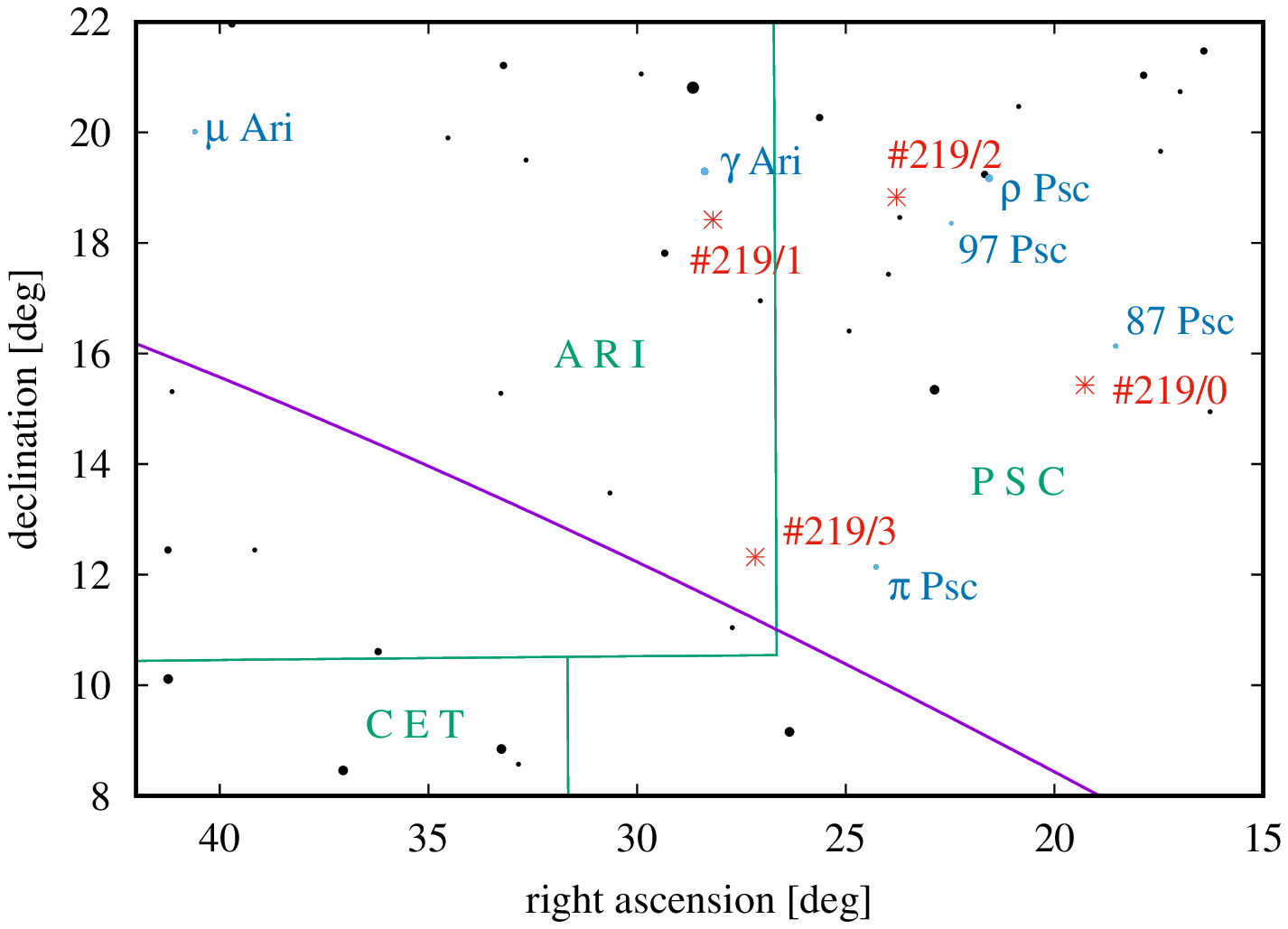}
    \caption{
    The positions of the radiants of four solutions of shower 219/PAU September mu-Arietids (red asterisks). Stars $\mu$-Arietis, $\gamma$-Arietis, $\rho$-Piscium, 97-Piscium, 87-Piscium, and $\pi$-Piscium are shown with the cyan circles, the other stars are depicted with the black circles. The purple curve indicates the ecliptic.}
    \label{fig:219radiants}
\end{figure}

{\bf 219/SAR September mu-Arietids.} 
The shower is represented by 
four solutions, all obtained from observations of radio meteors; one solution (219/SAR/00) comes from the work of \citep{1964AuJPh..17..205N}, the other three were given by \citep{1976Icar...27..265S}. As we read in the original papers, Nillson (in Table 4) named his shower as Gr.61.9.3; Sekanina considered the three solutions (219/01, 219/02 and 219/03) to be separate streams and named them (in Table 6) gamma-Arietids, rho-Piscis-Arietids and Arietids-Piscids, respectively. 
In Figure~\ref{fig:219radiants}, Nillson's 219/SAR/00 solution has a radiant located deep within the constellation Pisces near the star 87-Piscium. Sekanina's 219/SAR/01 solution with a radiant near the star $\gamma$-Arietis has a name in line with the old rules for naming meteoroid streams, \citep{2023NewAR..9601671J}.  The 219/SAR/02 solution has a radiant near the star 97-Piscium, roughly in the middle, between the stars $\rho$-Piscium and $\gamma$-Arietis. The third solution given by Sekanina, 219/SAR/03 has a radiant in the constellation Aries, but the closest star to this radiant is $\pi$-Piscium. Neither of these corresponds to the name September mu-Arietids given for these solutions in \cite{2006mspc.book.....J}. 

For this group we have $DH_{Min}=0.373$ and $DH_{MI}=0.099$, suggesting that we are dealing here with an inappropriate classification of duplicates. 
The DH-values of all pairs of orbits of 219/SAR range from $0.217$ to $0.580$, see Table~\ref{tab:DH}. Faced with an acceptable threshold value of $DH_{MI}=0.059$ for pairs in the studied set of orbits, this means that, according to the methodology used in this work, it is impossible to identify these four solutions as duplicates of the same meteoroid stream. We therefore recommend that they be  asigned the status of autonomous streams on the MDC Working list, with names given by their discoverers or new names together with stream codes set by the WG. Simultaneously, shower 219/SAR with a name September mu-Arietids will be moved to the List of removed showers. 
\subsection{MSS with two members}
\label{mss-2}
In this subsection, we discuss the MSS given in the third section of Table~\ref{tab:false-duplicates} for which we have only two solutions in our data sample. The threshold values of the DH functions involved here ($DH_{MI}=0.039$ for orbits with $i\leq 10^{\circ}$, $DH_{MI}=0.059$ for $i > 10^{\circ}$) do not depend on the cluster analysis method used. Therefore, as mentioned above, based on the $DH_P$ values from Table~\ref{tab:DH}, and the $DH_{MI}$ values of Table~\ref{tab:false-duplicates}, it can be categorically stated that we have a misclassification among these MSS.

However, using only $DH_{MI}$ and $DH_P$, we cannot conclude which of these solutions is a false duplicate. 
Therefore, in order to identify a solution that may be a false duplicate, we used additional information, such as: the inconsistency between geocentric and heliocentric shower data, the number of meteoroids identified in a given solution, and the difference in observation epochs of members of a given solution. We also took into account the averaged position of the shower radiants and the fact of a solution's association with a potentially parent comet. \\

{\bf 25/NOA Northern October delta-Arietids.}
%
The $DH_{Min}=0.210$ and $DH_{MI}=0.039$. The solutions 25/00 and 25/01 differ significantly in $V_g$,  ($\sim6\,$km$\,$s$^{-1}$) and in the inclination of the orbits ($\sim$$7^{\circ}$). The two solutions are separated by $\sim$$40$ years in observation epochs, which may account for the sizeable value $DH_{Min}$. 

For the 25/01 solution, using the two cluster analysis methods used in Paper I (single linking method and maximum sigma method), it was found to belong to a complex structure of 62 solutions, involving more than a dozen streams, among which the Northern and Southern Taurids dominate (17/NTA, 2/STA).
So, because the 25/01 radiant has positive ecliptic latitude, we propose that this solution be included in 17/NTA. 
An analogous suggestion is given in \citep{2006mspc.book.....J, 2024book...Jen}.

In Paper I, we did not find that 25/00 belongs to any of the streams collected in the MDC database; therefore, we propose that this solution be classified as an autonomous stream in the MDC. In the original article by \citep{1967SCoA...11..183K} in Tab 4.8, this shower was designated as No.161. Hence, we propose that the 25/00 solution remain in MDC while retaining the current codes and name.\\

{\bf 32/DLM  December Leonis Minorids.} 
Both solutions found internal data inconsistencies at the limit of our acceptable tolerance.
Based on the values of $DH_{Min}=0.321$ and $DH_{MI}=0.059$, both 32/DLM solutions should be considered autonomous.  
However, after inspecting the entire MDC catalog of stream data, it turned out that the 20/06/COM and 32/01/DLM solutions are exactly the same. This error was recognized earlier by \citep{2020eMetN...5...93K}.
So, after correcting this mistake, the problem of misclassification of stream 32/DLM was solved. The erroneous solution 32/01 with the appropriate annotation will be moved to the List of removed showers. \\

{\bf 40/ZCY zeta-Cygnids.} 
For this MSS we have $DH_{Min}=0.259$ and $DH_{MI}=0.059$. 
The time interval separating the observation epochs of the two solutions is about 40 years. 30 members of the 40/00 solution among nearly 20,000 radio observations identified \citep{1976Icar...27..265S}; 64 members of the 40/02 solution among $\sim$$110000$ video orbits identified \citep{2016Icar..266..331J}. The $e$ and $q$ parameters of the orbits of both solutions are very similar, but they differ significantly in the inclination of the orbits, by $\sim$$8^{\circ}$, making $DH_{in}=0.250$ in Table~\ref{tab:DH}. Also for this pair \citep{2020eMetN...5...93K} noticed the large difference in mean solar longitude, $\Delta \lambda_{S}=\sim$$13^{\circ}$.

We are unable to determine the reason for such a large difference in the inclination of the orbits. Hence, we propose that solution 40/00 named April Cygnids by its discoverer (\citep{1976Icar...27..265S}, Tab. VI, p. 276) be given autonomous status, and solution 40/02 be left unchanged in the MDC. 
However, we have a second option for this MSS. Taking into account the 40/04 solution recently provided by Shiba, we obtained $DH_{Min}=0.175$, which, together with the $D_{MI}=0.136$ value for the three orbits, allows to leave all 40/ZCY solutions in the MDC database without making changes. \\

{\bf 76/KAQ kappa-Aquariids.}   
The corresponding DH values are $DH_{Min}=0.142$ and $DH_{MI}=0.039$.
Both solutions, 76/KAQ/00 and 76/KAQ/01 were identified among the photographic orbits by \citep{1994PSS...42..151P}  and \citep{1971SCoA...12....1L}, respectively.  The numbers of identified members are very small $3$ and $4$; and the identifications were not made in independent samples of orbits --- there was some overlap between the samples of orbits tested in \citep{1994PSS...42..151P} and \citep{1971SCoA...12....1L}. 
The main orbital difference between the solutions 76/KAQ/00 and 76/KAQ/01 is due to the orientation of the orbits' apse lines. In Table~\ref{tab:DH}, the value  $DH_{\Pi}=0.121$, clearly brings the largest contribution to $DH_{Min}=0.142$. \citep{2020eMetN...5...93K} points out the large mutual angular distance of the radiants of the two solutions. In particular, he stated that the solutions are indistinguishable from the sporadic background.

As can be seen in Table~\ref{tab:MSS-1}, the parameters of the 76/KAQ/01 solution are not internally consistent. The reason for this is the value of the geocentric velocity of the meteoroid given in the original article \citep{1971SCoA...12....1L} as  $V_g=19\,$km$\,$s$^{-1}$. Lindblad also lists the catalogue numbers of the meteoroids included in this stream. Taking advantage of this special circumstance, we recalculated 
the arithmetic mean of the geocentric velocity of this solution, which yielded the value $V_g=14.23\,$km$\,$s$^{-1}$. And its application removed the inconsistency in the parameters of the 76/KAQ/01 stream. However, the corrected $V_g$ mean value does not affect the poor similarity of the orbits of the 76/KAQ/00 and 76/KAQ/01 solutions. 
  
\begin{figure}
    \centering 
\includegraphics[width=0.34\textwidth, angle=-90]{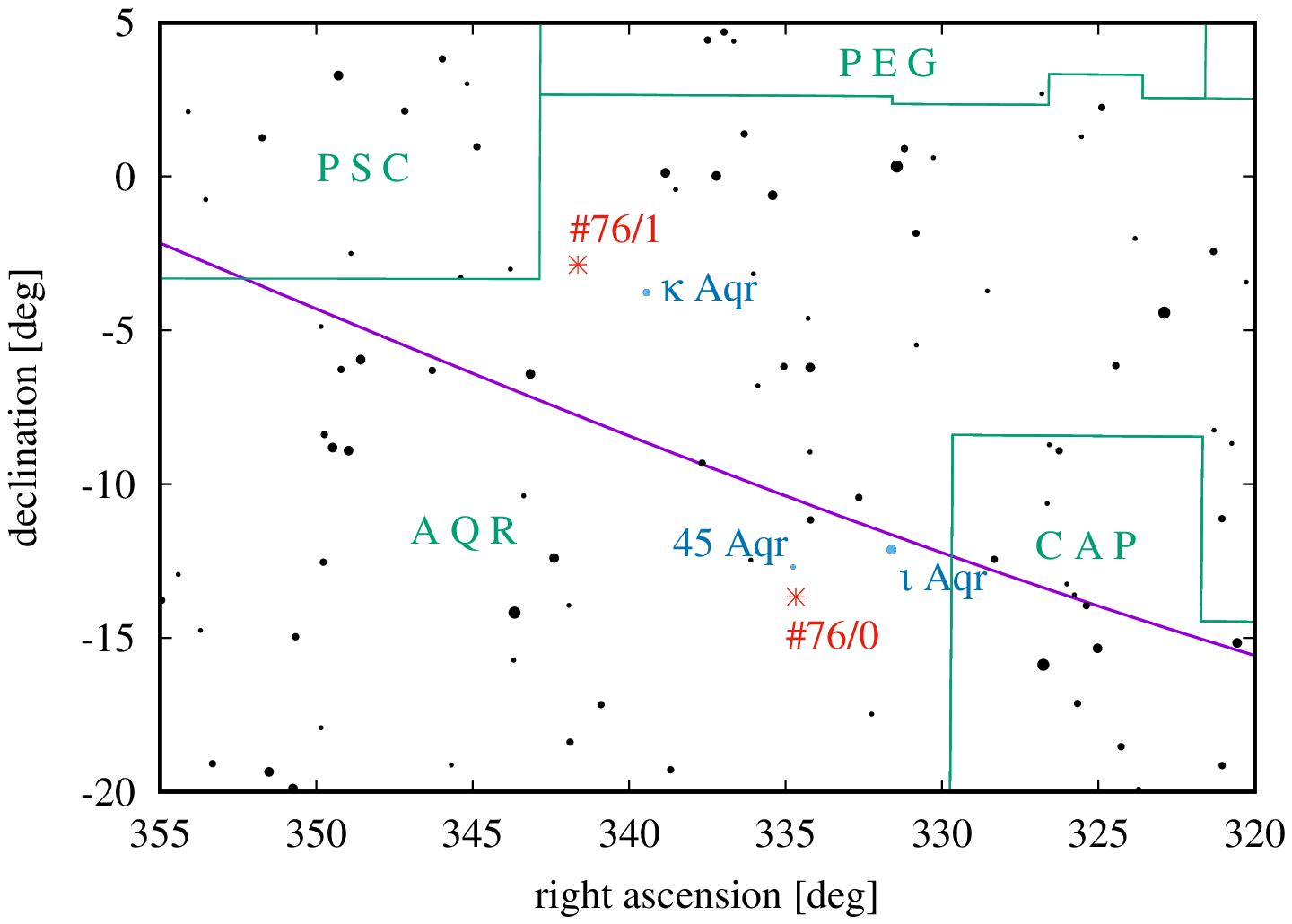}
    \caption{
    The positions of the radiants of two solutions of shower 76/KAQ (red asterisks). The closest star to the 76/KAQ/00 solution is the 45-Aquarii and the 76/KAQ/01 is the $\kappa$-Aquarii (cyan full circles). The violet curve shows the ecliptic. The radiants of the solutions are located on both, northern and southern, ecliptic celestial hemispheres.  }
    \label{fig:76KAQ}
\end{figure}

Moreover, as stated in \citep{2023P&SS..23505737N} and as can also be seen in Table~\ref{tab:MSS-1}, both solutions involve two branches of the 76/KAQ stream --- the ecliptic latitudes of their radians differ in sign, see Figure~\ref{fig:76KAQ}. This means that their status in the MDC will have to be changed. 
In  \citep{1971SCoA...12....1L} and \citep{1994PSS...42..151P} these streams were named kappa-Aquariids and September iota-Aquariids, respectively. We propose splitting the two solutions in the MDC and, in accordance with the suggestion by \citep{2023P&SS..23505737N}, renaming solution 76/00 to September iota-Aquariids, as originally
proposed by Porub\v{c}an and Gavajdov\'{a} and assigning it a new numerical and letter code. The solution 76/01 would retain the current shower name kappa-Aquariids and codes 76/KAQ.\\

{\bf 88/ODR omicron-Draconids.}  
The $DH_{Min}=0.362$ and $DH_{MI}=0.059$ clearly indicate the misclassification of one of the solutions. 14 members of 88/ODR/00 were identified among radio meteors by \citep{1976Icar...27..265S} and 63 members of 88/ODR/01 were identified among video data by \citep{2016Icar..266..331J}. 
The internal inconsistency of the 88/ODR/01 solution, see Table~\ref{tab:MSS-1}, concerns the longitude of the ascending node of the orbit and is at the limit of the tolerance adopted in \citep{2023P&SS..23505737N}. Probably it is due to the fact that \citep{2016Icar..266..331J} reports medians as mean values of stream parameters. 

The geocentric velocities of the two 88/ODR solutions differ by about $10\,$km$\,$s$^{-1}$ which is the reason of the large components $DH_e=0.131$, $DH_{in}=0.3$ and $DH_{\Pi}=0.156$, see Table~\ref{tab:DH}. Taking this into account, and because of the very large value of $DH_{Min}=0.362$, we propose that both solutions be given the status of autonomous streams. 
Sekanina named his stream $o$-Draconids, therefore we keep the old name/number/code of the shower and assign a new name/number/code only to the solution 88/01.
\\

\begin{figure}
    \centering 
\includegraphics[width=0.34\textwidth, angle=-90]{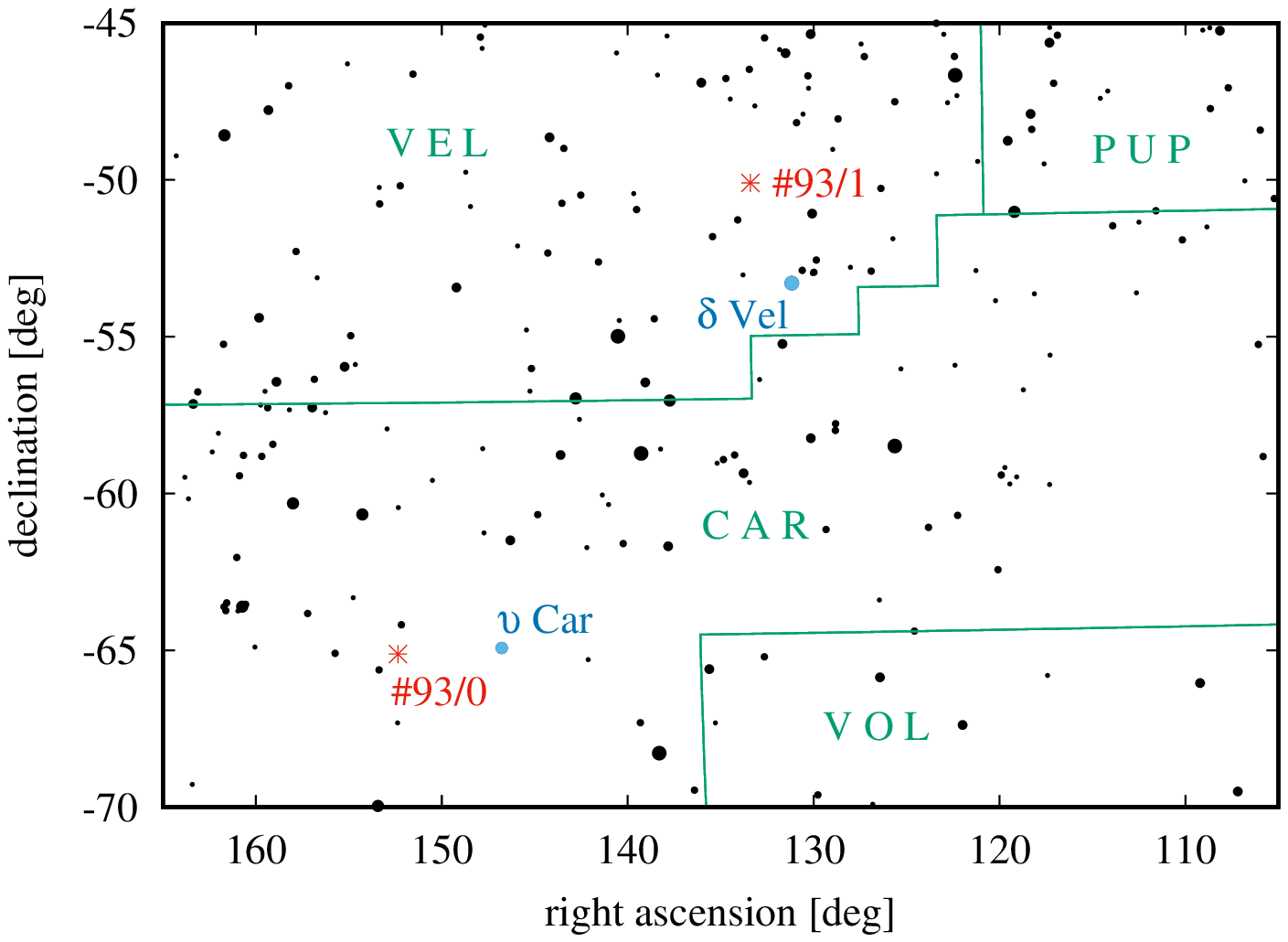}
    \caption{
    The positions of the radiants of two solutions of shower 93/VEL (red asterisks). The closest star to the 93/VEL/00 solution is the upsilon Carinae and to the the 92/VEL/01 is the $\delta$-Velorum (cyan full circles).}
    \label{fig:93radiants}
\end{figure}

{\bf 93/VEL Puppid-Velid II Complex.} 
In Table 1 of \cite{1975AuJPh..28..591G} these streams were designed code 2.13 and 2.09, respectively, they were identified among 1667 radio meteoroid orbits. In \citep{2006mspc.book.....J}, they are considered components of Puppids-Velids II. In view of the decidedly large value of $DH_{Min}=0.412$, for this MSS, which consists of contributions in both eccentricity, inclination and orientation of the lines of the apses, $DH_e=0.33$, $DH_{in}=0.213$, $DH_{Pi}=0.118$, respectively, the 93/VEL/00 and 93/VEL/01 solutions should be standalone streams in the MDC.
There are also large differences in the geocentric parameters of these solutions. The meteors of these  showers were observed in the same month of the same year, and yet the difference in the ecliptic longitudes of their radiants in the rotating reference system is nearly 40 degrees! Undoubtedly, this has some relation to the large difference in the position of the radiants of the two solutions, \citep{2020eMetN...5...93K} noted that the radiants differ by 19 degrees in the right ascension and by 15 degrees in the declination; see Figure~\ref{fig:93radiants}.  
Hence, we believe that the 93/00 and 93/01 solutions should be standalone streams in the MDC. Specifically, solution 93/VEL/01 with the radiant in the constellation Vela, near the border with the constellation Puppis, should remain unchanged in the MDC, except for omitting the Roman enumeration II and world ``Complex'' in its name (it is a shower, not a complex) 
and solution 93/VEL/00 should obtain a new IAU No., code, and name; its radiant is situated in constellation Carina. \\

{\bf 100/XSA Daytime xi-Sagittariids.} 
The values of $DH_{Min}=0.209$ and $DH_{MI}=0.039$ point to the misclassification of solutions of this MSS. Both solutions, 100/00 and 100/01, \citep{1976Icar...27..265S} identified among $\sim20000$ orbits of meteoroids observed from December 1968 to December 1969 (so-called synotpic-year sample). 
In Sekanina's work, these solutions are separate streams, their discoverer gave them the names January Sagittariids and xi-Sagittariids (Tab. VI, p. 274), respectively. 
Their fusion as representatives of a single shower was introduced in \citep{2006mspc.book.....J}.
The mean orbits of these solutions differ mainly in $q$ and ${\Pi}$ (see Table~\ref{tab:DH}), the respective values of $DH_q=0.147$ and $DH_{\Pi}=0.131$ clearly exceed the acceptable threshold of orbital similarity $DH_{MI}=0.039$ for orbits with low inclination. Thus,  we suggest that 
the two solutions be treated as standalone showers on the MDC Working list, with solution 100/01 retaining is current IAU MDC designation. For solution 100/00, we suggest assigning a new code and restoring the name as proposed by the discoverer.

At the same time, we propose moving 100/XSA to the MDC Working list. Introducing an established status to this shower has clearly failed. 
As in all cases, the final decision will be made by the Working Group on Meteor Shower Nomenclature.\\

{\bf 105/OCN  Centaurid I Complex.} 
Both solutions were identified in radio orbit samples, \citep{1975AuJPh..28..591G} searched 1667 orbits, \citep{1964PhDT........28N}  2101 orbits. \citep{1975AuJPh..28..591G} named their stream as Carinids. How Nilsson labeled his solution we could not determine. 
In \citep{1988msdc.book.....K} book, the stream was called eta-Carinids. 
However, in view of the small number of members of these solutions (3 members each), their reality is supported by the fact that they were discovered in two observation campaigns. The first in January 1961 and the second in January 1969. It is worth mentioning that the 150/00 solution was obtained from meteor observations observed from January 21-23, 1969. In January, outside this interval, the radar equipment was not working. 
This means that the meteoroids observed over the three days will have similar values for the length of ascending nodes. Thus, here we have a reduction, from four to three dimensions, of the issue of similarity of the determined orbits.
\begin{figure}
    \centering 
\includegraphics[width=0.34\textwidth, angle=-90]{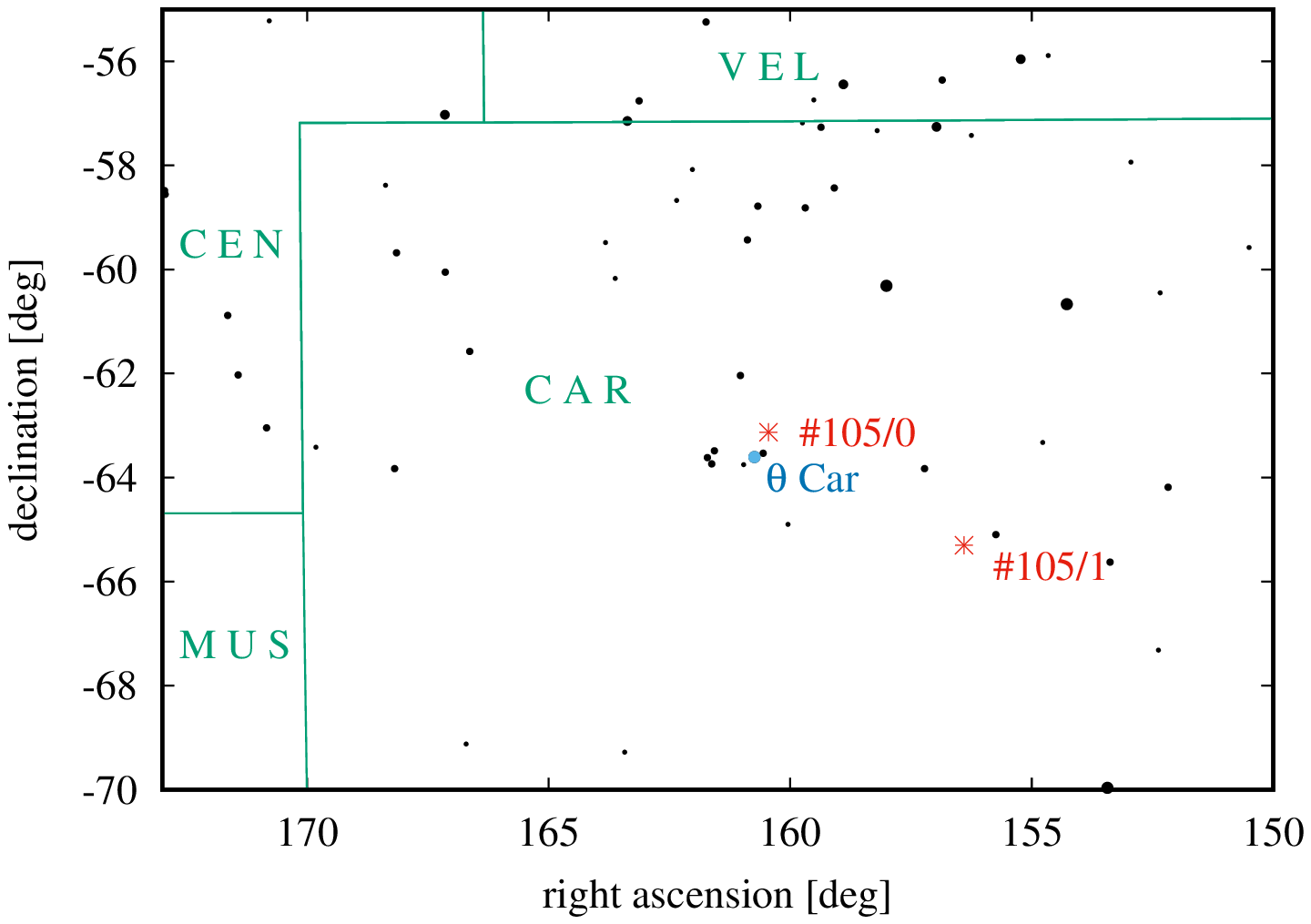}
    \caption{
    The radiants of two solutions of shower 105/OCN (red asterisks). The closest star to both radians is $\theta$-Carinae.} 
    \label{fig:105radiants}
\end{figure}

The flag visible in the first column of Table~\ref{tab:MSS-1} indicating that the data of the 105/OCN/01 solution are inconsistent is due to the erroneously reported $\lambda_s=323.4^{\circ}$ in the IAU MDC. In view of the large value of the orbit inclination of this stream, this value should be  $\lambda_s=299.7^{\circ}$, which is almost equal to the value of the longitude of the ascending node of the orbit, increased by $180^{\circ}$. 
The erroneous value of $\lambda_S$ does not enter into the calculated DH value. For the pair 105/00 and 105/01, it is $DH_{Min}=0.134$, which, in view of the threshold value of $DH_{MI}=0.059$, may indicate that these solutions are unjustifiably considered to represent the same meteoroid stream. 

The abundance of both solutions, N=3, is therefore very low which, in the absence of other orbital data of this shower in MDC, may indicate its spurious nature. Admittedly (in a sample of orbits that is a combination of those  used by Nielsson and Gartrell, Elford), \citep{1999md98.conf..307J} identified 6 Carinids with similar parameters to the solutions given by Nielsson and Gartrell and Elford. However, the solution obtained by \citep{1999md98.conf..307J} cannot be considered an independent confirmation of solutions 105/00 and 105/01. 
Hence, in view of the above, and due to quite large value of $DH_{Min}$$=$$0.134$ for this MSS, we believe that we have clear basis for considering both its solutions as autonomous. 

We believe the best course of action in this case is to move the 105/OCN Centaurid I Complex to the List of Removed Showers while keeping solution 105/00 on the Working List with a new IAU MDC code and a new name based on its radiant position. The radiant is located near the same star, theta Carinae; see Figure \ref{fig:105radiants}. 

At the same time, both solutions, under the current shower name and code, will be moved to the List of removed showers.\\ 

\begin{figure}
    \centering 
\includegraphics[width=0.34\textwidth, angle=-90]{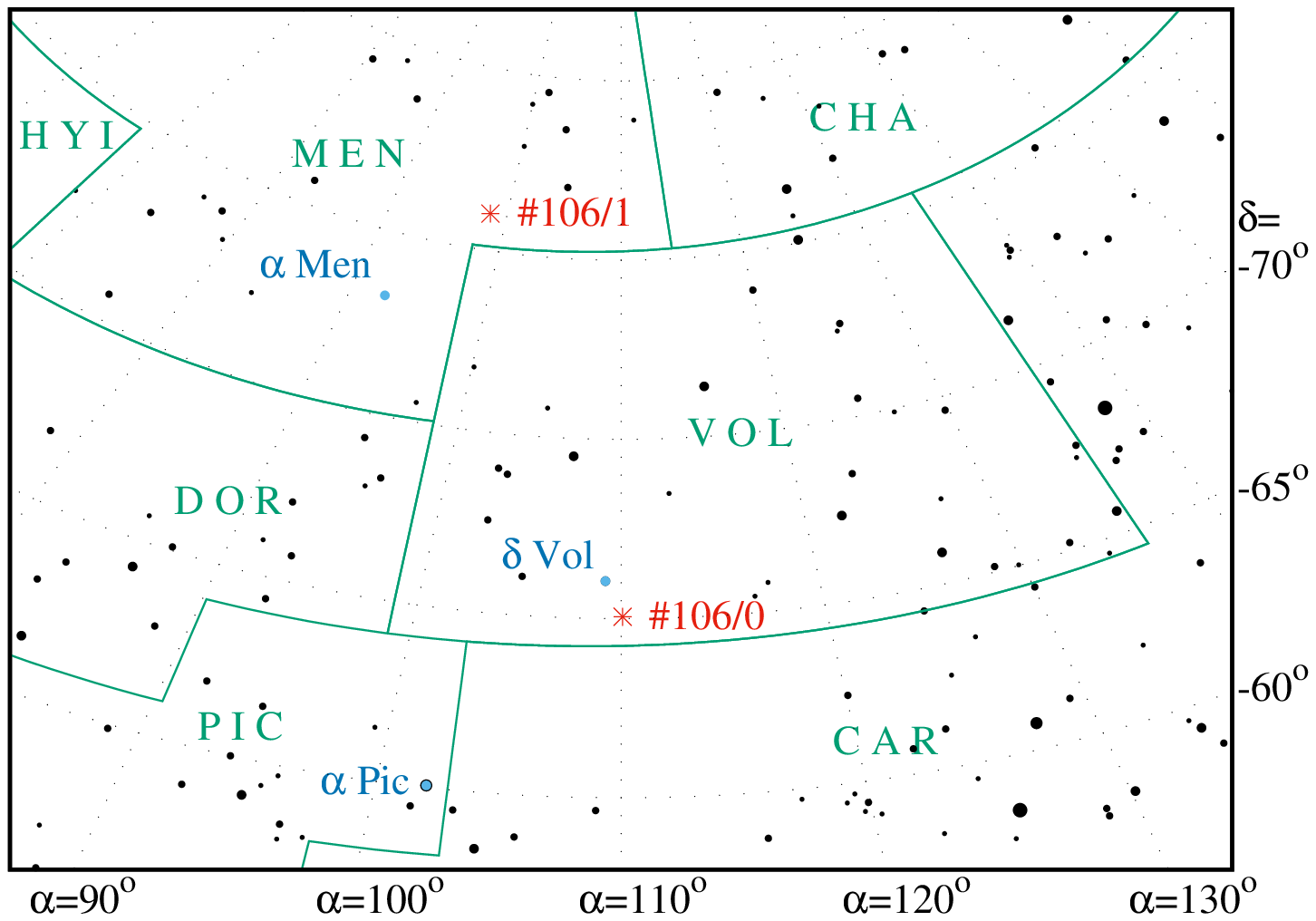}
    \caption{
    The radiants of two solutions of shower 106/API (red asterisks). The radiant of 106/API/00 shower is close to star $\delta$-Volantis, the radiant of 106/API/01 is close to the star $\alpha$-Mensae. }
    \label{fig:106radiants}
\end{figure}

{\bf 106/API alpha-Pictorids.} 
The values of $DH_{Min}=0.329$ and $DH_{MI}=0.059$ indicate the misclassification of these solutions as belonging to the same shower. 
The dominant contributors to such a high $DH_{Min}$ value are the differences in the eccentricity and orientation of the two orbits, $DH_e=0.22$ and $DH_{\Pi}=0.244$, respectively. \citep{2020eMetN...5...93K} stated an angular distance of radiant larger than $10^{\circ}$, see Fig. \ref{fig:106radiants}. Such difference cannot be justified by the daytime motion of the radiant because meteors of both solutions were observed within a few days of February in the same year 1969.
The streams were identified by \citep{1975AuJPh..28..591G} radio sample of 1667 orbits. Streams 106/API/00 and 106/API/01 included a small number of 3 and 5 members, respectively.
In Table 1 of Gartrell and Elford publication, the authors only proposed designations for their findings, namely 2.10 and 2.11.

Regardless of the questionable nature of the 106/API solutions, we propose that they remain on the MDC working list as standalone solutions. For both solutions, we propose new IAU MDC codes and names in accordance with the old nomenclature rules (see Fig. \ref{fig:106radiants}). Alternatively, one of them could retain its current designation. The WG will finalize the decision.
 \\

{\bf 108/BTU beta-Tucanids.} 
The values of $DH_{Min}=0.344$ and $DH_{MI}=0.059$ indicate the misclassification of at least one of the solutions 108/00 or 108/01.
Their discoverers, \citep{1975AuJPh..28..591G}, proposed 3.05 and 3.04 as designations for their findings. Ten  and eleven  members of these streams were identified among 1667 radio orbits. 
Both solutions have similar orbits, except for the eccentricities, in Table~\ref{tab:DH} we have D$H_e=0.34$, the other contributions have irrelevant influence. 
Apart from the two solutions mentioned, no information has been provided to the IAU MDC on further 
identification of this stream. 
However, in the literature, an outburst observed by the SAAMER radar in 2020 was claimed by 
\citep{2021Bruzzone} match two showers, 
108/BTU and 130/DME. In 2021, \citep{2021Jenniskens} reported the detection of 29 meteors belonging to 108/BTU.
\citep{2020eMetN...5...93K} also found an association of 108/BTU and 130/DME. However, shower 130/DME, the delta-Mensids, was moved to the list of removed showers because of the lack of complete data.

\begin{figure}
    \centering 
\includegraphics[width=0.34\textwidth, angle=-90]{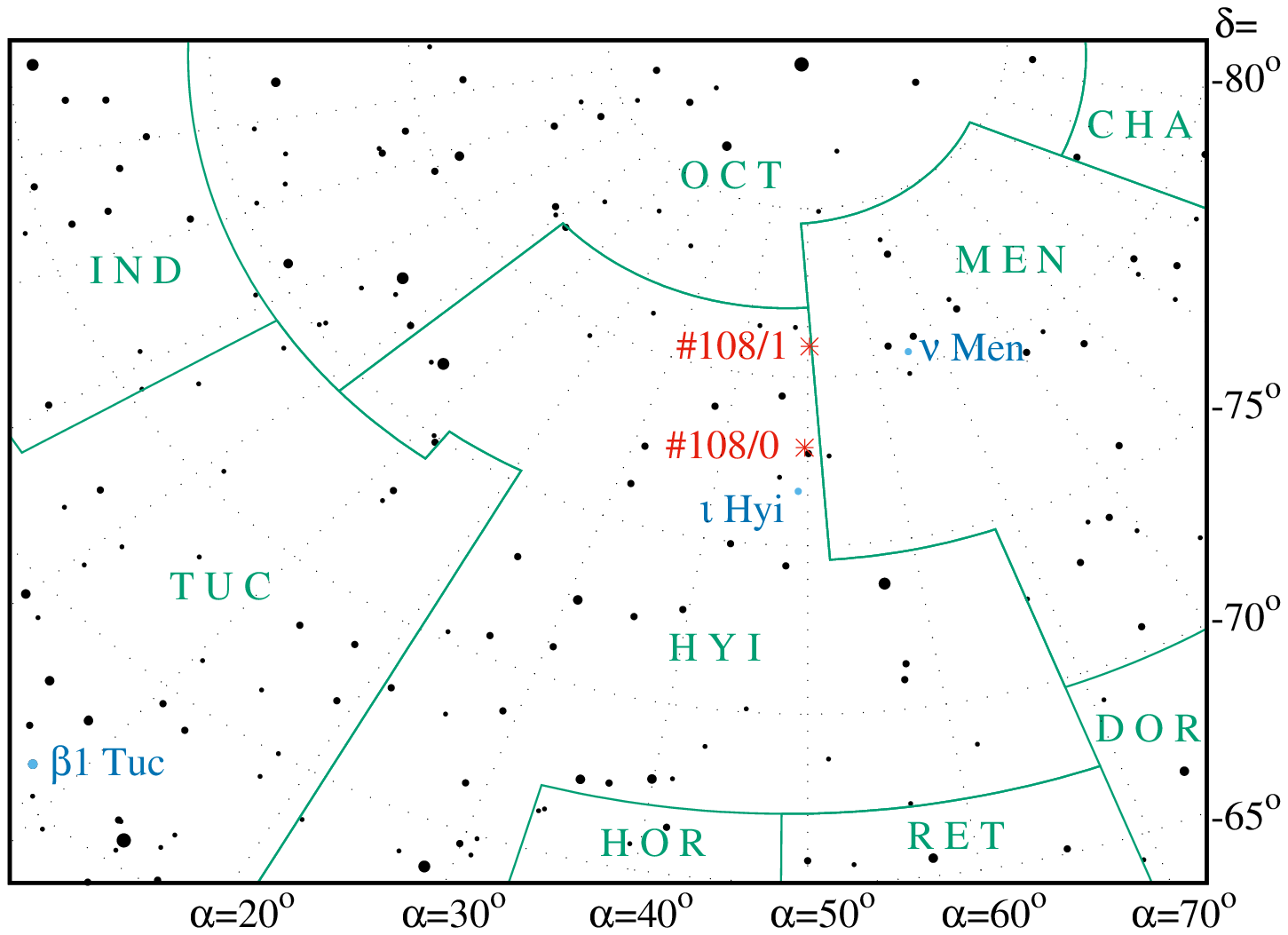}
    \caption{
    The radiants of two solutions of shower 108/BTU (red asterisks). The radiant of 108/BTU/00 shower is close to star $\iota$-Hydri, the radiant of 108/BTU/01 is close to the star $\nu$-Mensae.  }
    \label{fig:108radiants}
\end{figure}

Despite the similarity of some parameters, the two solutions of this MSS in the face of a significant difference, $\Delta V_g=4.2\,$km$\,$s$^{-1}$, in geocentric velocity and eccentricity value, $\Delta e=0.34$, should be considered autonomous. 
We propose, 
due to its use in the literature, keeping the codes 108/BTU and name beta-Tucanids for one of the solutions, despite the fact that it was not assigned to the shower in accordance with the nomenclature rules. The other solution will be given new codes and name corresponding to its radiant,
see Figure~\ref{fig:108radiants}.
The decision will require thorough discussion within the WG.\\

{\bf 113/SDL Southern delta-Leonids.} 
The values of $DH_{Min}=0.172$ and $DH_{MI}=0.039$ indicate the misclassification of at least one of the solutions. Solution 113/SDL/00 was named, in the source publication by \citep{1989JIMO...17..242T} as $\alpha$-Cnc (c), alpha-Cancrids. 
According to the test we performed, the parameters of this solution are not internally consistent, which was also found by \citep{2016JIMO...44..151K}. Koseki suggests that the declination value of the radiant given by Terentjeva is incorrect and should be $\delta=7.8^{\circ}$. With such a declination value of the radiant, its ecliptic latitude would have a negative value, which would remove the second problem of having together northern and southern branches in a single shower. The radians of both solutions 113/00/SDL and 113/01/SLD would then belong to the southern branch of this stream. 
However, such a correction does not change the orbital similarity of the two solutions. Their orbits differ mainly in the distance of the perihelion and the orientation of the apses line, $DH_q=0.114$, $DH_{Pi}=0.122$.  
Solution 113/01/SDL \citep{1976Icar...27..265S} was identified among 1968-69 synoptic-year sample of $\sim 20000$ radio meteors. \citep{1989JIMO...17..242T} identified her solution among $554$ fireballs observed by the US and Canada fireball networks during 1963-1984. 
\begin{figure}
    \centering 
\includegraphics[width=0.34\textwidth, angle=-90]{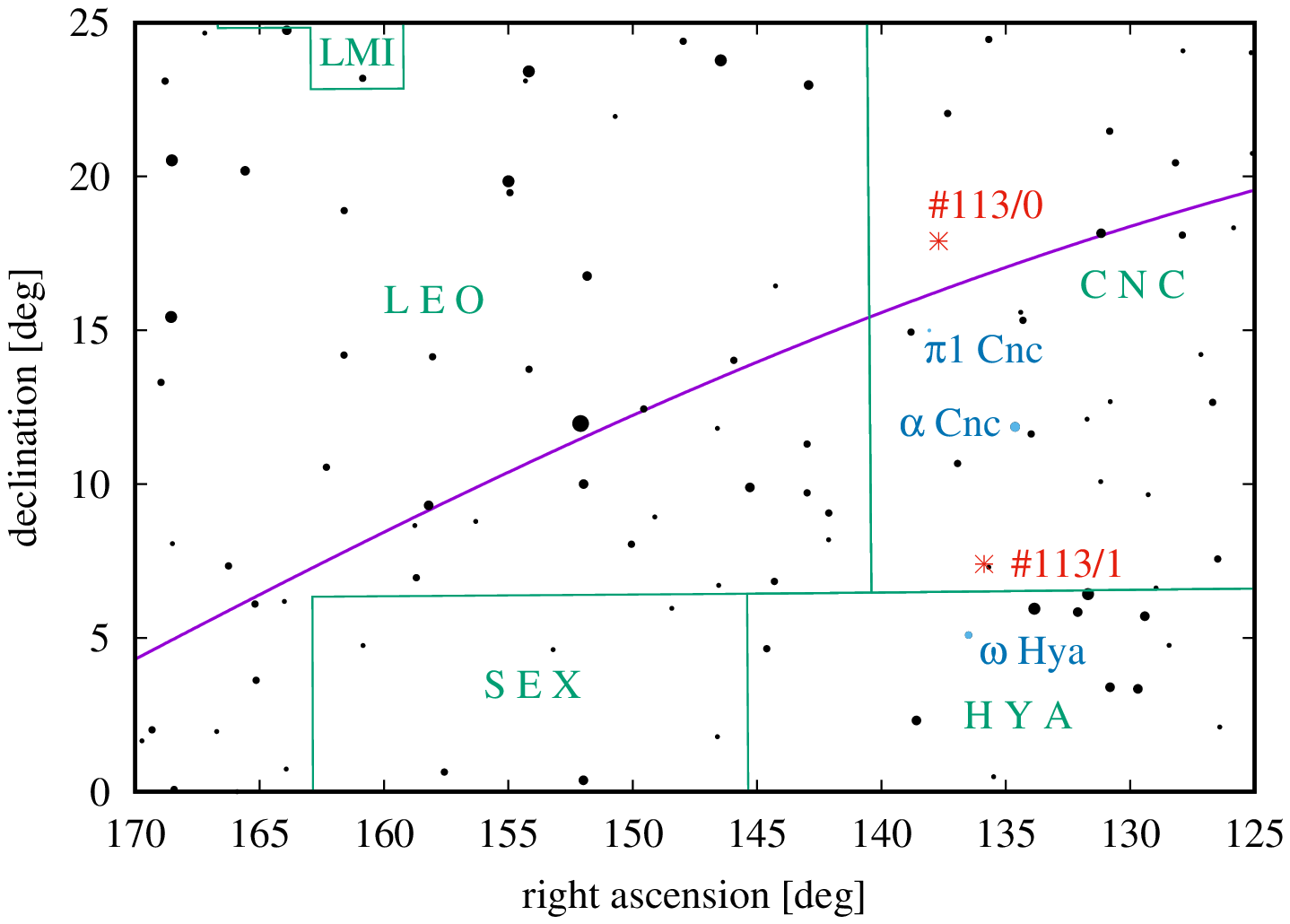}
    \caption{
    The radiants of two solutions of shower 113/SDL (red asterisks). The radiant of 113/SDL/00 shower is close to star $\pi$1-Cancri, the radiant of 113/SDL/01 is close to the star $\omega$-Hydrae. The purple curve is the ecliptic.}
    \label{fig:113radiants}
\end{figure}

In view of the significant value of $DH_{Min}$$=$$0.172$, we believe that both streams should be treated as autonomous in the IAU MDC.  Since the original name from Terentieva's work, is already in use by another stream, 266/00/ACC, alpha-Cancrids, identified by \citep{1994PSS...42..151P}, we propose that this solution be named pi1-Cancrids (see Figure~\ref{fig:113radiants}), with new codes assigned. Since none of the two solutions has a radiant
consistent with the shower designation, for the solution given by Sekanina, we suggest assigning a new IAU MDC desigantion.
\\

{\bf 118/GNO gamma-Normids.} 
The extremely large value of $DH_{Min}=0.798$ in comparison to $DH_{MI}=0.059$ unquestionably indicates the misclassification of at least one of the solutions for this stream. Both analyzed solutions were identified by \citep{1975AuJPh..28..591G}, determined among a sample of $1667$ radio meteors observed from December 1968 to October 1969. The numbers of members of these solutions are low, 118/00/GNO was obtained from averaging 3 orbits, 118/01/GNO from 6 orbits. In the second solution, we found inconsistency among the geo-helio data, albeit at the borderline of the applied maximum-error tolerance.  
In Table~1 of the paper~\citep{1975AuJPh..28..591G}, the authors designate these streams with the codes '3.14' and '3.15', and later by \citep{2006mspc.book.....J} these solutions were proposed as duplicates of the 118/GNO stream.

In the MDC database, we obtained two other submissions for this stream, provided by \citep{1994A&A...287..990J} and \citep{2014JIMO...42...68M}, in which, however, the full set of orbital elements is not given. Therefore, the similarity of their orbits cannot be verified. Moreover, as stated in \citep{2024MNRAS.535.3661D}, on the $inc$ versus $1/Q$ plane (inclination of the orbit versus the inverse of the aphelion distance), the points representing this stream are in an area where there are no currently known comets or NEAs. These facts support the claim that both solutions 118/00 and 118/01 could have been identified as meteoroid streams by chance only. The solutions of 118/GNO were regarded as an unreasonable combination of orbits by \citep{2020eMetN...5...93K}. Thus, the most appropriate recommendation would be to exclude solutions 118/00 and 118/01 from the Working list of showers, and place them on the List of removed shower's data. \\

{\bf 124/SVI Southern March Virginids. } 
The $DH_{Min}=0.226$ and $DH_{MI}=0.039$ values for this stream indicate the misclassification of at least one of the solutions. Both solutions were identified among the radio orbits of meteors observed during the 1961/65, solution 124/SVI/00, and during 1959/60, solution 124/SVI/01. The first one identified by \citep{1973Icar...18..253S} includes 13 members and the second one identified by \citep{1967SCoA...11..183K} includes only 5 members. 
Their orbits differ mainly in the inclination and orientation of the apses line: $DH_{in}=0.105$ and $DH_{\Pi}=0.177$ respectively, see Table~\ref{tab:DH}.
We suggest that these solutions should not be considered as representations of the same stream. The stream he discovered, Sekanina called  Southern eta-Virginids, so we would prefer to keep the current name ``Southern March Virginids'' and codes for this solution: 124/SVI/00.  

On the other hand, the stream identified by \citep{1967SCoA...11..183K} proved to be similar to 223/GVI/00, Daytime gamma-Virginids, also discovered by \citep{1976Icar...27..265S} among the radio data. Indeed, we found the orbital similarity of this pair, $DH=0.04$,  is just slightly over the set limit of 0.039, so both solutions, 124/SVI/01 and 223/GVI/00, can be treated as representing the same stream. 

We propose to include 124/SVI/01 as the second solution of the 223/GVI stream. 
The 'old' 124/SVI/01 solution should be added to the List of removed showers to ensure all solutions remain traceable and identifiable.\\ 

\begin{figure}
    \centering 
\includegraphics[width=0.34\textwidth, angle=-90]{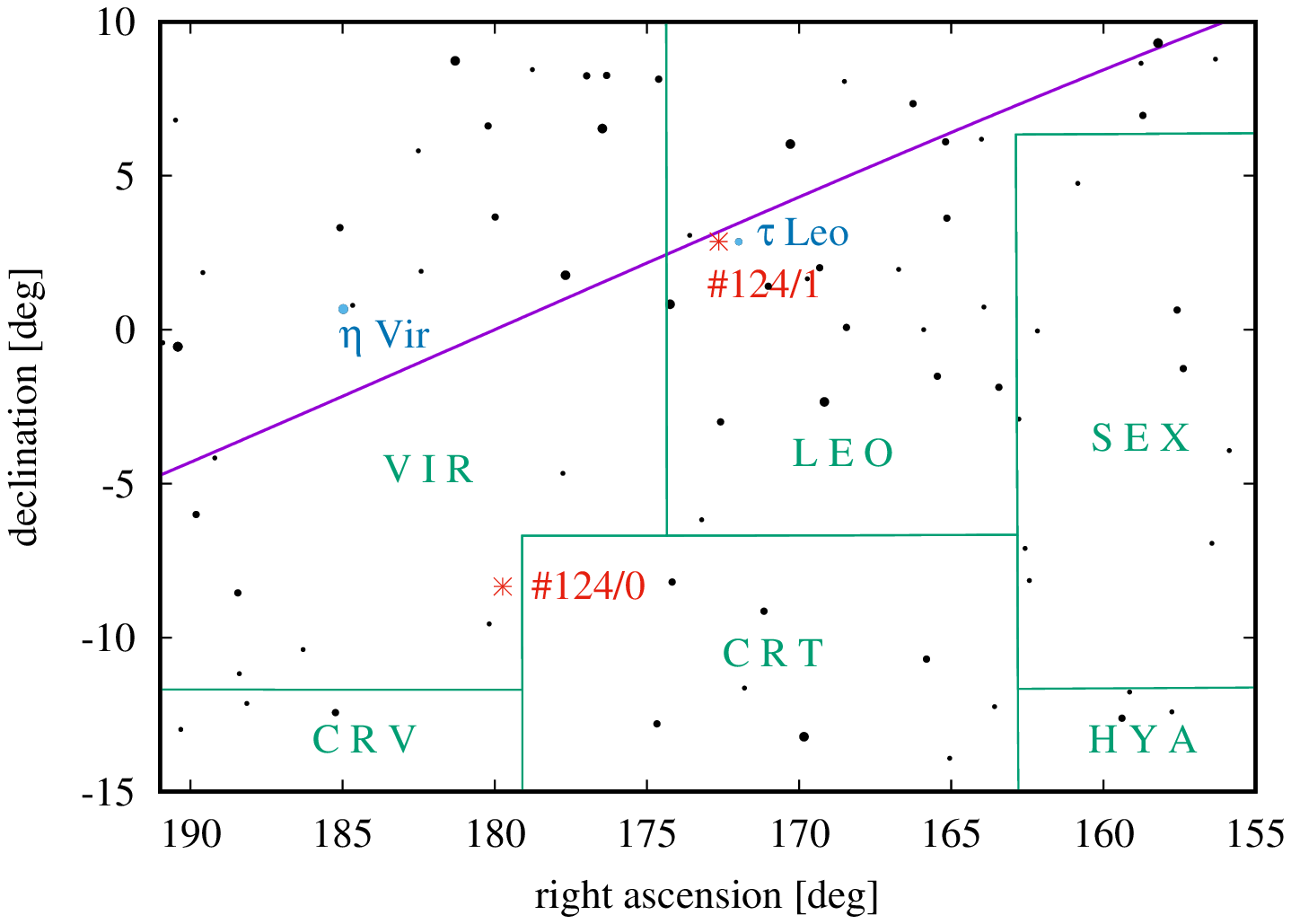}
    \caption{
    The radiants of two solutions of shower 124/SVI (red asterisks). The radiant of 124/00/SVI shower is close to star $\eta$-Virginis, the radiant of 124/01/SVI is close to the star $\tau$-Leonis.}
    \label{fig:124radiants}
\end{figure}

{\bf 128/MKA Daytime kappa-Aquariids.} 
The values of $DH_{Min}=0.315$ and $DH_{MI}=0.039$ for this stream indicate that the misclassification also occurred in this case. Both solutions were identified among radio meteors observed with radar at the Adeleide station. \citep{1975AuJPh..28..591G} among 1667 orbits identified 7 meteoroids, solution 128/MKA/00. \citep{1964AuJPh..17..205N} among 2101 orbits identified only 3 members of the 128/MKA/01 stream. The parameters reported by Nilsson are internally inconsistent but at the limit of our accepted error tolerance. The orbits of the 128/MKA differ significantly in the perihelion distance and the orientation of the apses line, $DH_q=0.25$, $DH_{\Pi}=0.189$.
\citep{1999md98.conf..307J}, in a total sample of 1667+2101 orbits, identified an alpha-Aquariids stream of 16 orbits that appears likely to contain members of the 128/MKA/00 and 128/MKA/01 solutions. 
In the MDC database, the values of the mean perihelia distances of the streams identified by Nilsson were necessarily calculated using values of the eccentricities and inverses of the semi-major axes. Using the uncertainties of these quantities given by Nielsson, in the case of jet 128/MKA/01, the perihelion distance could just as well have been $q=0.2653\,$au, instead of the $q=0.3\,$au used in our calculations. This, in turn, entails a decrease of $DH_{Min}=0.27$ and $DH_q=0.191$ for the 128/MKA. 
Unfortunately, we do not have the full information to proceed with this kind of calculation.  
We therefore have no way of confirming or rejecting the hypothesis that these solutions involve the same meteoroid stream. 
 
For these reasons, and due to the close mutual positions of the radiant of the solutions 128/00 and 128/01 and also that obtained by \citep{1999md98.conf..307J} ($RA= 339^{\circ}$, $DE=-7^{\circ}$) and by \citep{2010Icar..207...66B} ($RA =332^{\circ}$, $DE =-8.4^{\circ}$), 
we consider it appropriate to leave 128/00/ and 128/01 solutions in the MDC database without any changes. 

{\bf 133/PUM April psi-Ursae Majorids. } 
The $DH_{Min}=0.139$ and $DH_{MI}=0.039$ values for this stream indicate the misclassification of one of the solutions. 
Following \citep{2006mspc.book.....J}, two solutions of the 133/PUM given in the MDC are 133/00 identified by \citep{1989JIMO...17..242T} among 554 bolide orbits, and 133/01 identified among $\sim$$20000$ radio meteors by \citep{1976Icar...27..265S}. 
However, the discoverers of these streams, in their works, give other names to them: Terentjeva gave the name Lyn II (Lyncids II), Sekanina gave the name April Ursids. 
The mean orbits of these solutions differ mainly in the eccentricity and orientation of the apsidal lines: $DH_e=0.089$, $DH_{{\Pi}}=0.096$.  
However, these solutions differ significantly in the position of the meteor radiant, their right ascensions  differing by about $30^{\circ}$.  Also a large difference of $\lambda_{RSC}=20^{\circ}$, see Table A, occurs for the ecliptic longitudes of the radiants measured in the Sun-centered reference system. In this situation, the cause of these differences cannot be the diurnal motion of the radiant. Therefore, we consider that both solutions should be given autonomous status.    
\begin{figure}
    \centering
\includegraphics[width=0.34\textwidth, angle=-90]{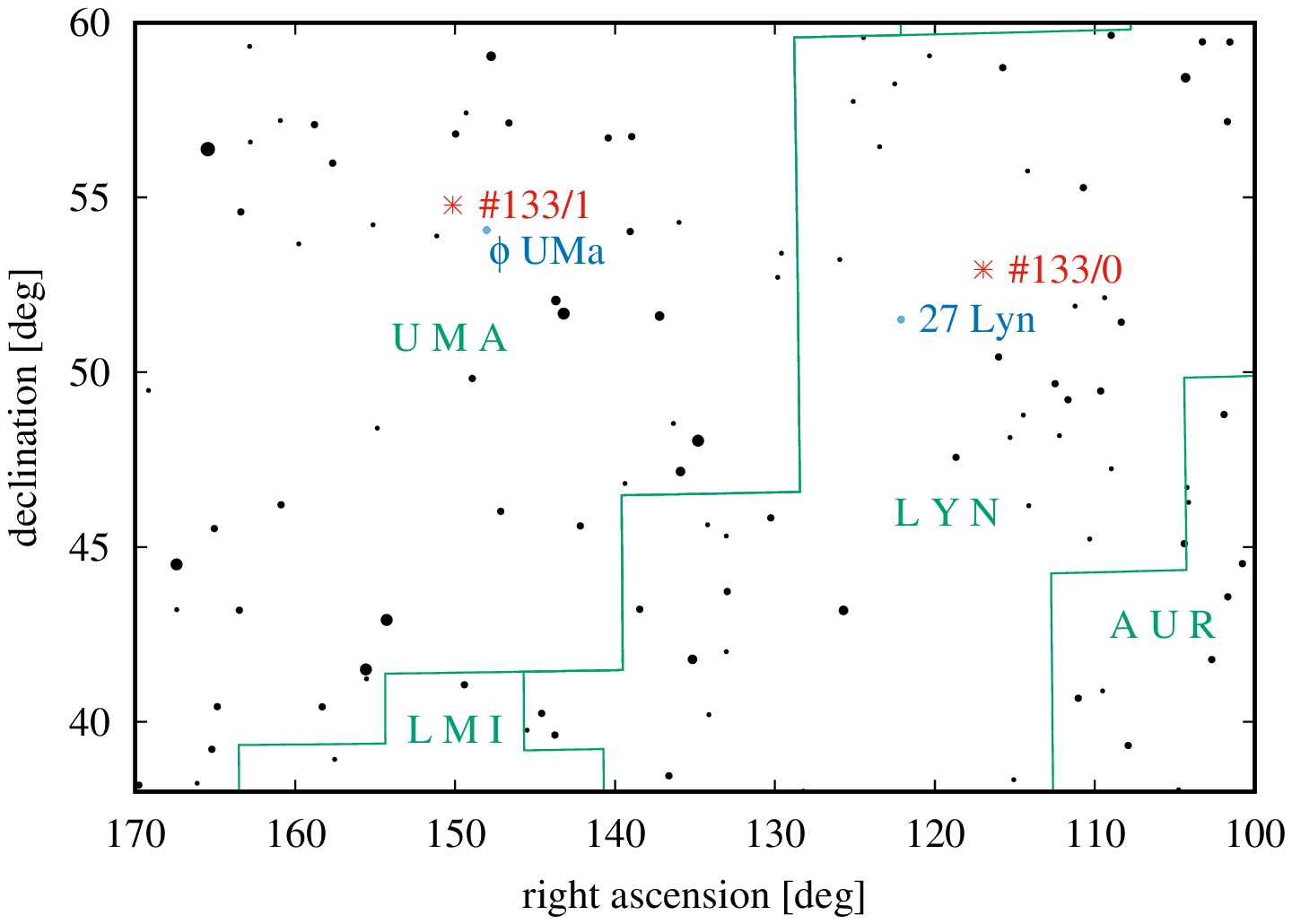}
    \caption{
    The radiants of solutions of 133/PUM shower (red asterisks). The radiant of 133/0 shower is located in the constellation Lynx, close to star 27-Lyncis. The radiant of 133/1, located in constellation Ursae Majoris is close to the star $\phi$-Ursae Maioris. }
    \label{fig:133radiants}
\end{figure}
%
Based on Figure~\ref{fig:133radiants}, we propose assigning a new name and codes for the stream discovered by Terentjeva, while keeping the name April phi-Ursae Majorids
for the Sekanina stream, 133/PUM/01.  
To ensure that the 133/PUM/00 solution 
remains traceable it will be additionally listed in the List of removed showers under 
its old designation.\\

{\bf 150/SOP Southern May Ophiuchids. } 
The values of $DH_{Min}=0.343$ and $DH_{MI}=0.039$ for this stream clearly indicate the misclassification of one of the solutions. Both solutions were identified among meteors observed with video techniques. \citep{2010MNRAS.404..867J}, among 231 orbits, 7 members of the 150/SOP/01 stream were found. They were observed in New Zealand campaign on 2002 May. \citep{2022JIMO...50...38S} identified 115 members of the 150/SOP/08 solution among 298687 meteors from SonotaCo network observed during 2007-2020. As we can see in Table B1, the solutions differ markedly only in the orientation of the $DH_{Pi}=0.329$ apsidal line. This may indicate, unknown for us, a systematic factor differentiating these solutions. Therefore, we propose to give them the status of autonomous streams in the IAU MDC. 

\begin{figure}
    \centering
\includegraphics[width=0.34\textwidth, angle=-90]{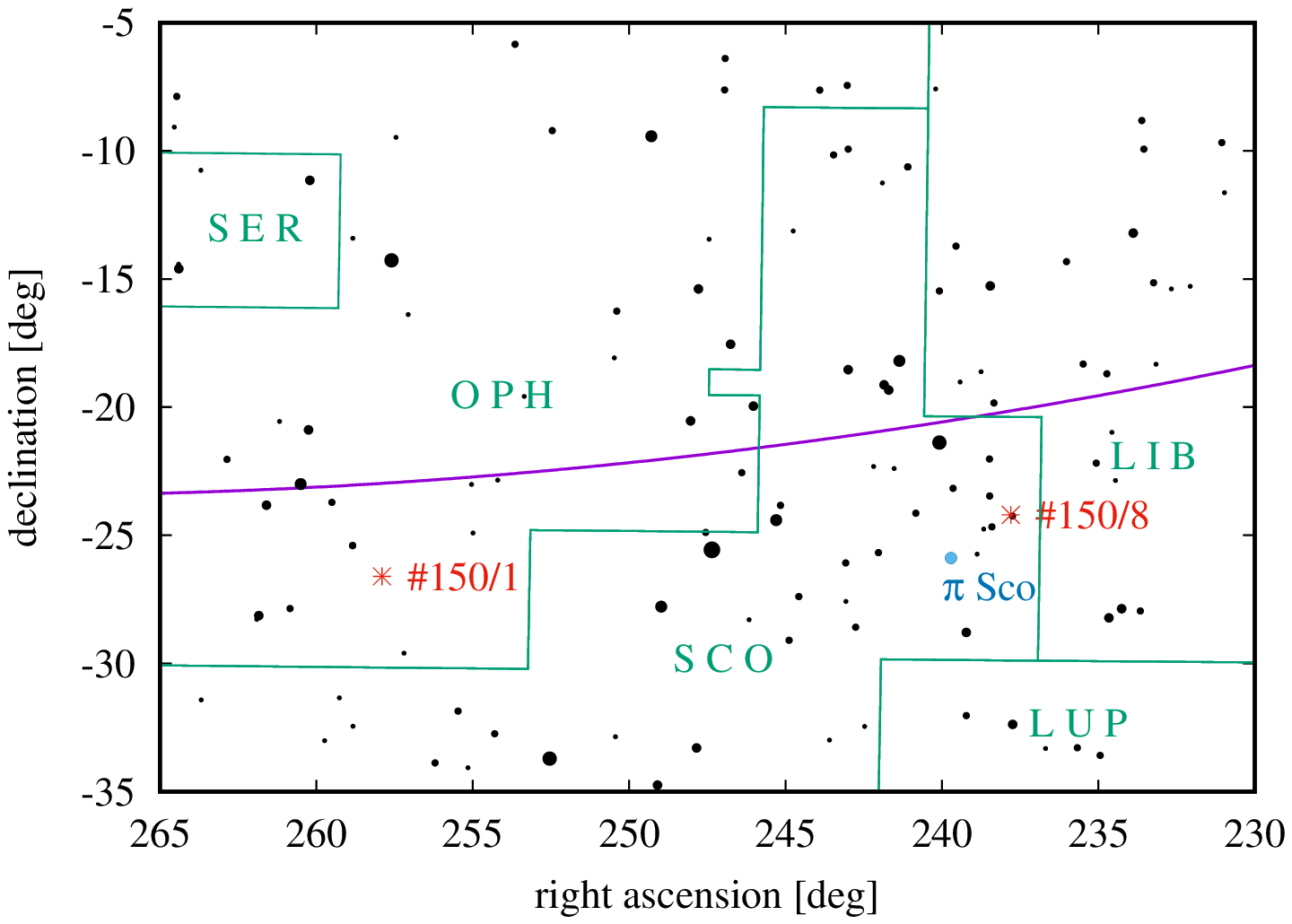}
    \caption{
    The radiants of solutions of 150/SOP shower (red asterisks). The radiant of 150/01/SOP shower is located in the constellation Ophiuchus. The radiant of 150/08/SOP, located in constellation Scorpius is close to the star $\pi$-Scorpii.} 
    \label{fig:150radiants}
\end{figure}

Since the radiant of the 150/SOP/01 solution is located in the constellation Ophiuchus (see Figure~\ref{fig:150radiants}) and the members of the stream were observed within 10 days in May, we propose to keep the codes and name of this solution. The members of the 150/SOP/08 solution were observed in different years (from 2008 to 2020) at different times of the year (the ecliptic longitudes of the Sun differed by 30 days), hence, their observed radians could take values from a relatively wide range. As we can see in Figure~\ref{fig:150radiants}, the mean radiant of this solution lies in the constellation Scorpius near the star pi-Scorpii. Due to its recent delivery to the MDC, we propose 
naming this solution in accordance with the new nomenclature rules (valid at the time of this submission), namely M2022-W1.\\

{\bf 151/EAU epsilon-Aquilids.}
The values of  $DH_{Min}=0.155$ and $DH_{MI}=0.059$ indicate a possible misclassification of the 151/00 or 151/01 solutions.
17 members of the first solution identified \citep{1976Icar...27..265S} among $\sim$$20000$ radio meteors; 11 members of the 151/01 solution identified \citep{2016Icar..266..331J} in a sample of $\sim$$110000$ meteors observed with video cameras. We do not have any other solutions of this stream in the MDC, which may imply the lack of its annual activity and even, the random nature of their identification in samples of orbits determined at mutually distant epochs of osculation.
As indicated in Table~\ref{tab:DH}, all contributors to $DH_{Min}=0.155$ are of the same order, unlike, for example, for the 150/SOP stream, whose orbits differ mainly in the orientation of the abse lines.  
At this stage, we cannot show with certainty, but we are inclined to the opinion that similar causal values in the $DH_{Min}$ value testify to the random nature of the difference in the parameters of the two 151/EAU stream solutions, which would support their autonomous nature. An additional argument for this conclusion is the nearly 10-degree difference in the right ascensions of the two radiants and the similar value of the difference in the $\lambda_{RCS}$  --- the ecliptic longitude of the radiant in the rotating reference frame. 
We propose that solution 151/00/EAU remain unchanged, while solution 151/01/EAU be 
given a new IAU name and codes. At the same time, we propose changing the status of 151/EAU and moving it from the List of established showers back to the Working list.\\
\begin{figure}
    \centering 
    \includegraphics[width=0.34\textwidth, angle=-90]{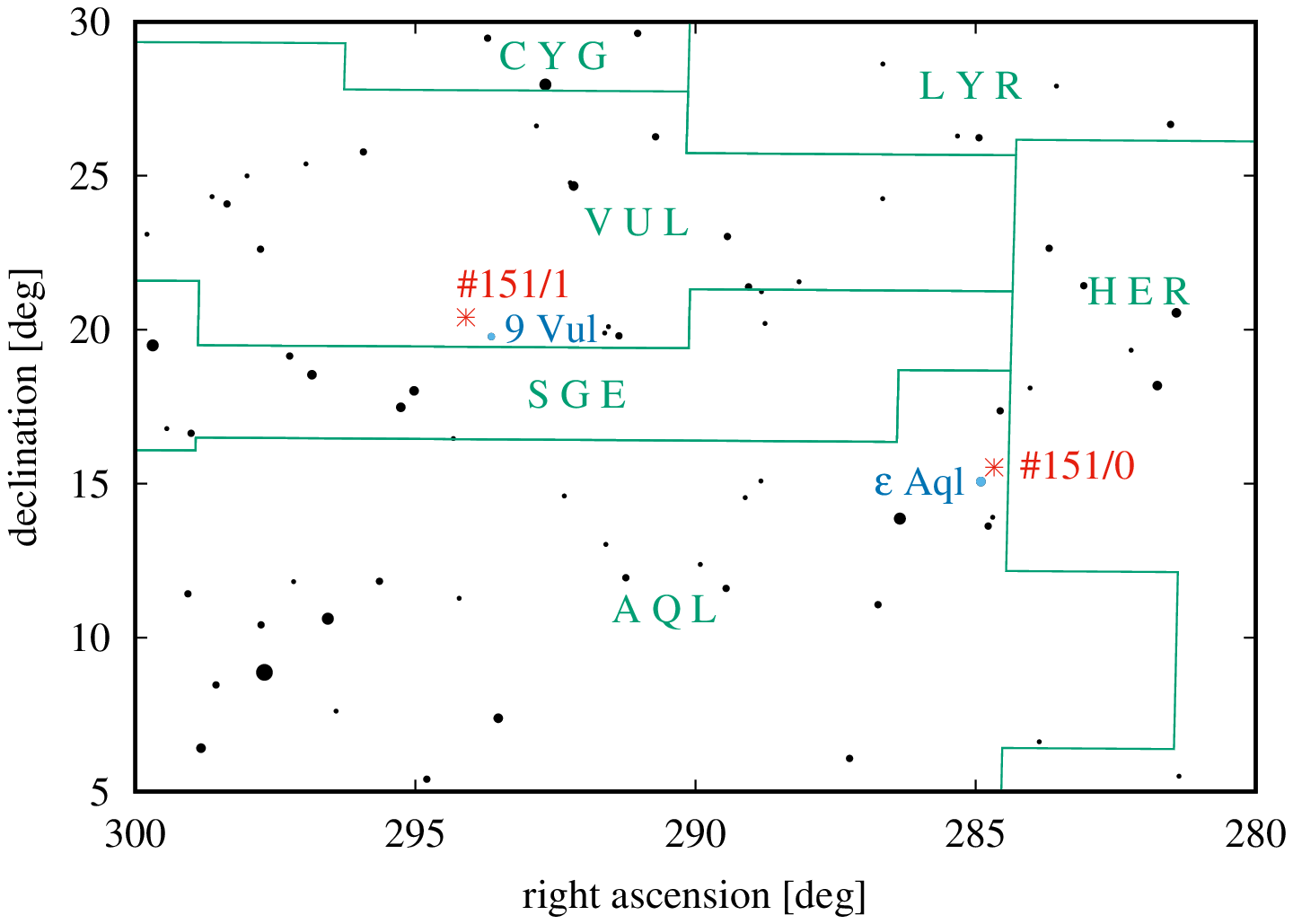}
    \caption{
    The radiants of solutions of 151/EAU shower (red asterisks). The radiant of 151/00/EAU shower is located in the constellation Aquila near the star $\epsilon$-Aquilae. The radiant of 151/01/EAU, located in constellation Vulpecula is close to the star 9-Vulpeculae.} 
    \label{fig:151radiants}
\end{figure}

{\bf 154/DEA Daytime epsilon-Arietids.}
The values of  $DH_{Min}=0.178$ and $DH_{MI}=0.039$ indicate a possible misclassification of the 154/00 or 154/01 solutions. Both were identified among radio meteors, by \citep{1976Icar...27..265S} and \citep{1964PhDT........28N}, respectively. Sekanina identified 25 members of the stream, Nilsson only 6 orbits. 
In Table~\ref{tab:DH}, we can see that, in principle, the solutions differ only in the orientation of the line of apsides, $DH_{\Pi}=0.177$. Sekanina analyzed $\sim$$20000$ orbits observed during the period 1968-1969, Nilsson 2101 orbits determined during the radio survey in 1961. 
\citep{2020eMetN...5...93K} pointed to a 15-degree difference in the Sun's ecliptic longitude for these solutions. But in view of the almost identical values of $\lambda_{RSC}$ ($359.9^{\circ}$ and $359.3^{\circ}$, see Table A1), the differences in $\lambda_S$ do not necessarily mean that we are dealing with different streams.
Hence, we believe that both solutions involve the same stream, observed at different epochs.\\

\begin{figure}
    \centering
  \includegraphics[width=0.34\textwidth, angle=-90]{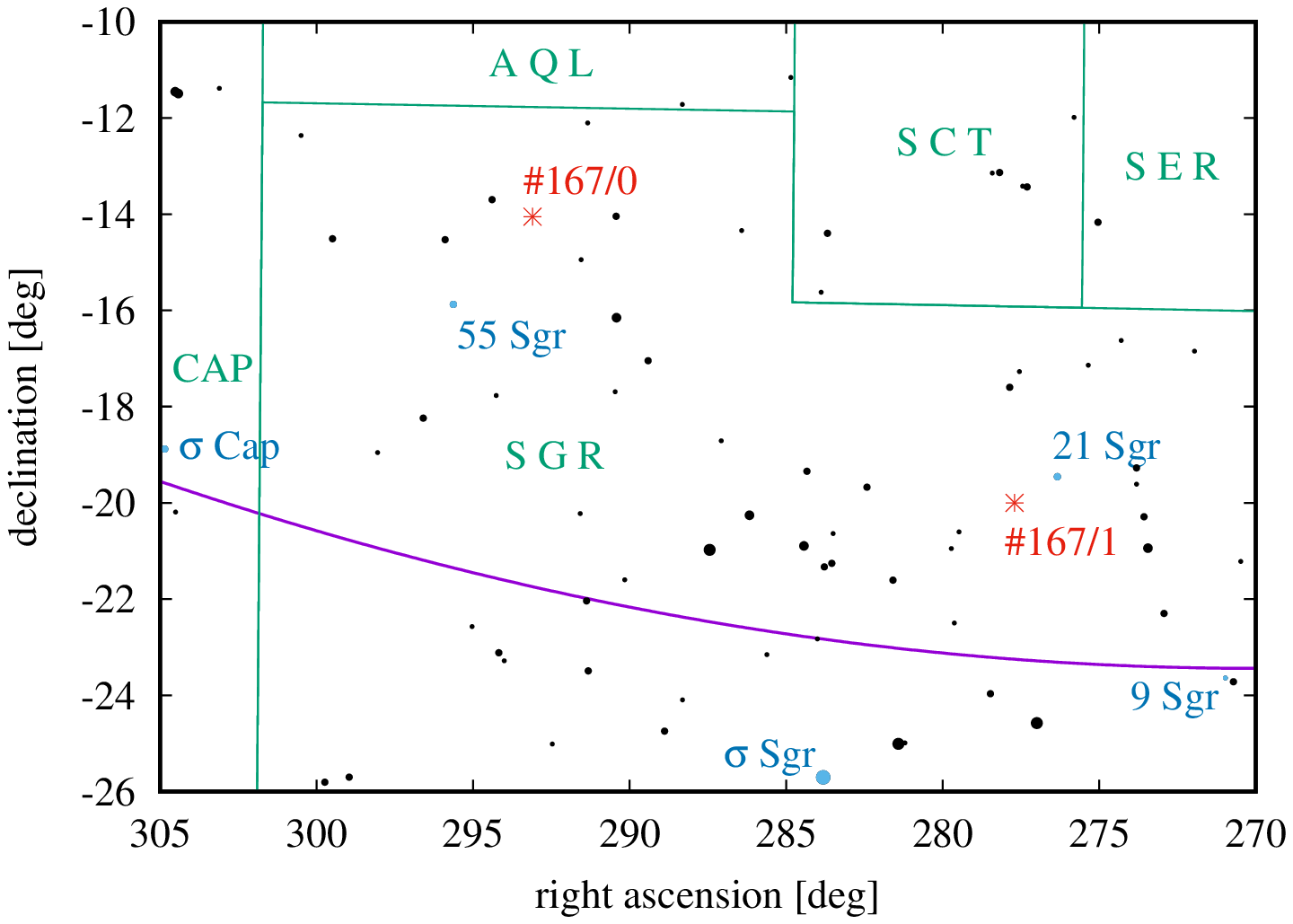}
    \caption{
    The radiants of solutions of 167/NSS shower (red asterisks). The radiant of 167/00 shower is located in the constellation Sagittarius near the star 55 Sagittarii. The radiant of 167/01 is located in the same constellation close to the star 21 Sagittarii.}
    \label{fig:167radiants}
\end{figure}

{\bf 167/NSS Northern sigma-Sagittariids.}
The values of  $DH_{Min}=0.282$ and $DH_{MI}=0.039$ clearly indicate a misclassification of the 167/NSS solutions.
Both solutions were identified in the same radio samples, by the same authors as in the previous case: \citep{1976Icar...27..265S} identified 45 members, \citep{1964AuJPh..17..205N} only 4 members of 167/NSS stream. 
Sekanina gave his discovery the name sigma-Capricornids (which is inconsistent with old naming rules, see Figure~\ref{fig:167radiants}), while Nilsson referred to it only as shower 61.6.6, see  Table 4 of his paper.
The reason for such a large value of $DH_{Min}$ is the different orientation of the apse lines $DH_{\Pi}=0.256$, see Table~\ref{tab:DH}. Other contributors to the $DH_{Min}$ value are moderate, as is the case with the 154/DEA stream. 
However, the two solutions differ markedly in the radiants $\lambda_{RSC}$ coordinate, the difference being $\sim$$8^{\circ}$. This difference cannot be explained by compensating for the perturbed motion of stream members over an interval of 8 years, equal to the difference in the epochs of observation of the two streams.
In the absence of arguments in favor, we propose that these solutions be given the status of autonomous streams. 

We propose retaining the current shower name for the stream identified by Nilsson (167/01) and assigning a new IAU MDC desigantion to the Sekanina stream (167/00), see Figure~\ref{fig:167radiants}.\\

{\bf 170/JBO  June Bootids.}
The values of  $DH_{Min}=0.220$ and $DH_{MI}=0.039$ indicate a misclassification of the 170/JBO solutions.
The 170/00 obtained by \citep{1976Icar...27..265S} among the $\sim$$20000$ radio orbits (54 members of the stream) has an internal inconsistency in the eccentricity. However, it can be considered acceptable within slightly expanded tolerance limits accepted in our calculations. In the second solution, 170/04 given by \cite{2022JIMO...50...38S}, the 9 members identified among $\sim$$298687$ video orbits,  differ from the first mainly in the orientation of the abse line, $DH_{PI}=0.203$.  
The two solutions differ quite significantly in the equatorial coordinates of the radiant, $\Delta RA=18^{\circ}$, $\Delta DE=16^{\circ}$. These differences cannot be explained by the diurnal motion of the radiant, because the difference in the values of the ecliptic longitudes of the radiant in the rotating reference system, $\Delta\lambda_{RSC}=40^{\circ}$, is far too great.
We therefore propose separating the two solutions.

\begin{figure}
    \centering
  \includegraphics[width=0.34\textwidth, angle=-90]{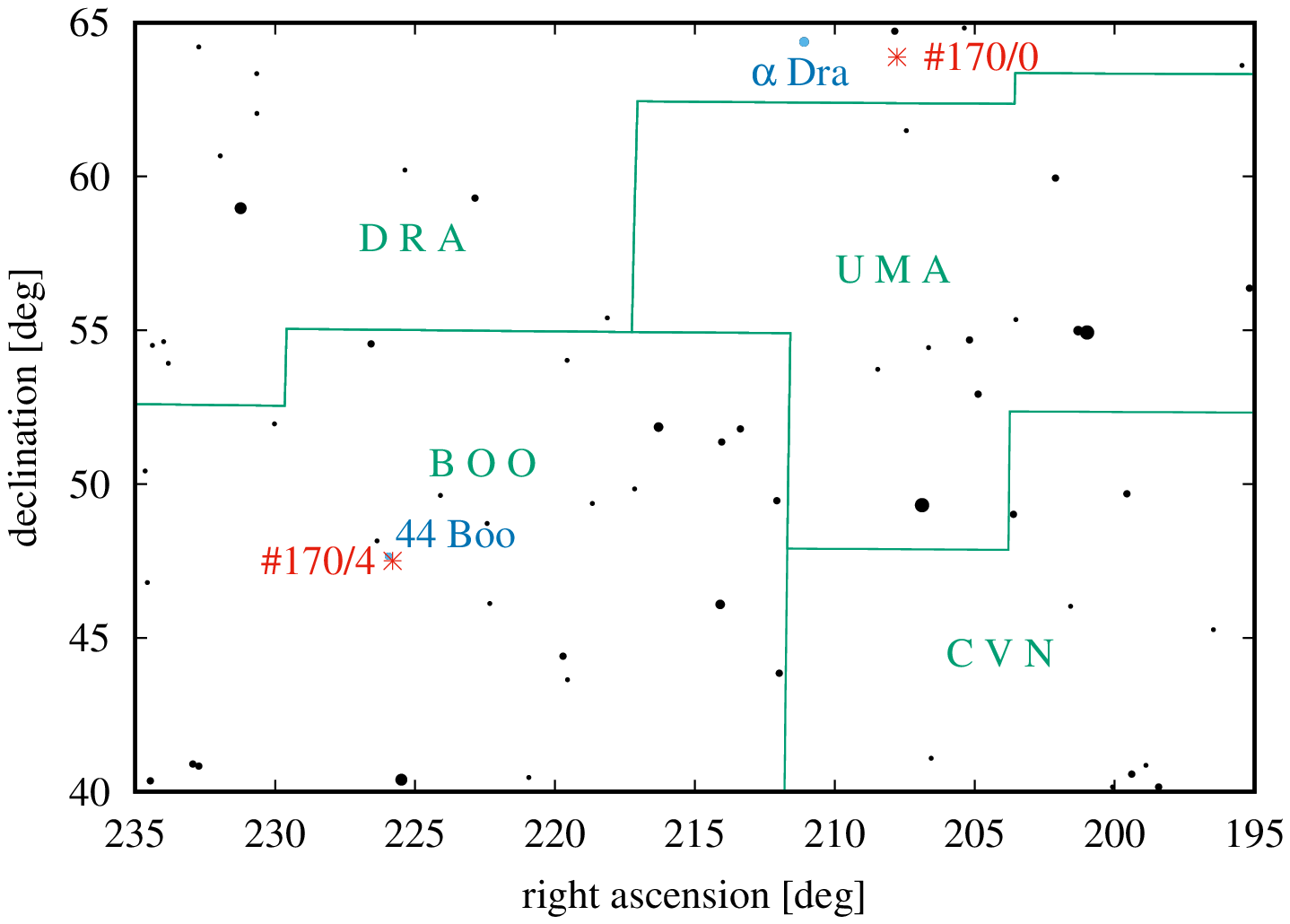}
    \caption{
    The radiants of solutions of 170/JBO shower (red asterisks). The radiant of 170/00 shower is located in the constellation Draconis near the star $\alpha$-Draconis. The radiant of 170/04 is located in the Bootes constellation close to the star 44-Bootis.}
    \label{fig:170radiants}
\end{figure}

As can be seen in Figure~\ref{fig:170radiants}, the Shiba solution has a radiant inside the Bootes constellation, so we propose not to change its codes and name. 
For the solution 170/00, with a radiant close to the alpha-Draconis star, we propose restoring the name given to it by its discoverer, June alpha-Draconids \citep{1976Icar...27..265S}. This shower will be listed in the MDC as an autonomous shower, with new codes assigned. 

Since 170/JBO is on the list of established showers, its status must be reconsidered after excluding the 170/00 solution. The reconsideration will include three new solutions for 170/JBO recently submitted to the MDC by Roggemans et al. (2023). \\

{\bf 179/SCA sigma-Capricornids.}
The values of $DH_{Min}=0.243$ and $DH_{MI}=0.039$ indicate a misclassification of the 179/SCA solutions.
In the IAU MDC database, this stream is represented by 2 solutions: 179/00 identified by \citep{1976Icar...27..265S} among $\sim$$20000$ radio meteors and 179/01 identified by \citep{2016Icar..266..355J} among $\sim$$110000$ video meteors observed after about 40 years.
There is an inconsistency between the geocentric and heliocentric data for the 179/01 solution, most likely due to the way the averaged parameters were determined. In \citep{2016Icar..266..355J} the authors took the medians of the parameters as average values. 

The largest contributors to $DH_{Min}=0.234$ are $DH_e=0.119$ and $DH_q=0.172$, which is undoubtedly due to the large difference in the geocentric velocities of the two solutions, $\Delta V_g=7.2\,$km$\,$s$^{-1}$. 
The meteoroids of the solution 179/01 observed by the video technique by \citep{2016Icar..266..355J} proved to be noticeably faster than those observed 40 years earlier by the radio technique of the solution 179/00. However, in his new publication, \citep{2024book...Jen} reports two solutions for this shower over different periods of activity, observing a decrease in velocity with decreasing $\lambda_s$. Unfortunately, the new solution has not been submitted to the MDC.
Nor we have the data of the members of the 179/00 and 179/01 solutions at our disposal, which makes further discussion of this case difficult. In view of the significant value of $DH_{Min}=0.234$, we favour giving both solutions the status of autonomous streams.
For the first solution, 179/00, we suggest keeping the current designation of the shower, while for the solution 179/01, a new IAU designation should be assigned.  \\
\begin{figure}
    \centering 
  \includegraphics[width=0.34\textwidth, angle=-90]{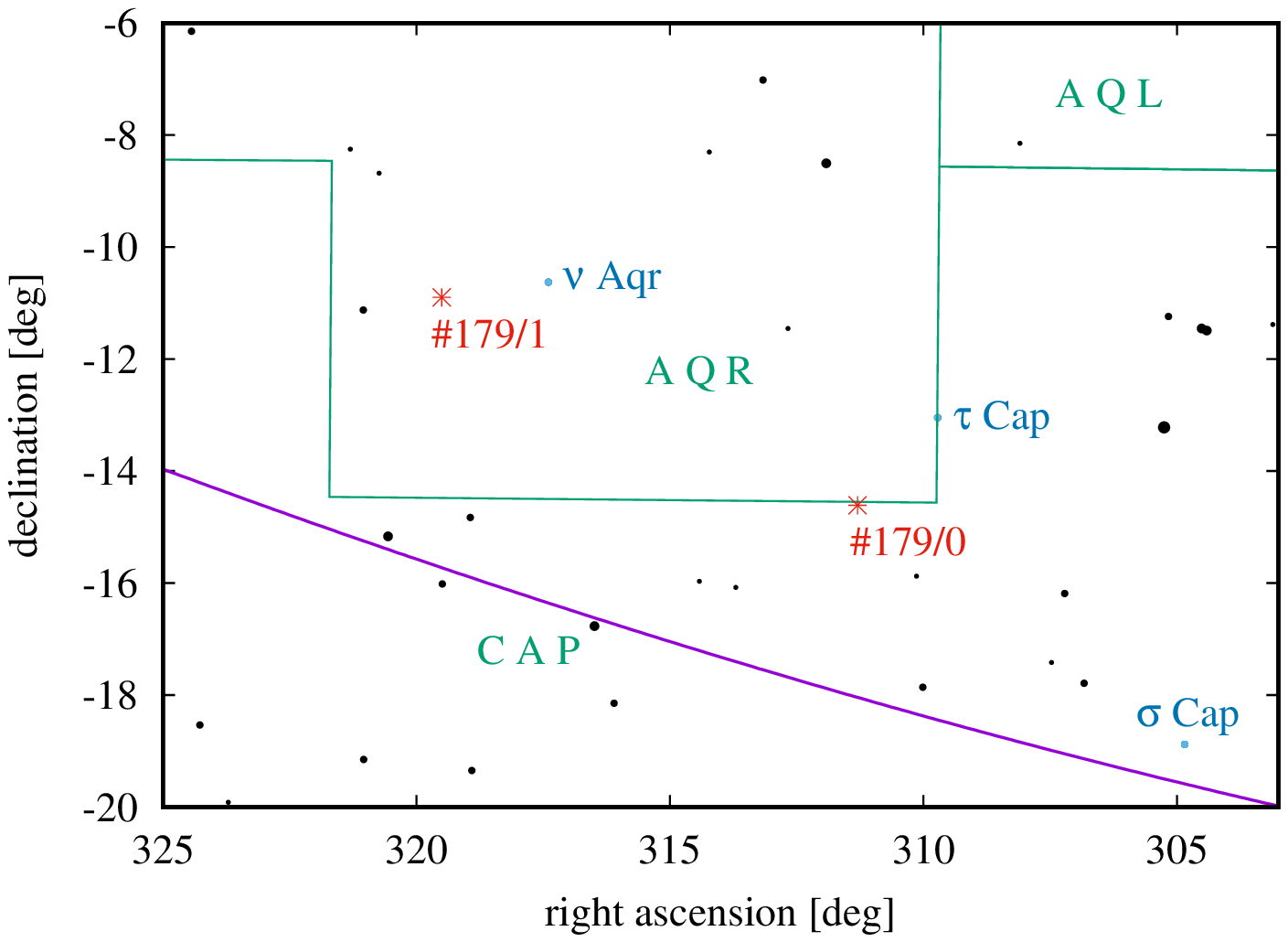}
    \caption{
    The radiants of solutions of 179/SCA shower (red asterisks). The radiant of 179/00 shower is located in the constellation Capricornus near the star $\tau$-Capricorni. The radiant of 179/01 is located in the Aquarius constellation close to the star $\nu$-Aquarii. }
    \label{fig:179radiants}
\end{figure}

{\bf 186/EUM epsilon-Ursae Majorids.}
The values of  $DH_{Min}=0.194$ and $DH_{MI}=0.039$ indicate that we are dealing with two different streams. 
Solution 186/00 was identified by \citep{1989JIMO...17..242T} among 554 photographic fireballs, solution 186/01 was found by \citep{2016Icar..266..355J} among $\sim$$110000$ video orbits.

The solutions differ mainly in the inclination of the orbital planes $DH_{in}=0.109$ and in the orientation of the apse lines $DH_{\Pi}=0.154$. However, there is a distinctly large difference in the equatorial coordinates of the radians, $41^{\circ}$ in right ascension, and in the rotating ecliptic reference system, $\Delta_{\lambda_{RSC}}=63^{\circ}$. 
We do not have geocentric data as well as orbital data of the members of these solutions, which prevents further discussion of this case. Therefore, due to the large value of $DH_{Min}=0.194$, we propose to give autonomous status to both solutions. 
\begin{figure}
    \centering 
  \includegraphics[width=0.34\textwidth, angle=-90]{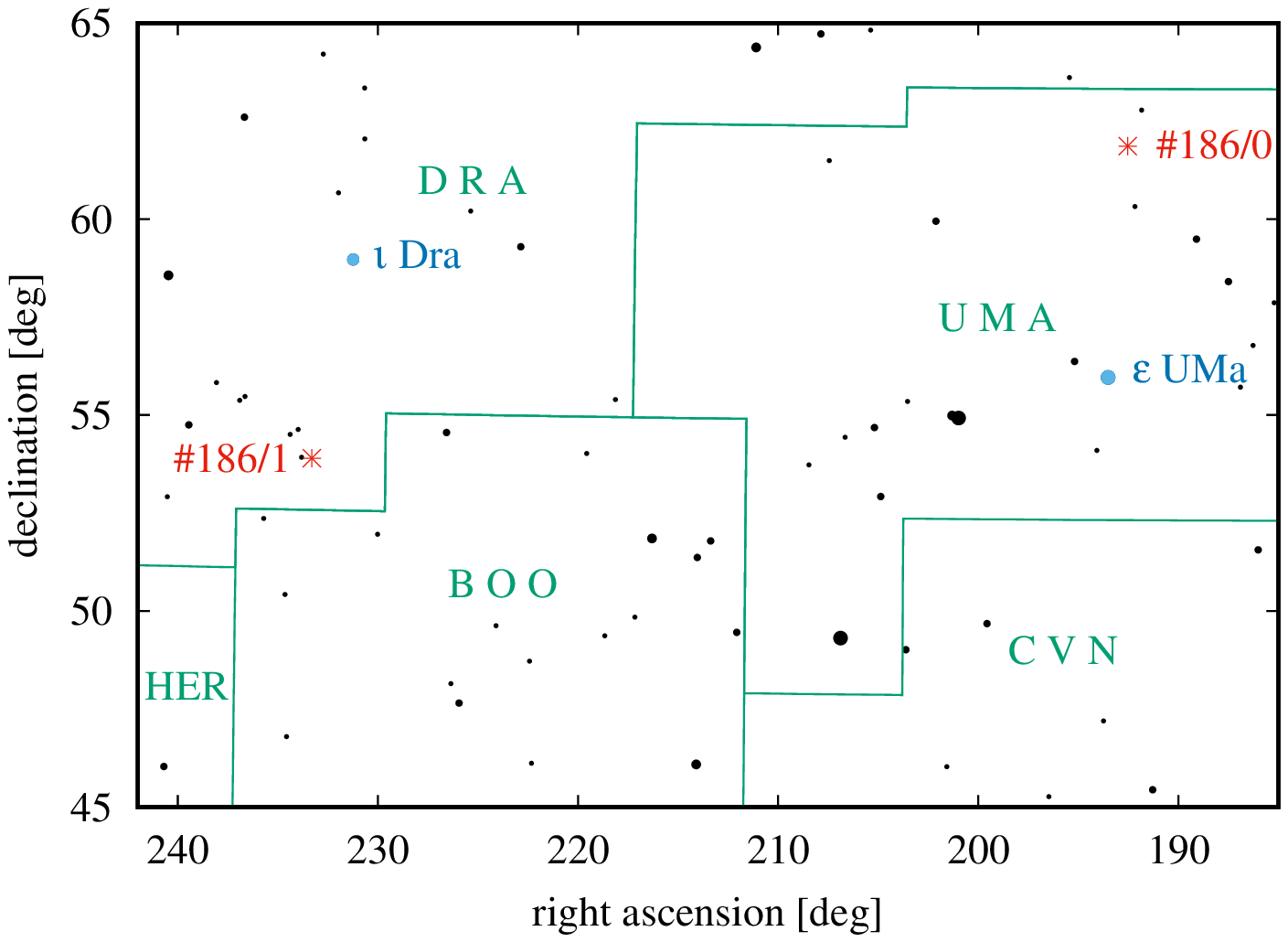}
    \caption{
    The radiants of solutions of 186/EUM shower (red asterisks). The radiant of 186/00 shower is located near the star $\epsilon$-Ursae-Maioris. The radiant of 186/01 is located in the Aquarius constellation Draco close to the star $\iota$-Draconis }
    \label{fig:186radiants}
\end{figure}
The solution 186/00 given by \citep{1989JIMO...17..242T} could retain its current shower name and codes, while 
for solution 186/01, we propose a new name and new MDC codes assigned.
\citep{2020eMetN...5...93K} pointed out a neighbourhood of 186/1 with shower 170/JBO.  
For 170/04 and 186/01, DH=0.079, however the MSA test provided in Paper I has failed. \\

{\bf 189/DMC Daytime mu-Cancrids.}
The values of  $DH_{Min}=0.454$ and $DH_{MI}=0.039$ indicate a misclassification of the 189/00 or 189/01 solution identified among the radio meteors observed in the 1960s by \citep{1976Icar...27..265S} and \citep{1964AuJPh..17..205N}, respectively.
The orbits of the two solutions differ markedly in the orientation of the apse lines, $DH_{\Pi}=0.317$, in the distance of the perihelia $DH_q=0.278$ and in eccentricity $DH_e=0.158$. The reason for such large values in the DH function is the different values of the geocentric velocities $V_g=24.3\,$km$\,$s$^{-1}$ and $V_g=33.1\,$km$\,$s$^{-1}$, and the different equatorial positions of the radiants not compensated for in the ecliptic rotating system $\Delta \lambda_{RSC}=8.5^{\circ}$. 

\begin{figure}
    \centering
  \includegraphics[width=0.34\textwidth, angle=-90]{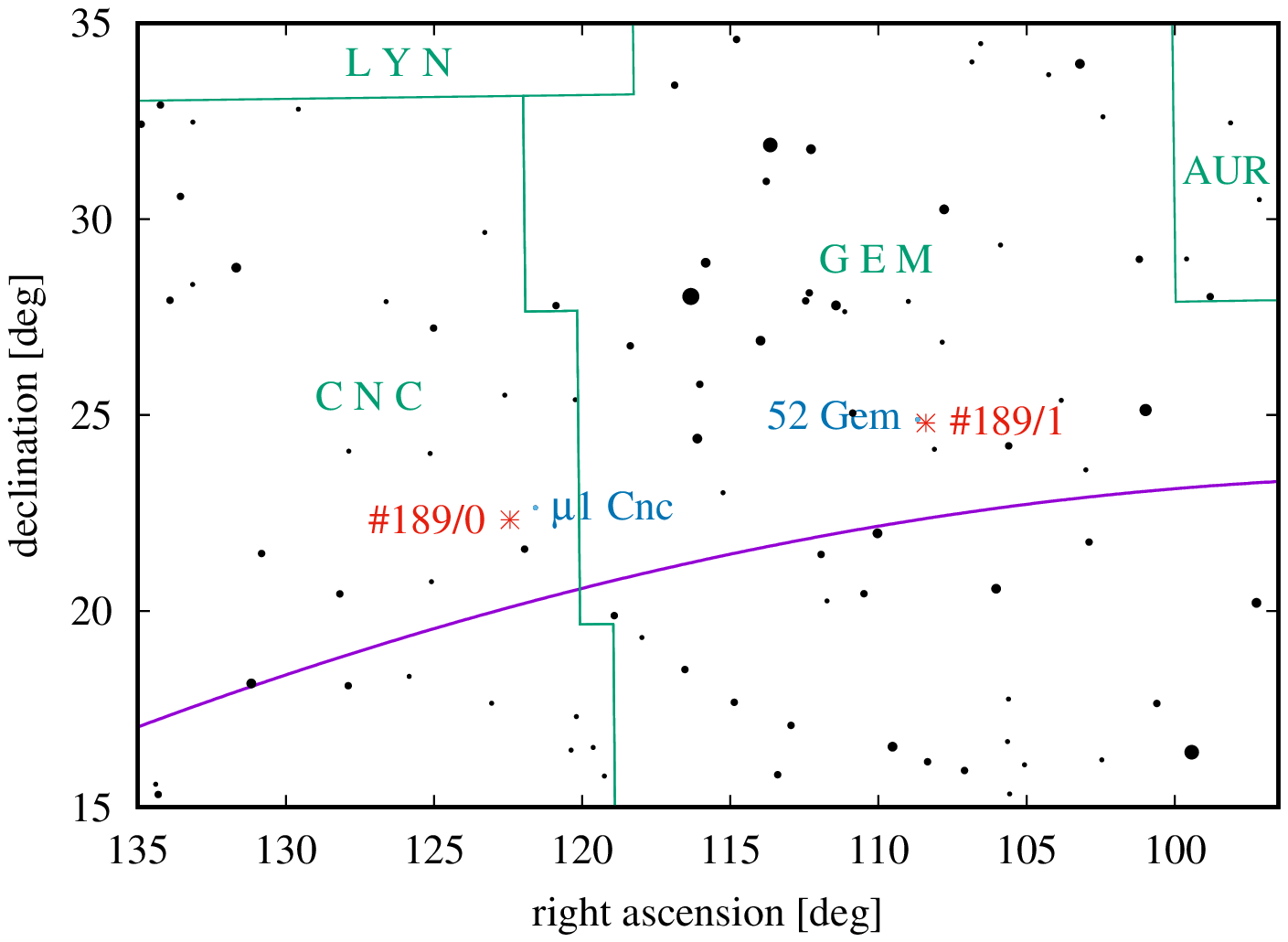}
    \caption{
    The radiants of solutions of 189/DMC shower (red asterisks). The radiant of 189/00 shower is located in the constellation Cancer near the star $\mu1$-Cancri. The radiant of 189/01 is located in the Gemini constellation close to the star 52-Geminorum. }
    \label{fig:189radiants}
\end{figure}

We propose that the two solutions be given the status of separate streams; stream 189/DMC/00 should be left on the Working list retaining the current shower name and codes, while solution 189/01 with radiant near the star 52-Geminorum, see Figure~\ref{fig:189radiants}, should be left on this list with new codes and designation. \\

{\bf 197/AUD August Draconids.}
The values of  $DH_{Min}=0.316$ and $DH_{MI}=0.039$ indicate that we are dealing with two different streams. 
\citep{1976Icar...27..265S} identified solution 197/00 with 54 members among $\sim$$20000$ radio orbits, and \citep{2016Icar..266..331J} found solution 197/01 with 17 members among $\sim$$110000$ video orbits. The main reason for such a large difference in DH value are eccentricities of the two orbits. Sekanina reported $e=0.335$, Jenniskens et al. gave $e=0.644$, which is probably due to the considerable difference in the geocentric velocity of the solutions, $\Delta V_g=\sim 4\,$km$\,$s$^{-1}$. 
Another discrepancy between these solutions is the large difference in the ecliptic longitudes of the radians, amounting to $\Delta \lambda_{RSC}=27^{\circ}$. The equatorial radians of both solutions lie in the constellation Draco. We suggest to exclude the Sekanina solution from the 197/AUD.  Unfortunately, the name phi-Draconids originally suggested by Sekanina has been used for another shower 45/PDF. Hence, we suggest giving it a new name and codes. 
As for the 197/01 solution, however, in order to keep the number of changes made to the MDC as small as possible, we propose that its name and codes be left unchanged in the MDC. \\

{\bf 199/ADC August delta-Capricornids.}
The values of $DH_{Min}=0.107$ and $DH_{MI}=0.039$ indicate that we are dealing with two different streams.
Regarding the previously discussed cases, the value of DH is not large, the main contributor comes from the $DH_{in}=0.08$ related to difference in the inclination of the orbits of the two solutions. The smallest contribution comes from the orientation of the lines absides of the orbits, $DH_{\Pi}=0.006$. 
Solution 199/ADC/00 was determined by \citep{1994PSS...42..151P}, and solution 199/ADC/02 was announced by \citep{2022eMetN...7..293R}.

However, 
the ecliptic latitudes of the two radians have opposite signs. Which means that solutions 199/00 and 199/02 belong to separate branches of the same stream, and according to tradition, this property should be reflected in their names and codes. Therefore, we propose rename solution 199/00 to Southern delta-Capricornics, and solution 199/02 to Northern delta-Capricornids. The relevant codes will be determined by the WG. 
Alternatively, the northern branch retains the original shower name and code. The WG will make the final decision. \\

{\bf 202/ZCA Daytime zeta-Cancrids.}
The values of  $DH_{Min}=0.434$ and $DH_{MI}=0.039$ clearly indicate that we are dealing with two different streams. 
Solutions 202/00 and 202/01 are the only ones we have in the MDC, coming from observations that are separated by an interval of $\sim 40$ years. 
The parameters of the 202/00 solution given by \citep{1964AuJPh..17..205N} are not internally consistent. However, using the methodology proposed by \citep{2024CoSka..54a..57N}, we found a large difference only in the calculated value of the orbit inclination. Nilsson reports $i=21.1^{\circ}$ and according to our calculations $i=7^{\circ}$. \citep{2016JIMO...44..151K} also found similar inconsistency in the parameters of the 202/00 solution, however, he states that in addition to the orbit inclination, the inconsistency also occurs for the argument of the perihelion orbits. 
Due to the lack of access to the data of individual members of both solutions, we cannot determine the reason for the inconsistency of solution 202/00. In view of the particularly large $DH=0.434$ and its internal data inconsistency we propose to move solution 202/00 to the List of removed showers. Consequently, we propose cancelling the established status of 202/ZCA. 
\\

{\bf 220/NDR nu-Draconids.}
The only solutions given in the MDC of this stream are 220/00 and 220/01.  
The first was identified by \citep{1976Icar...27..265S} among $\sim$20000 radio meteors, the second by \citep{2016Icar..266..355J} among $\sim$11000 meteors observed by video technique. 
The value of $DH_{Min}=0.144$ exceeds the acceptable value of $DH_{MI}=0.059$, and the solutions differ the most in the inclination, $DH_{in}=0.101$ and the orientation of the lines of the apses, $H_{\Pi}=0.093$. 
Solutions were observed with different techniques, in different eras (40 years difference), so in view of the comments given in Section~\ref{sec:metodologia}, $DH_{Min}=0.144$ is not critical. Therefore, we do not propose to change the 220/00 and 220/01 solution status in the MDC.\\

{\bf 233/OCC October Capricornids.}
The values of $DH_{Min}=0.160$ and $DH_{MI}=0.039$ indicate a possible misclassification of the 233/00 or 233/01 solutions. 
These are the only solutions for which orbital data is provided in the MDC, and the 233/OCC stream has an 'established' status with an IAU-approved name. 
The orbits of the two solutions differ practically only in the orientation of the apse lines, $DH_{PI}=0.151$. They were determined by visual observations by \citep{1988JIMO...16..191W} and photographic observations by \citep{1989JIMO...17..242T}. The parameters of the 233/01 solution were obtained from averaging a number of (unfortunately unknown to us) orbits. The orbit of 233/00 solution was obtained by means of estimating the position of the radiant and the off-atmospheric velocity, for the date corresponding to the maximum of shower activity.
     
As described by \citep{1988JIMO...16..191W}, the existence of the 233/OCC stream is not in doubt, its activity was observed in 1971--1987. Another confirmation of its nature, especially the relatively high activity in 1972, was the return of comet Haneda-Campos (D/1978 R1) before its observation in 1978. The similarity of its orbit to that of the 233/00 solution was noted by \citep{1988JIMO...16..191W}. However, it is not the only object associated with this stream. \citep{2024MNRAS.535.3661D} confirmed the similarity of 233/00 to the orbit of comet D/1978 R1, but also pointed out the similarity to the orbits of comets 15P and 103P. And according to \citep{2017A&A...607A...5D}, the orbit of 233/OCC is also similar to the orbits of many near-Earth asteroids. Undoubtedly, the reason for so many potential parent bodies of 233/OCC streams is that its orbit has a slight inclination to the plane of the ecliptic. 
In view of the above, we propose that the current status of jet 233/OCC in the MDC database be maintained.\\

{\bf 253/CMI December Canis Minorids.}
The values of $DH_{Min}=0.347$ and $DH_{MI}=0.059$ indicate a misclassification of the 253/00 or 253/03 solutions. Solutions were obtained from video observations, 253/00 by \citep{2016Icar..266..355J},  253/03 by \citep{2022JIMO...50...38S}. Both were identified in large orbital samples and they are the only solutions available in the MDC.  
The high value of the DH-function mainly originates from the orientation of the solutions' orbits: $DH_{in}=0.195$, $DH_{\Pi}=0.221$. 
\begin{figure}
    \centering
  \includegraphics[width=0.34\textwidth, angle=-90]{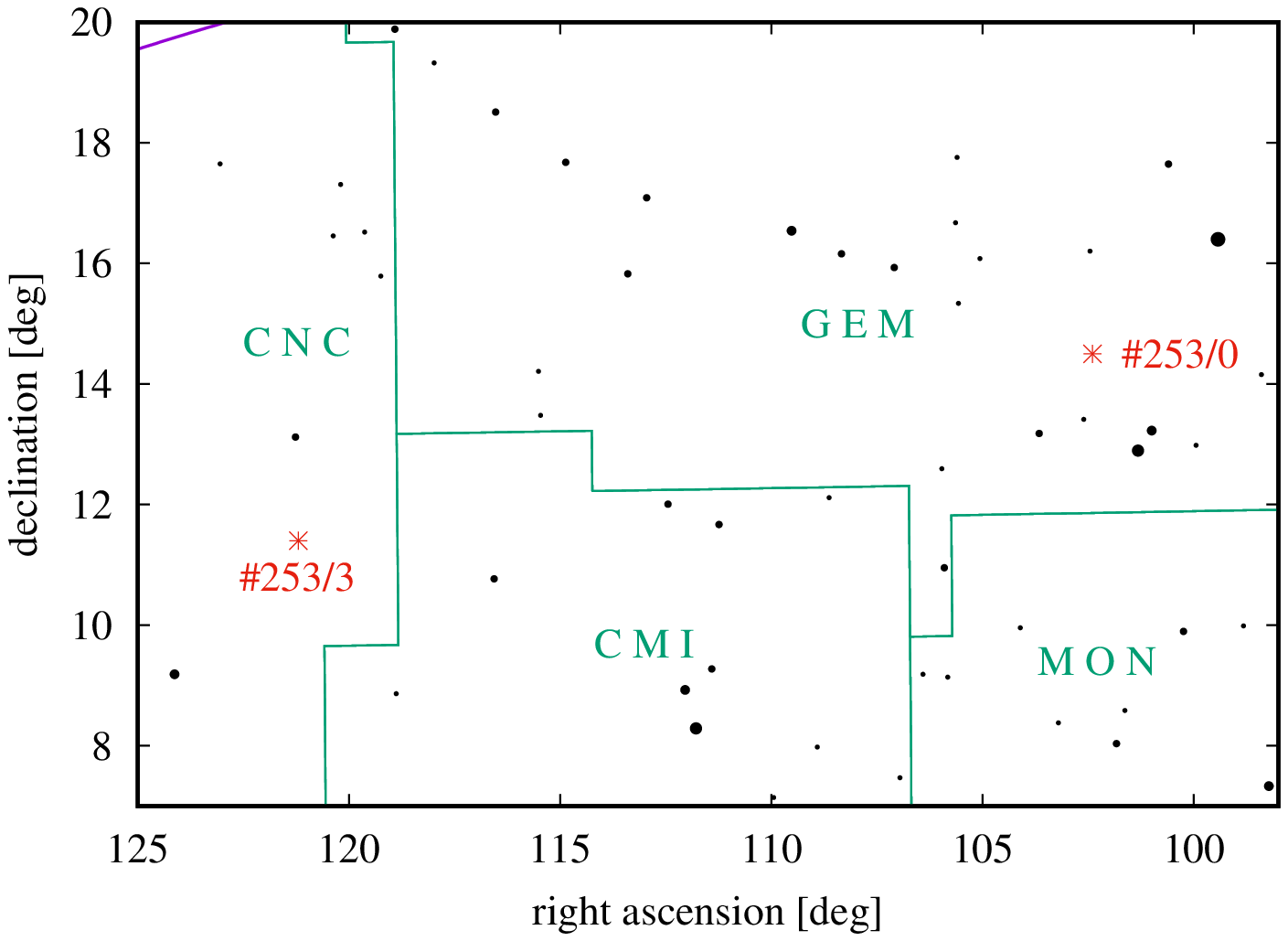}
    \caption{
    The radiants of solutions of 253/CMI shower (red asterisks). The radiant of 253/00 shower is located in the constellation Gemini, the radiant of 253/03 is located in the  constellation Cancer. }
    \label{fig:253radiants}
\end{figure}
As we can see in Figure~\ref{fig:253radiants}, the name Canis Minorids of this MSS is inconsistent with the old stream nomenclature rules. Both radiants are located outside of the constellation Canis Minor.  
However, in order to make as few changes to the MDC as possible, we recommend that both solutions remain in the MDC as separate streams: solution 253/00 with no changes in codes and name; solution 253/03 with new codes and name assigned by the WG.\\

{\bf 254/PHO Phoenicids.}
The values of $DH_{Min}=0.199$ and $DH_{MI}=0.039$ indicate a possible misclassification of the 254/00 or 254/01 solutions. Phoenicids in MDC are represented by two solutions and have established status. 
 
For the MDC, 254/00 solution was taken from \citep{2006mspc.book.....J}, who used the two orbits given by \citep{1973NASSP.319..183C}. Cook, in turn, used results given by \citep{1962R,1963MNSSA..22...42R} who used visual observations and  gave two orbits. The first was calculated as parabolic and the second was based on the similarity of the parabolic orbit with the orbit of comet 289P (1819 Blanpain). This elliptical orbit was calculated  assuming that its orbital period is 5.1 years, the same as the orbit period of comet 289P. So, it is not unusual that the similarity of the orbits of solution 254/00 and comet 289P was confirmed by \citep{2024MNRAS.535.3661D}. 
The 254/PHO stream parameters given by \citep{2006mspc.book.....J}, more likely obtained by averaging the data given by \citep{1973NASSP.319..183C}. Unfortunately, as noted in Table \ref{tab:MSS-1}, they are internally inconsistent.

The 254/00 solution was given by \citep{1973NASSP.319..183C} on the basis of visual observations. Solution 254/01 came from \citep{2022JIMO...50...38S}, who identified only 9 members between $\sim$300,000 video orbits. 
The orbits of these solutions differ markedly in inclination, $DH_{in}=0.184$. But they are also clearly different in the radiant declination $\Delta DE=37^{\circ}$ and the radiant ecliptic longitude in the rotating reference frame $\Delta\lambda_{RSC}=33^{\circ}$.

Ultimately, we propose to give autonomous status: to solution 254/00 with no changes in codes and name; to solution 254/01 with new codes and name assigned by the WG.

After excluding one solution, shower 254/PHO will not fulfil the condition for established status; therefore, we also suggest changing its status from established to working.\\

{\bf 324/EPR epsilon-Perseids.}
The values of $DH_{Min}=0.250$ and $DH_{MI}=0.059$ indicate a possible misclassification of the 324/00 or 324/01 solution.
\citep{2008Icar..195..317B} identified 203 members of the 324/00 solution among more than 2 million orbits obtained over the period 2000–2006 radio survey. In contrast, \citep{2016Icar..266..331J} identified only 4 members of the 324/01 solution among $\sim$110000 video orbits. The largest contribution to the $DH_{Min}=0.250$ comes from the difference in the inclination of the orbits, $DH_{in}=0.211$.  
The identification of the 324/00 stream is not in doubt, which cannot be said of the 324/01 stream. 
This MSS is represented by two solutions only, and have established status in the MDC. However, in the absence of other arguments, we propose that both 324/EPR stream solutions be given standalone status in the MDC. Solution 324/00 with an unchanged name and codes, solution 324/01 with a new name and codes fixed by the WG. Consequently, we propose moving 324/EPR to the Working list.\\

{\bf 326/EPG epsilon-Pegasids.}
This MSS is represented by two solutions and has an established status in the MDC. 
The values of $DH_{Min}=0.175$ and $DH_{MI}=0.059$  indicate a problem with the correct clarification of solutions 326/00 and 326/01. The first solution was identified by \citep{2008Icar..195..317B} (62 members) among more than three million radio orbits, the second was given by \citep{2016Icar..266..331J} (33 members) analyzing $\sim$ 110.000 video orbits. The orbits of these solutions differ mainly in inclination, $DH_{in}=0.125$. 
In the absence of other arguments, we propose that both 326/EPR stream solutions be given a standalone status in the MDC. We propose that solution 326/00 retain its current name and codes, while a new name and code be assigned to solution 326/01 by the WG. After separation, the conditions for the shower established status will not be sufficient, so both solutions will be moved to the Working list. \\

\begin{figure}
    \centering
  \includegraphics[width=0.34\textwidth, angle=-90]{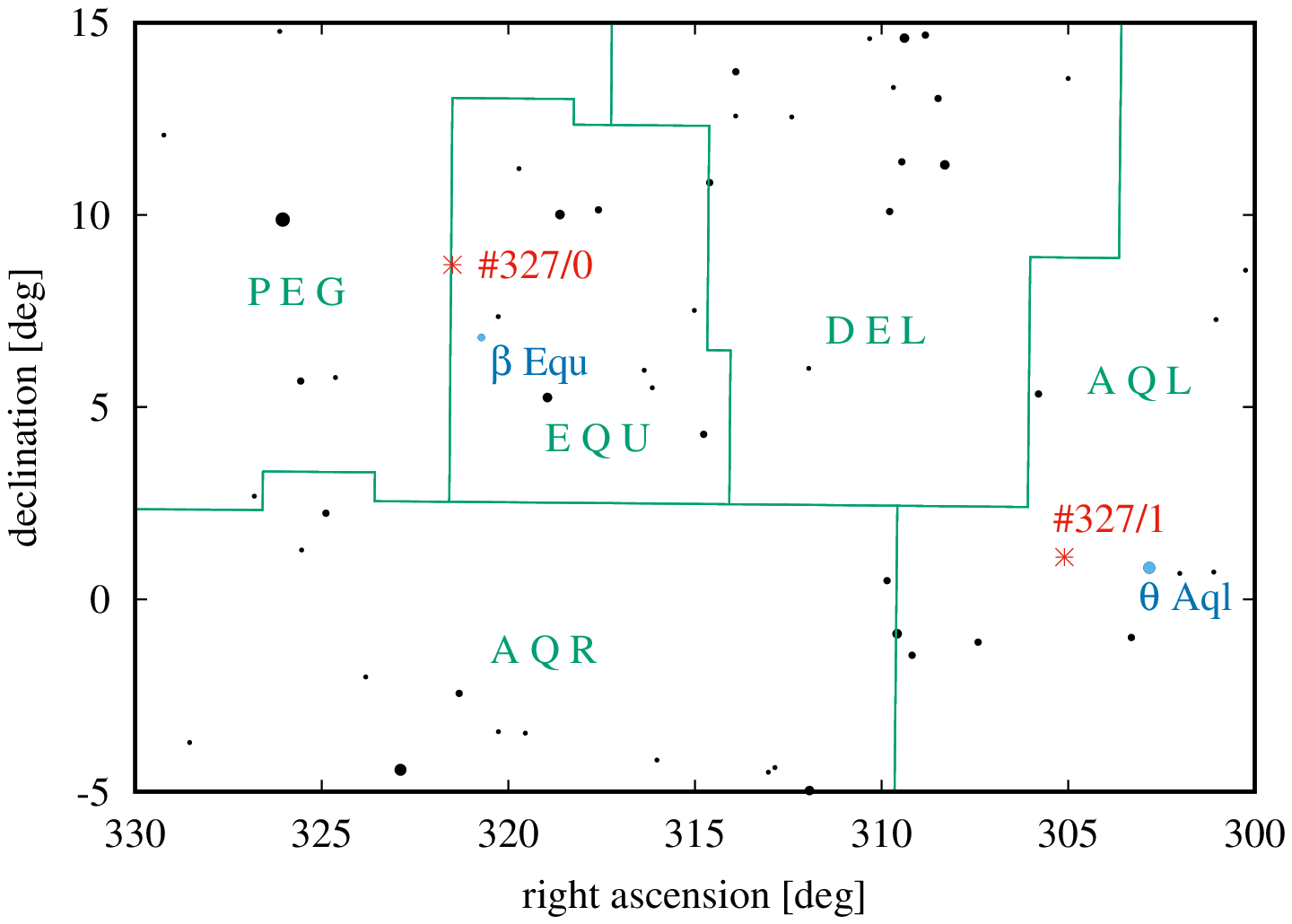}
    \caption{
    The radiants of solutions of 327/BEQ shower (red asterisks). The radiant of 327/00 shower is located in the constellation Equuleus, the radiant of 327/01 is located in the  constellation Aquila. }
    \label{fig:327radiants}
\end{figure}

{\bf 327/BEQ beta-Equuleids.}
The values of $DH_{Min}=0.377$ and $DH_{MI}=0.059$ indicate misclassification of one of the 327/00 or 327/01 solutions. They are the only solutions of the beta-Equuleids shower included in the MDC database. 
327/00 identified by \citep{2008Icar..195..317B}, 89 members from three million radio data, has a radiant located near the star beta-Equulei. Solution 327/01 given by \citep{2016Icar..266..331J}, 38 members from $\sim$$110000$ video orbits, has a radiant in the constellation Aquila, near the star theta-Aquilae, see Figure~\ref{fig:327radiants}. 
The orbits of these solutions differ mainly in inclination, $DH_{in}=0.282$, and orientation of the apsidal line, $DH_{\Pi}=0.247$. 
In addition, their radians are located at a considerable distance from each other, which is also noted by \citep{2020eMetN...5...93K}. However, in the case of 327/BEQ solutions, the components of diurnal motion of the radians given in the source documents do not allow to reconcile the positions of these radians. The reason lies in the opposite signs of the radiant's diurnal motion in the declination component.  
The large values $DH_{Min}=0.377$ of the discussed shower indicate the absence of a genetic connection for these solutions, for which we also failed to find the parent comet; \citep[see][]{2024MNRAS.535.3661D}. We therefore propose that both solutions be given stand-alone status, 327/00 without the code and name change, while 327/01 with a new code and designation. Both solutions will be moved to the Working list as, being standalone showers, they will no longer meet the conditions for the established status. \\  

{\bf 334/DAD December alpha-Draconids.}
The values of $DH_{Min}=0.110$ and $DH_{MI}=0.059$ indicate some problem to classify the solutions of this stream.  Solution 334/01 identified \citep{2016Icar..266..331J}, 47 members among $\sim$$100000$ video data, 334/03 found \citep{2022JIMO...50...38S}, 487 members among $\sim$$230000$ video orbits. 
The mean orbits differ mainly in the inclination and orientation of the apse line, $DH_{in}=0.068$ and $DH_{\Pi}=0.078$. But these are not big differences. The radians of both solutions are located in the constellation of Ursa Maior. 

Another solution to this stream, 334/00 obtained by \citep{2009JIMO...37...55S}, 145 members among 240000 video observations, is given in MDC. However, \citep{2009JIMO...37...55S} did not provide the mean orbital data of this shower. But for the purposes of our study we calculated this orbit using the geocentric data of this solution. We applied an approach analogous to that used by \citep{2008Icar..195..317B}, who computed the mean orbits using the radiant and velocity observed at the time of maximum shower activity.
The calculated orbit of the 334/00 solution turned out to be more similar to that of 334/03, $DH_P=0.084$, also the application of the MMS method yielded a positive result for this pair.  Repeating the cluster analysis calculations for the three members of 334/DAD also yielded a positive result, $DH_{Min}=0.108$, with an acceptable threshold value of $DH_{MI}=0.136$. 
This means that we have no grounds to change anything in the MSS 334/DAD status.\\

{\bf 347/BPG beta-Pegasids.}
The values of $DH_{Min}=0.230$ and $DH_{MI}=0.059$ indicate a problem with the correct clarification of solutions 347/00 or 347/01. 
These are the only solutions for this stream in the MDC. The first was identified among three million radio orbits by \citep{2010Icar..207...66B}, 1105 members. The second was reported by \citep{2016Icar..266..355J}, 11 members, identified amongst $\sim$$110000$ video orbits.  The orbits of these solutions differ mainly in inclination $DH_{in}=0.150$ and orientation of the abse line $DH_{\Pi}=0.131$. Both radiants are located in the constellation Pegasus. 
In view of the large DH=0.230 value, our recommendation is to split this MSS into two standalone solutions, with 347/00 keeping its current name and codes, and 347/01 receiving new ones.  \\

{\bf 372/PPS phi-Piscids.}
The stream has an 'established' status officially granted by the IAU. However, the values of $DH_{Min}=0.350$ and $DH_{MI}=0.059$ indicate a misclassification of its solutions 372/00 or 372/02. 
In our study, 372/PPS is represented by two solutions obtained from radio observations by \citep{2010Icar..207...66B}, 1395 members, and from video observations by \citep{2016Icar..266..331J}, 379 members. 
The orbits of these solutions differ in eccentricity and orientation of the apse line, $DH_e=0.292$ and $DH_{\Pi}=0.187$. 
Following the completion of our research published in Paper I, \citep{Shiba2023}
identified 358 members among 126184 video orbits belonging to this shower; the solution was added to the MDC with the codes 372/PPS/04. It is closer to solution 372/02, $DH_{P}=0.085$. However, when the 372/04 solution was incorporated into the 372/00, 372/02 pair, it became apparent that the identification of this group as representing the 372/PPS stream still required a value of $DH_{Min}=0.350$. This justifies the claim that solution 372/00 should be recognised as a separate stream, while solutions 372/02 and 372/04 represent the same 372/PPS stream.

\begin{figure*}
    \centering
  \includegraphics[width=0.8\textwidth, angle=0]{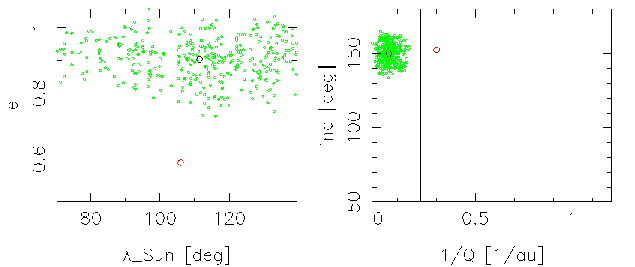} 
    \caption{Distributions of $e$ versus $\lambda_S$ and $1/Q$ versus $inc$ (inverse of the distance of the aphelion versus the inclination of the orbit), $358$ members of the 372/04 stream (green) and two points corresponding to the average values of the 372/04 (black) and 372/00 (red) solution.   
    The mean values of the eccentricities of the two solutions are definitely different from each other. In the figure on the right, solution 372/00 belongs to the area to which known comets belong.  Solution 372/00 is located in an area where known comets do not exist. }
    \label{fig:372PPS}
\end{figure*}

The meteoroids included in solutions 372/02 and 372/04 were observed, in different years, for almost 2.5 months, from early June to first half August. In the case of 372/00, observations were made on several days in early July only. This may suggest that the radio sample of 1395 orbits was obtained from a different part of the 372/PPS stream than the video observations. 
With the recently introduced requirement to provide the MDC with stream averaged parameters (see \citep{2020P&SS..18204821J}), data of its members, we have access to the geocentric parameters of the members of the 372/04 solution. Using these data, we calculated the orbital elements of the members of this solution, the two distributions of these elements are illustrated in Figure~\ref{fig:372PPS}, where the points corresponding to the members of the 372/04 stream and the average values of the 372/00 solution are plotted. 
This figure leads to the conclusion that the 372/00 solution does not represent the same stream as the 372/04 solution. In particular, the $1/Q$ versus $inc$ diagram (inverse of the distance of the aphelion versus the inclination of the orbit) shows that these solutions have different parent bodies. Stream 372/04 lies in an area dominated by comets (see~\citep{2024A&A...682A.159J}) while stream 372/00 lies in an area devoid of comets.

Hence, we propose that solution 372/00 be given standalone status with a new name and codes, while solutions 372/02 and 372/04 be left unchanged in the MDC. We also suggest reconsidering the established status of 372/PPS.\\

{\bf 386/OBC  October beta-Camelopardalids. }
The values of $DH_{Min}=0.266$ and $DH_{MI}=0.059$ indicate misclassification of its only solutions 386/00 or 386/01. 
Solution 386/00 obtained from radio observations by \citep{2010Icar..207...66B}, 355 members, and 386/01 from video observations by \citep{2016Icar..266..355J}, 28 members.
The radiant of solution 386/01 is situated in constellation of Perseus.
The mean orbits of the two solutions differ mainly in their inclination to the plane of the ecliptic, $DH_{in}=0.260$. Other differences are negligible contributors. 
A considerable difference of about 6 degrees is in $\lambda_{SRC}$. All of this argues in favour of giving both solutions standalone stream status in the MDC, with 368/00 retaining its current name and codes. Solution 386/01 will be assigned new IAU MDC codes and a new name corresponding to its radiant position.\\

{\bf 392/NID November i-Draconids. }
\citep{2010Icar..207...66B}, among radio orbits, identified 2059 members of the 392/00 stream. \citep{2016Icar..266..331J}, among video orbits, found 65 members of the 392/01 solution. 
The values $DH_{Min}=0.317$ and $DH_{MI}=0.059$ suggest that these solutions do not represent the same stream. The reason for this lies in the large value of the components $DH_{in}=0.227$ and $DH_{\Pi}=0.222$, the differences in the inclination of the orbits and the orientation of the apse lines.

The geo-helio parameters of the 392/01 solution are inconsistent, problematic is the value of the longitude of the ascending node $\Omega=254.4^{\circ}$. This inconsistency was also noted by \citep{2020eMetN...5...93K}. Using $\Omega=242.0^{\circ}$ removes the incompatibility, however, the calculated new value of $DH_{Min}=0.182$ is still too large to consider both solutions as representing the same stream. Also the MSM method test resulted in the same conclusion. 

However, it is still a problem of rather large difference of $9^{\circ}$ in the $\lambda_{RSC}$ of the two solutions, which in this case cannot be explained by the diurnal motion of the two radians. The values of the Sun's ecliptic longitudes $\lambda_S$ differ by only 1 degree. 

The 392/NID stream was discovered and sent to MDC by \citep{2010Icar..207...66B} who noted that it has similar parameters to the 10/QUA stream. The authors write ' {\it ... a surprising
finding from our survey is an apparently new shower (NID) that
appears to be directly associated with the Quadrantids. The NID
has the same radiant location (in ecliptic coordinates) and
speed as the QUA and both have overlapping periods of activity.
Indeed, our automatic algorithm linked portions of the showers
as though they were a single long-duration shower, extending
the QUA period of activity into November. From our observations, we interpret the NID as simply an early extension of the QUA.}' 
This remark suggests that we may be dealing with a more complex structure in the case of 392/NID. A similar comment is made by Koseki, who believes that the stream may be an earlier activity of 334/DAD. \citep{2024book...Jen} consider 392/NID as the central  part of a larger complex including showers 334/DAD and 336/DKD.
However, this problem is beyond the scope of our current research. Therefore, we propose that both solutions 392/00 and 392/01, should represent separate streams in MDC, with the first solution retaining the name and codes. The WG's final decision may require a detailed discussion. \\

{\bf 490/DGE December delta-Eridanids.}
Both solutions come from video-observed meteor sightings.
490/00 was identified among a combined sample of orbits (64,650 + 40,744) determined by the SonotaCo consortium in Japan from 2007-2009 and by the Cameras for All-sky Meteor Surveillance (CAMS) project in California from October 2009 December 2011.
\citep{2014me13.conf..217R}  identified 7 members of the stream. The 490/01 solution (20 members) was obtained by \citep{2016Icar..266..371J}, among an enlarged combined sample of 168,000 SonotaCo and 238,000 CAMS orbits. 
 It is clear from \citep{2016Icar..266..371J} that the subsequent augmented sample of orbits included the data used to identify the 490/00 solution. This means that the two solutions cannot be considered independent representations of the 490/DGE stream.
The $DH_{Min}=0.149$ and $DH_{MI}=0.039$, the main difference takes place in the orientation of the apse lines of the two orbits, $DH_{\Pi}=0.137$.  We are unable to pinpoint the reason for this difference, hence, according to the adopted methodology, these solutions should be considered autonomous.  However, in view of the fact that both come from samples of orbits one of which is contained in the other, and especially the moderate value of the DH-function  we leave the problem of justifying a proper recommendation as open. \\

{\bf 507/UAN  upsilon-Andromedids.}
The first 507/00 solution was obtained from video observations (CAMS) over a two-month period from June 2 to August 2011. \citep{2013JIMO...41...43H} identified 13 members of the stream. The second solution 507/001, obtained from 28 members, \citep{2016Icar..266..384J} identified among 110,000 orbits determined from CAMS observations from October 2010 to March 2013.  Thus, it appears that solution 507/00 was obtained from the sample included in the sample in which solution 507/01 was identified. 

The $DH_{Min}=0.326$ and $DH_{MI}=0.039$ clearly suggest incorrect classification of at least one of the solutions. The mean orbits differ mainly by perihelion distances $DH_q=0.106$ and by orientation of the line of abse $DH_{\Pi}=0.299$. The reason for such a large value of $DH_{\Pi}$ is the difference in the argument of the perihelium of the two orbits, $\Delta\omega\sim20^{\circ}$. Unfortunately, we do not have the data of the members of these solutions at our disposal, so it is difficult to give reasons for this discrepancy. Hence, due to the extremely large value of $DH_{Min}=0.326$, we propose that they will obtain a status of autonomous streams in MDC, with solution 507/00 retaining the current name and codes.  \\

{\bf 512/RPU rho-Puppids.}
Stream 512/00, with 16 members, was identified by \citep{2013JIMO...41...70S} among 10645 video meteors observed from 2007-2010 by the Croatian Meteor Network and among 168xxx video meteors observed from 2007-2011 by the SonotaCo consortium. 
Stream 512/01, with 22 members, was identified by \citep{2016Icar..266..355J} among 110,000 meteors observed from 2010-2013 by the CAMS project. 
\\
For these solutions, $DH_{Min}=0.370$, $DH_{MI}=0.059$ which suggests a misclassification of one of them. The mean orbits differ mainly in the inclination $DH_{in}=0.131$, and in the orientation of the apse line $DH_{Pi}=0.346$. 

After the end of December 2022, \citep{Shiba2023} delivered another solution 512/03 to the MDC. Shiba searched the video orbits from the SonotaCo consortium catalogues. After incorporating this solution with two others, it turned out that they could be identified at $DH_{Min}=0.202$ and the 512/03 solution is more similar to 512/00 than to 512/01. The corresponding value $DH_{\Pi}=0.174$, with the greatest contribution from the orientation in the apse line $DH_{\Pi}=0.152$. 
We do not have access to the orbital data of the members of the 512/01 solution, which would allow for a more detailed analysis of the reason for its difference from the other solutions. But we think that some reason may be that \citep{2016Icar..266..355J} reported the median values as the mean parameters of his solution, and \citep{2013JIMO...41...70S} and \citep{Shiba2023} reported the arithmetic mean values of the stream members' parameters.

According to \citep{2013JIMO...41...70S},  the parent body of the 512/RPU  is most likely comet C/1879 M1. This conclusion was confirmed by \citep{2024MNRAS.535.3661D} for the solution 512/00 not for 512/01.
It is an important argument when it comes to the feasibility of this stream. Because of this we propose an autonomous status for the 512/01 solution, whose radiant is located in the Pyxis constellation, and retaining the 512/00 solution as the representative of the 512/RPU shower. In spite of excluding one of the shower solutions, due to the recent submission by \citep{Shiba2023}, the established status of the shower does not need to be reconsidered.\\

{\bf 555/OCP October gamma-Camelopardalids.}
The values of $DH_{Min}=0.282$ and $DH_{MI}=0.059$ indicate an incorrect classification of at least one of the solutions. Both solutions 555/00 and 555/01 were identified among the meteors observed by the video technique using the same orbital samples as for the 512/RPU stream discussed earlier. 
Stream 555/00, with 16 members, was identified by \citep{2014JIMO...42...90A},  
stream 555/01, with 14 members, was identified by \citep{2016Icar..266..355J}. The solutions differ practically only in the orientation of the apse line, $DH_{\Pi}=0.273$. 
This large $DH_{\Pi}$ value is the result of a 17-degree difference in the perihelion argument of the mean orbits. 
As in the case discussed earlier, the mean values of the parameters of these solutions were obtained by different methods, in solution 555/00 the arithmetic means were taken, in solution 555/01 the medians were taken. In the case of the asymmetric nature of the distribution of the parameters, this can lead to marked differences in the averaged values.
We do not have access to the parameter values of the members of the two solutions, which prevents further analysis of this case.  We therefore propose that solutions 555/00 and 555/01 be given autonomous status in MDC, with 555/00 retaining its current name and codes, while 555/01 being assigned new IAU MDC codes and a new name corresponding to its radiant position.\\

{\bf 574/GMA gamma-Ursae Majorids.}
Both solutions were identified among the meteors observed with video technology. \citep{2014JIMO...42..132G} identified 21 members of solution 574/00 among orbits observed from 2007 to 2010 by Croatian Meteor Network and among orbits observed from 2007 to 2011 by SonotaCo consortium. 
\citep{2022JIMO...50...38S} found 15 members of solution 574/01 among 298,689 orbits observed in the interval 2007--2020 by SonotaCo Network. 
Although \citep{2014JIMO...42..132G} does not specify the size of the sample searched, it can be deduced that they searched  among more than 100,000 SonotaCo orbits. So the two obtained solutions of the 574/GAM stream, probably are not fully independent. 
In view of the small number of members identified, we can conclude that the 574/GMA solutions indicate that we are dealing with a very faint stream.

$DH_{Min}=0.480$ and $DH_{MI}=0.059$, so there is undoubtedly a misclassification of at least one of the solutions.  
The large value of $DH_{Min}$ for these orbits is the result of large contributors in $DH_e=0.186$, $Dh_{in}=0.117$ and especially in $DH_{\Pi}=0.425$, i.e. due to significant differences in eccentricity, inclination and in the orientation of their apse lines. 

We propose that both solutions should be considered as solutions of different streams. Solution 574/00 has a radiant close star gamma-Ursae Majoris, so we can leave it in the MDC with unchanged codes and name. On the other hand, solution 574/1, with a radiant near the star 5-Canum Venaticorum must be given stand-alone status with a new designation.
\\

{\bf 644/JLL January lambda-Leonids.}
$DH_{Min}=0.256$ and $DH_{MI}=0.039$, thus, the correctness of classification test of the 644/00 and 644/02 solutions of this stream turned out to be negative. 
Both solutions were determined using video meteor observations. In the first case, \citep{2016Icar..266..384J} identified 24 meteoroids among 110000 orbits, while in the second case, \citep{2022JIMO...50...38S} identified 172 members among 298000 orbits.

The internal data consistency test of the 644/00 solution came out negative. 
 The critical contribution in the assessment of orbital similarity is due to the orientation of the apse lines of the two orbits $DH_{Pi}=0.228$. Which, in turn, has its source in the incompatibility of the ecliptic longitude of the Sun, $\lambda_S=288.0^{\circ}$, and the length of the ascending node $\Omega=277.7^{\circ}$ in the 644/00. In our opinion, as also noted by \citep{2016JIMO...44..151K} , the length of the ascending node should be $287.7^{\circ}$ instead of $277.7^{\circ}$. Substituting this value in the intrinsic consistency test yielded a positive result for this solution. 
The introduced correction of the length of the ascending node clearly decreased the value $DH_{Min}=0.121$. 

We do not have access to the individual data of members of the 644/00 and 644/02 streams to make a direct comparison and assess the impact of averaging methods on their orbital similarity.  In the 644/00 solution, \citep{2016Icar..266..384J} reports the medians of the parameters. \citep{2022JIMO...50...38S} calculated the arithmetic averages. The 644/00 stream activity lasted about 25 days, which means that the median values of the 24 members may not be sufficiently representative. Hence, it can be assumed that the $DH_{Min}=0.121$ value has some degree of variation due to the method used to average the stream parameters.
Thus, we believe that the 644/00 and 644/02 solutions do not require any change in MDC.\\

{\bf 709/LCM lambda-Canis Majorids.}
Both solutions to this stream were determined by the same authors and in the same video data samples as in the previous case. \citep{2016Icar..266..384J} identified 18 members in 709/00, while \citep{2022JIMO...50...38S} found 192 in 709/01. 
The values $DH_{Min}=0.142$, $DH_{Mi}=0.039$, formally, suggest a problem in the correctness of the classification of the two solutions. The greatest contribution comes from the inclination of the orbits, $DH_{in}=0.132$.
For both solutions, the internal data compatibility test was very positive. Hence, without access to the data of the members of the streams, we are unable to pinpoint the reason for the divergence of their mean orbits. 
In view of the rather small discrepancy, we propose leaving 709/00 and 709/01 unchanged in the MDC despite the fact that radiant 709/01 is located in the constellation Lepus, relatively far from the star $\lambda$ Canis Majoris.\\

{\bf 1048/JAS January psi-Scorpiids.}
$DH_{Min}=0.150$ and $DH_{Mi}=0.059$ indicate a discrepancy between solutions 1048/00 and 1048/01. 
Among those discussed previously, the case of the 1048/JAS stream is an exception. 
Both solutions were obtained using predictions of the stream parameters obtained from studies of the dynamic evolution of comet C/1936 O1, see \citep{2021PSS..19505152H}. The authors compared the parameters of the orbits of the hypothetical streams with the orbits of individual meteoroids collected in the IAU MDC Orbital Database
\footnote{The orbital part of the IAU MDC database is available at \url{https://ceres.ta3.sk/iaumdcdb/}}. 
Two groups of orbits were identified, the parent body of which could be comet C/1936 O1. Solution 1048/00 has ten members found by Hajdukov\'{a} \& Neslu\v{s}an in the CAMS \citep{Gural2011, Jenniskens_etal2011, Jenniskens_etal2016a, Jenniskens_etal2016b, Jenniskens_etal2016c, Jenniskens_Nenon2016} 
 and 1048/01 eight members in the SonotaCo database \citep{2009JIMO...37...55S, 2016JIMO...44...42S, SonotaCo2017, SonotaCo_etal2021}, IAU MDC version 2013.  The discrepancy in the parameters of the orbits of these solutions has its origin mainly in the inclination of the orbital planes $DH_{in}=0.120$.

Since 1048/JAS is a daytime shower (the angular distance of its mean radiant from the Sun at maximum activity is $52^{\circ}$) with a high mean inclination ($i \sim 144^{\circ}$ to $\sim$$151^{\circ}$), the meteoroids having a large geocentric speed of $60\,$km$\,$s$^{-1}$ \citep{2021PSS..19505152H}, and the significant difference of various parts of the stream was predicted by the simulation performed by these authors, we suggest to keep both solutions unchanged.

\section{False duplicates and search for the parent bodies}
\label{sec:parent_bodies}
It is worth noting that false duplicates cause a problem when we search for the parent body of the meteoroid stream. It may occur that a comet or asteroid can be identified as the parent associated with some solutions of stream, but cannot be identified with the other solution or solutions of the stream. Usually, this problematic identification is caused by false duplicates. 

In our previous paper \citep{2024MNRAS.535.3661D}, we searched for the parent bodies of known meteor showers, with known mean orbits, and found several examples of problematic parent-body identifications. These examples include not only the new associations discovered by the method we used, but also the suggestions published in the literature.  Several illustrative examples of the problematic identification are given below.

Besides these examples, the parent comets of 56 investigated problematic MSS are given as found for the individual solutions in the last column of Table~\ref{tab:MSS-1}. In some cases, a given parent comet was found for one solution, but not confirmed for other solution. This circumstance can also help to distinguish between the duplicate and false duplicate showers.\\

{\bf 11/EVI.} \citep{2006mspc.book.....J} suggested that comet D/1766 G1 is the parent body of shower $\eta$-Virginids, 11/EVI. We confirmed this suggestion with our method, but only for solution AdNo=1; see  \citep{2024MNRAS.535.3661D}. The solutions AdNo=0 and  AdNo=6 were not found to be associated with the comet; see Table~\ref{tab:MSS-1}. Which, in our opinion, supports the finding that the solutions given in 
Table~\ref{tab:MSS-1} not represent the same 11/EVI stream.  \\

{\bf 113/SDL.} In \citep{2024MNRAS.535.3661D} solution 113/00  was associated with comet 26P, but the comet does not share a relation with 113/01, 
which further supports our decision to consider the two solutions as separate showers.\\

{\bf 133/PUM.}
Here again, the proposed decision of splitting the solution of shower 133/PUM is supported by our parent body search, as only solution AdNo = 0 was associated with comet 79P. Upon splitting the solutions, the association with the comet is valid only for the Terenjeva stream. \\

{\bf 152/NOC.} 
For 152/NOC, \citep{2006mspc.book.....J} suggested an association with C/2003 Q1 for this MSS. However, in \citep{2024MNRAS.535.3661D} we have not confirmed this relationship for any of the three solutions of the 152/NOC stream.\\

{\bf 199/ADC.} Considering shower 199/ADC, the comet 45P was identified in \citep{2024MNRAS.535.3661D} as a parent body for solutions 199/00 and 199/02. This finding confirms the hypothesis of a common origin of the two 199/ACD solutions, while at the same time not denying that solutions 199/00 and 199/02 correspond to two branches of this stream, see Section~\ref{mss-2}.\\

{\bf 233/OCC.} For the October Capricornids we are dealing with a more complex case. We confirmed the relationship between 233/00 solution and comet D/1978 R1, already proposed by \citep{ 1988JIMO...16..191W}. But this comet was not associated with solution 233/01. Moreover, we found that solution 233/00 is also related with comet 15P/Finlay. Solution 233/01 is not associated with this comet, but is associated with comets 11P/Tempel-Swift-LINEAR, 85D/Boethin and 103P/Hartley 2; see \citep{2024MNRAS.535.3661D}. 
We also note that \citep{2017A&A...607A...5D} has identified many near-Earth asteroids associated with the 233/OCC stream, which undoubtedly means that we have a particularly complex group of objects here. \\

{\bf 254/PHO.} Also the case of stream 254/PHO is more complicated.
\citep{2024MNRAS.535.3661D} showed that the orbits of three comets are similar to the orbits of the 254/PHO solutions: 289P, 46P and 104P, see Table~\ref{tab:MSS-1}.
Thus, we have an unusual situation: basically, the 254/PHO stream, its both autonomous (in our opinion) solutions come from three different comets, or from any of these comets. However, resolving this question goes beyond the framework outlined in this study, it requires simulating the formation of the stream and its dynamical evolution.\\

{\bf 512/RPU.} 
Solutions 512/00 and of the $\rho$-Puppids, 512/RPU, were confirmed to be associated with comet C/1879 M1, but this association was not confirmed for solution 512/01.

\section{Conclusion and future action}
%
\label{sec:wnioski}
The research we conducted in Paper I revealed that some of 1182 meteor shower solutions given in the MDC database were unfortunate classified. The presence of misclassified meteor showers  degrade the database's rank, and the use of erroneous data can lead to the generation of artifacts. 
The methodology used in Paper I allowed us to conclude that among the 56 meteor 
showers represented by two or more solutions, most likely, there is a misclassification of solutions. 
The reason for this is, among other things, the lack of criteria to properly decide whether the new meteoroid data provided to the IAU MDC represent a newly discovered or another representation of an already known meteoroid stream. \\
In the present work, we examined the results from Paper I in greater detail and assessed the individual cases from various aspects. 
We relied on comparing the averaged geocentric and heliocentric parameters of meteoroid streams and also the relationship of the streams with their parent comets. 
As a result of the methodology used, taking into account its limitations listed in Section~\ref{sec:metodologia}, we considered 13 of 56 MSS streams to be correctly classified. We identified 43 streams as requiring reclassification, including formal assignment of new names and codes to them by the IAU WG. Among them are 11 cases of streams with an established status. These will require requalification back to working status.  Table~\ref{tab:proposals} 
lists the streams and their solutions that require such corrections.

In our research, we also encountered typographical errors in the data, some of them already noted by other authors. Table~\ref{tab:proposals} lists such cases, while giving the correct values of the erroneous data to be entered into the IAU MDC database.

Although our study is not “definitive” it clearly shows that in the past there has been repeated misclassification of new meteoroid streams delivered to the MDC. To avoid such errors, we propose that, based on the methods used in Paper I, a tool be made available on the MDC website to compare the new stream parameters provided with the contents of the database.

In Section~\ref{sec:metodologia} we have listed some of the reasons why two sets of averaged parameters of the same meteoroid stream may differ from each other. Large values of differences in parameters unquestionably indicate misclassification of at least one of the compared solutions. However, as we have seen in our research, assessing the correctness of the classification of meteoroid streams based only on their averaged parameters is not always possible. In such cases, it is invaluable to access the parameter values of all members of the stream. Therefore, the requirement (see \citep{2023A&A...671A.155H}) that when new averaged data are provided to the IAU MDC, their authors should also provide data of the members of the stream, we consider a very useful move. Having at one's disposal the parameter values of individual meteoroids allows a more objective classification of the stream they represent.  

Finally, we would like to expressly emphasize that the research results presented here only deal with the problem of misclassification of different solutions of the same stream. Thus, the results of our research should not be considered as prejudging the existence or non-existence of a given stream. 
%
\include{Tables/summary5}
%
\section{Acknowledgements}
This work was supported, in part, by VEGA - the Slovak Grant Agency for Science, grant No. 2/0009/22.
The research has also made use of the Astrophysics Data System, funded by NASA under Cooperative Agreement 80NSSC21M00561. 
Excerpts from this work were translated with DeepL.com (free version).

\bibliographystyle{unsrt} 
\bibliography{Shower-duplicity-PII} 
\appendix
\section{Meteoroid stream parameters }
Table~\ref{tab:MSS-1} lists the geocentric and heliocentric parameters of the multi-solution meteoroid streams from the IAU MDC database, among which false duplicates are possible.
The data was grouped into three sections. The first contains data of 3 streams for which the possibility of false duplicates is only of a formal nature; see Section \ref{sec:metodologia}. 
%
\newpage
\include{Tables/tabelaA1-v-3}
\newpage
\section{Values of the DH  orbital similarity function }
Table~\ref{tab:DH} gives the values of the $DH$ function of the orbital similarity, and their individual components for pairs of orbits listed in Table~\ref{tab:MSS-1}.
Given the Keplerian elements of the orbits S1 and S2 --- eccentricity $e$, perihelion distance $q$, inclination $i$, argument of perihelion $\omega$, and longitude of the ascending node $\Omega$ ---
the corresponding orbital similarity value $DH_P$ is given by the formula:
\begin{eqnarray}
\label{dhjopek}
{DH_P}^{2} & = &  {DH_e}^2 + {DH_q}^2 + {DH_{in}}^2 + {DH_{\Pi}}^2 = \nonumber \\ \nonumber \\
 & =&  [e_{1}-e_{2}]^{2}+\left[\frac{q_{1}-q_{2}}{q_{1}+q_{2}}\right]^{2}+\left[2\cdot\sin \frac{I_{12}}{2}\right]^{2} \nonumber \\ \nonumber \\
& & +\left[\frac{e_{1}+e_{2}}{2}\right]^{2}\left[2\cdot \sin \frac{\Pi_{12}}{2}\right]^{2}\, .
\end{eqnarray}
Where $I_{12}$ is the angle between the two orbital planes and $\Pi_{12}$ is the difference of the longitudes of perihelia measured from the intersection point of the orbits, namely:
\begin{equation}
\left[2 \sin \frac{I_{12}}{2} \right]^2 =\left[2 \sin \frac{i_1-i_2}{2} \right]^2 + \sin i_1 \sin i_2 \left[ 2 \sin \frac{\Omega_1-\Omega_2}{2} \right ]^2\,,
\nonumber
\label{anI12}
\end{equation}
\begin{equation}
\Pi_{12}=\left( \omega_1-\omega_2 \right)+
2 \arcsin \left[ \cos \frac{i_1+i_2}{2} \cdot \sin \frac{\Omega_1-\Omega_2}{2} \cdot
\sec \frac {I_{12}}{2} \right].
\label{PI12}
\end{equation}
 When  $\mid\Omega_1-\Omega_2\mid > 180^\circ$, the sign of  $\arcsin$ should be opposite.
\begin{table}
\begin{center}
\scriptsize
\caption{Values of the orbital similarity $DH_P$  and their components for $S1$-$S2$ solutions. 
$DH_e, DH_q, DH_{in}$ components of $DH_P$ in eccentricity, perihelion distance and mutual inclination of the two orbits. $DH_{\Pi}$ --- the component representing the orientation of the apsidal line measured from the common intersection node of the two orbits. }
\begin{tabular}{l c c c c c c }
\multicolumn{1}{c}{Shower Code} & \multicolumn{1}{c}{S1-S2} & \multicolumn{1}{c}{$DH_P$} & \multicolumn{1}{c}{$DH_e$} & \multicolumn{1}{c}{$DH_q$} &  \multicolumn{1}{c}{$DH_{in}$} &  \multicolumn{1}{c}{$DH_{\Pi}$}\\
\hline
\hline
 0003/SIA & 0-2 &  0.065 & 0.053 &  0.023 &  0.028 & 0.002 \\ 
 0127/MCA & 0-2 &  0.041 & 0.036 &  0.002 &  0.017 & 0.011 \\ 
 0321/TCB & 0-1 &  0.089 & 0.006 &  0.036 &  0.019 & 0.078 \\  
\hline
 0011/EVI & 0-1 &  0.322 & 0.061 &  0.240 &  0.056 & 0.197 \\ 
 0011/EVI & 0-6 &  0.326 & 0.090 &  0.309 &  0.046 & 0.032 \\ 
 0011/EVI & 1-6 &  0.203 & 0.029 &  0.074 &  0.101 & 0.157 \\ 
 0107/DCH & 0-1 &  0.526 & 0.490 &  0.011 &  0.146 & 0.124 \\ 
 0107/DCH & 0-2 &  0.358 & 0.330 &  0.005 &  0.085 & 0.111 \\ 
 0107/DCH & 1-2 &  0.279 & 0.160 &  0.005 &  0.228 & 0.018 \\ 
 0121/NHY & 0-2 &  0.201 & 0.003 &  0.075 &  0.159 & 0.097 \\ 
 0121/NHY & 0-3 &  0.289 & 0.196 &  0.022 &  0.043 & 0.207 \\ 
 0121/NHY & 2-3 &  0.269 & 0.199 &  0.096 &  0.119 & 0.097 \\ 
 0152/NOC & 0-1 &  0.788 & 0.041 &  0.223 &  0.557 & 0.509 \\ 
 0152/NOC & 0-2 &  0.173 & 0.037 &  0.044 &  0.139 & 0.085 \\ 
 0152/NOC & 0-4 &  0.200 & 0.067 &  0.059 &  0.074 & 0.164 \\ 
 0152/NOC & 1-2 &  0.644 & 0.004 &  0.181 &  0.429 & 0.445 \\ 
 0152/NOC & 1-4 &  0.687 & 0.026 &  0.278 &  0.522 & 0.348 \\ 
 0152/NOC & 2-4 &  0.193 & 0.030 &  0.103 &  0.138 & 0.083 \\ 
 0183/PAU & 0-1 &  0.401 & 0.001 &  0.164 &  0.332 & 0.152 \\ 
 0183/PAU & 0-4 &  0.294 & 0.020 &  0.115 &  0.189 & 0.193 \\ 
 0183/PAU & 1-4 &  0.241 & 0.019 &  0.051 &  0.234 & 0.018 \\ 
 0188/XRI & 0-1 &  0.534 & 0.010 &  0.385 &  0.293 & 0.226 \\ 
 0188/XRI & 0-2 &  0.268 & 0.003 &  0.250 &  0.096 & 0.017 \\ 
 0188/XRI & 1-2 &  0.687 & 0.013 &  0.579 &  0.313 & 0.196 \\ 
 0219/SAR & 0-1 &  0.580 & 0.003 &  0.452 &  0.143 & 0.334 \\ 
 0219/SAR & 0-2 &  0.217 & 0.074 &  0.104 &  0.043 & 0.171 \\ 
 0219/SAR & 0-3 &  0.461 & 0.024 &  0.328 &  0.195 & 0.257 \\ 
 0219/SAR & 1-2 &  0.414 & 0.071 &  0.364 &  0.101 & 0.152 \\ 
 0219/SAR & 1-3 &  0.373 & 0.021 &  0.145 &  0.335 & 0.072 \\ 
 0219/SAR & 2-3 &  0.344 & 0.050 &  0.232 &  0.235 & 0.080 \\ 
\hline
 0025/NOA & 0-1 &  0.210 & 0.093 &  0.109 &  0.147 & 0.044 \\ 
 0032/DLM & 0-1 &  0.321 & 0.143 &  0.123 &  0.074 & 0.249 \\ 
 0040/ZCY & 0-2 &  0.259 & 0.012 &  0.001 &  0.250 & 0.065 \\ 
 0076/KAQ & 0-1 &  0.142 & 0.050 &  0.010 &  0.054 & 0.121 \\ 
 0088/ODR & 0-1 &  0.362 & 0.131 &  0.003 &  0.300 & 0.156 \\ 
 0093/VEL & 0-1 &  0.412 & 0.330 &  0.039 &  0.213 & 0.118 \\ 
 0100/XSA & 0-1 &  0.209 & 0.044 &  0.147 &  0.056 & 0.131 \\ 
 0105/OCN & 0-1 &  0.134 & 0.052 &  0.000 &  0.090 & 0.085 \\ 
 0106/API & 0-1 &  0.329 & 0.220 &  0.000 &  0.021 & 0.244 \\ 
 0108/BTU & 0-1 &  0.344 & 0.340 &  0.000 &  0.052 & 0.000 \\ 
 0113/SDL & 0-1 &  0.172 & 0.012 &  0.114 &  0.041 & 0.122 \\ 
 0118/GNO & 0-1 &  0.798 & 0.190 &  0.195 &  0.275 & 0.698 \\ 
 0124/SVI & 0-1 &  0.226 & 0.068 &  0.062 &  0.105 & 0.177 \\ 
 0128/MKA & 0-1 &  0.315 & 0.030 &  0.250 &  0.013 & 0.189 \\ 
 0133/PUM & 0-1 &  0.139 & 0.089 &  0.005 &  0.046 & 0.096 \\ 
 0150/SOP & 1-8 &  0.343 & 0.057 &  0.073 &  0.023 & 0.329 \\ 
 0151/EAU & 0-1 &  0.155 & 0.081 &  0.067 &  0.099 & 0.057 \\ 
 0154/DEA & 0-1 &  0.178 & 0.002 &  0.007 &  0.013 & 0.177 \\ 
 0167/NSS & 0-1 &  0.282 & 0.073 &  0.067 &  0.066 & 0.256 \\ 
 0170/JBO & 0-4 &  0.220 & 0.067 &  0.006 &  0.049 & 0.203 \\ 
 0179/SCA & 0-1 &  0.234 & 0.119 &  0.172 &  0.068 & 0.078 \\ 
 
\hline
\hline
\end{tabular}
\label{tab:DH}
\end{center}
\normalsize
\end{table}
\addtocounter{table}{-1}
\begin{table*}
\begin{center}
\scriptsize
\caption{ Continuation.}
\begin{tabular}{l c c c c c c }
\multicolumn{1}{c}{Shower Code} & \multicolumn{1}{c}{S1-S2} & \multicolumn{1}{c}{$DH_P$} & \multicolumn{1}{c}{$DH_e$} & \multicolumn{1}{c}{$DH_q$} &  \multicolumn{1}{c}{$DH_{in}$} &  \multicolumn{1}{c}{$DH_{\Pi}$}\\
\hline
\hline
0186/EUM & 0-1 &  0.194 & 0.042 &  0.018 &  0.109 & 0.154 \\ 
 0189/DMC & 0-1 &  0.454 & 0.158 &  0.278 &  0.053 & 0.317 \\ 
 0197/AUD & 0-1 &  0.316 & 0.309 &  0.000 &  0.060 & 0.032 \\ 
 0199/ADC & 0-2 &  0.107 & 0.058 &  0.040 &  0.080 & 0.006 \\ 
 0202/ZCA & 0-1 &  0.434 & 0.009 &  0.277 &  0.107 & 0.316 \\ 
 0220/NDR & 0-1 &  0.144 & 0.045 &  0.000 &  0.101 & 0.093 \\ 
 0233/OCC & 0-1 &  0.160 & 0.038 &  0.002 &  0.035 & 0.151 \\ 
 0253/CMI & 0-3 &  0.347 & 0.012 &  0.182 &  0.195 & 0.221 \\ 
 0254/PHO & 0-1 &  0.199 & 0.038 &  0.023 &  0.184 & 0.063 \\ 
 0324/EPR & 0-1 &  0.250 & 0.011 &  0.125 &  0.211 & 0.047 \\ 
 0326/EPG & 0-1 &  0.175 & 0.035 &  0.091 &  0.125 & 0.075 \\ 
 0327/BEQ & 0-1 &  0.377 & 0.033 &  0.019 &  0.282 & 0.247 \\ 
 0334/DAD & 1-3 &  0.110 & 0.036 &  0.004 &  0.068 & 0.078 \\ 
 0347/BPG & 0-1 &  0.230 & 0.054 &  0.067 &  0.150 & 0.151 \\ 
 0372/PPS & 0-2 &  0.350 & 0.292 &  0.019 &  0.046 & 0.187 \\ 
 0386/OBC & 0-1 &  0.266 & 0.004 &  0.038 &  0.260 & 0.040 \\ 
 0392/NID & 0-1 &  0.317 & 0.003 &  0.007 &  0.227 & 0.222 \\ 
 0490/DGE & 0-1 &  0.149 & 0.033 &  0.036 &  0.035 & 0.137 \\ 
 0507/UAN & 0-1 &  0.326 & 0.058 &  0.105 &  0.053 & 0.299 \\ 
 0512/RPU & 0-1 &  0.370 & 0.002 &  0.001 &  0.131 & 0.346 \\ 
 0555/OCP & 0-1 &  0.282 & 0.006 &  0.055 &  0.046 & 0.273 \\ 
 0574/GMA & 0-1 &  0.480 & 0.186 &  0.036 &  0.117 & 0.425 \\ 
 0644/JLL & 0-2 &  0.256 & 0.010 &  0.095 &  0.069 & 0.228 \\ 
 0709/LCM & 0-1 &  0.142 & 0.003 &  0.033 &  0.132 & 0.038 \\ 
 1048/JAS & 0-1 &  0.150 & 0.012 &  0.030 &  0.120 & 0.085 \\  
\hline
\hline
\end{tabular}
\end{center}
\normalsize
\end{table*}

\end{document}

%% file: Tables/summary5.tex

\begin{table*}
\scriptsize
\caption{The proposed action  for individual shower solutions that require changes, and the consequent change in the shower status, to correct the IAU MDC List of Showers. Abbreviation NAS stands for ``new autonomous shower''; LRS - List of Removed Showers. Each new name for a NAS, when specified, corresponds to the original shower name given by the discoverers in their original publication.}
\begin{tabular}{rll}
\hline \hline
IAU No./AdNo. & action & remark/ main reason for the discrepancy \\
\hline
11/00 & NAS: a new IAU No., code, name as Northern Virginids & geocentric velocity off by $\sim6\,$km$\,$s$^{-1}$ \\
11/01 & move to the LRS & inconsistent parameters \\
11 & change the status from established to working & \\

107/00 & NAS: a new IAU No., code, and name & geocentric velocity, radiant\\
107/02 & NAS: a new IAU No., code, and name & geocentric velocity, radiant\\

121/00 & to correct R.A. ($168^{o}$),  & \\
 & NAS: a new IAU No., code, and name &  \\
121/02 & NAS: a new IAU No., code, name as Southern $\rho$-Leonids & a variation of the original name\\
121/03 & NAS: a new IAU No., code, and name &  \\
121 & move to the LRS  & None of the three solutions has a radiant\\
 &   &  consistent with the shower designation. \\

152/01 & NAS: a new IAU No., code, name as $\alpha$-Arietids,  & inclination; 152/01 is largely different \\
 &  & from the other three solutions of 152 \\

183/00 & NAS: a new IAU No., code, and name & inclination, geocentric velocity\\
188/01 & NAS: a new IAU No., code, and name & perihelion distance, inclination\\
188/02 & NAS: a new IAU No., code, and name & perihelion distance, inclination\\
188 & change the status from established to working & \\
219/00 & NAS: a new IAU No., code, and name & radiant, various orbital elements dependent \\
 &  & on the compared solution
\\
219/01 & NAS: a new IAU No. and code, name as $\gamma$-Arietids& various orbital elements \\

219/02 & NAS: a new IAU No. and code, name as $\rho$-Piscids-Arietids & various orbital elements \\ 
219/03 & NAS: a new IAU No. and code, name as Arietids-Piscids & inclination; too different from the other\\
 &  & solutions of 219\\ 
219 & move to the LRS  & None of the four solutions has a radiant\\
 &   &  consistent with the shower designation.\\
 
\hline

25/01 & move to 17/NTA as its solution with AdNo=12 & a small difference with solutions of 17/NTA;\\
 & & a large difference with 25/00 in geocentric\\ 
 & &  velocity $\sim6\,$km$\,$s$^{-1}$, in inclination $\sim7^{\circ}$\\
 
32/01 & move to the LRS & identical with 20/06 \\

76/00 & NAS: a new IAU No. and code, name as September $\iota$-Aquariids & souther branch of
 76 \\


88/01 & a new IAU No., code, and name &  geocentric velocity off by $\sim10\,$km$\,$s$^{-1}$,\\
 &  & inclination off by $\sim8^{\circ}$\\

93/00 & a new IAU No., code, and name &  radiant, difference in ecliptic longitudes of both\\
& & solutions in the rotating reference system \\ 
& & $\Delta_{\lambda_{RSC}}=40^{\circ}$\\

100/00 & NAS: retain IAU No. and code, name as $\xi$-Sagittariids & perihelion distance \\


 100 & change the status from established to working & \\

105/00 & NAS: a new IAU No. and code, name as Carinids &  \\

105/01 & to correct $\lambda_s=299.7^{\circ}$ & \\
 & NAS: a new IAU No., code, and name & \\

105 & move to the LRS & The existence of the complex is not\\
 &   &  substantiated.\\
 
106/00 & NAS: a new IAU No. and code, and name 
&  eccentricity
and orientation of the orbit\\

106/01 & NAS: a new IAU No., and code, and name 
&  \\

106 & move to the LRS & None of the two solutions has a radiant\\
 &  & consistent with the shower designation.\\

108/00 & NAS: a new IAU No. and code, and name 
& geocentric velocity $\sim4\,$km$\,$s$^{-1}$\\

\hline
\end{tabular}
\end{table*}

\addtocounter{table}{-1}
\begin{table*}
\scriptsize
\caption{$-$ continuation.}
\begin{tabular}{rll}
\hline
IAU No./AdNo. & action & remark \\
\hline
113/00 & correct the declination of radiant to $Decl. = 7.8^{\circ}$, & \\
 & NAS: a new IAU No. and code, name as $\pi$1-Cancrids & a variation of the original name \\

113/01 & NAS: a new IAU No., code, and name & perihelion distance and orientation of the orbit\\

113 & move to the LRS & None of the two solutions has a radiant\\
 &  & consistent with the shower designation.\\
 
118/00 & move to the LRS & identified as a stream by chance \\
118/01 & move to the LRS & identified as a stream by chance \\

124/01 & move to 223/GVI as its solution with AdNo=1  & $D_{h} = 0.040$ between 124/01 and 223/00\\

133/00 & NAS: a new IAU No., code, and name &  eccentricity
and orientation of the orbit, \\
 & & right ascensions off by about $30^{\circ}$\\
 
150/08 & NAS: 
designation M2022-W1 & submitted after the new nomenclature\\
 &  &  rules were accepted; orientation of the orbit\\

151/01 & NAS: a new IAU No., code and name & radiant, orientation of the orbit\\

151 & change the status from established to working & \\

167/00 & NAS: a new IAU No., code and name &  radiant\\

170/00 & NAS: a new IAU No., code, name as June $\alpha$-Draconids & radiant, $\Delta_{\lambda_{RSC}}=63^{\circ}$ \\
  
170 & reconsider the established status& \\

179/01 & NAS: a new IAU No., code and name 
& geocentric velocity\\

186/01 & NAS: a new IAU No., code, and name 
& $\Delta_{\lambda_{RSC}}=63^{\circ}$\\
 & 
 & radiant\\

189/01 & NAS: a new IAU No., code, and name & geocentric velocity off by $\sim10\,$km$\,$s$^{-1}$\\

197/00 & NAS: a new IAU No., code, and name & eccentricity differs by $\sim0.3$,\\
 &  & geocentric velocity off by $\sim4\,$km$\,$s$^{-1}$,\\
 &  & $\Delta \lambda_{RSC}=27^{\circ}$\\

197 & reconsider the established status& \\

199/00 & NAS: new IAU No. and code, 
& souther branch of the 199 stream\\

199/01 & NAS:  new IAU No. code, and name 
& northern branch of the 199 stream \\


202/00 & move to the LRS & inconsistent parameters\\
202 & change the status from established to working & \\


253/03 & NAS: a new IAU No., code, and name & orientation of the orbit\\

254/01 & NAS: a new IAU No., code, and name & radiant declination \\
254 & change the status from established to working & \\

324/01 & NAS: a new IAU No., code, and name & inclination\\ 
324 & change the status from established to working & \\
326/01 & NAS: a new IAU No., code, and name & inclination\\
326 & change the status from established to working & \\

327/01 & NAS: a new IAU No., code, and name & radiant, inclination\\
327 & change the status from established to working & \\

347/01 & NAS: a new IAU No., code, and name & \\

372/00 & NAS: a new IAU No., code, and name & eccentricity, orientation of the orbit\\
372 & reconsider the established status & \\

386/01 & NAS: a new IAU No., code, and name & radiant, inclination \\
392/00 & NAS: a new IAU No., code, and name & inclination, orientation of the orbit \\
507/01 & NAS: a new IAU No., code, and name & perihelion distance, difference in\\
 &  & argument of perihelion, $\Delta\omega\sim20^{\circ}$\\

512/01 & NAS: a new IAU No., code, and name & inclination, orientation of the orbit\\
555/01 & NAS: a new IAU No., code, and name & argument of perihelion differs by $\sim17^{\circ}$\\
574/01 & NAS: a new IAU No., code, and name & eccentricity, inclination,\\
& & orientation of their apse line\\
\hline \hline
\end{tabular}
\label{tab:proposals}
\end{table*}



%% file: Tables/tabelaA1-v-3.tex
\begin{table*}
\label{tab:MSS-1}
\begin{center}
\scriptsize
\caption{Heliocentric and geocentric parameters of $56$ problematic MSS. 
From the left, the columns include: the IAU multi solution shower code, the Keplerian orbital elements: the eccentricity, the perihelion distance [au], the perihelion argument, the length of the ascending node and the inclination of the orbit to the ecliptic. The following columns state geocentric parameters: the radiant coordinates right ascension and declination, the geocentric velocity [km/s], the ecliptic longitude of the Sun, the ecliptic Sun centered longitude of the radiant and the ecliptic latitude of the radiant.  
 The values of angular values given in degrees refer to J2000 reference system. $N$ --- lists the number of averaged orbits of each solution. An asterisk in the first column indicates that there is an inconsistency between geocentric and heliocentric data for this solution. $To$--- indicates the type of technique that was used to observe the members of the MSS in question: R -- radio, P -- photographic, V - Tv or video, V -- visual. In the last column 'nca' means 'none comet associated' according to \citep{2024MNRAS.535.3661D}. }
\begin{tabular}{@{ }p{0.1cm}@{ } l@{$\,$} c@{$\,\,\,\,$} c@{$\,\,\,\,$} c@{$\,\,\,\,$} c@{$\,\,\,\,$} c@{$\,\,\,\,$} c@{$\,\,\,\,$} c@{$\,\,\,\,$} c@{$\,\,\,\,$} c@{$\,\,\,\,$} c@{$\,\,\,\,$} c@{$\,\,\,\,$} c@{$\,\,\,\,$} c@{$\,\,\,\,$} c@{$\,\,\,\,$} l | l}
\multicolumn{2}{c}{Shower code} & \multicolumn{1}{c}{e} &\multicolumn{1}{c}{q} & \multicolumn{1}{c}{$\omega$} &  \multicolumn{1}{c}{$\Omega$} &  \multicolumn{1}{c}{inc} & 
\multicolumn{1}{c}{$\alpha_g$} &  \multicolumn{1}{c}{$\delta_g$} & \multicolumn{1}{c}{Vg}  & \multicolumn{1}{c}{$\theta$} &
\multicolumn{1}{c}{$\lambda_S$} &  \multicolumn{1}{c}{$\lambda_{RSC}$} & \multicolumn{1}{c }{$\beta_R$}  &
\multicolumn{1}{c}{N} & To &   References & Parent comet\\
\hline
\hline
 & 3/00/SIA &  0.912 &  0.208 &  131.8 &  311.7 &    6.9 &  334.0 &  -14.6  &  33.8 & 108.9 & 131.7 &  198.9 &   -3.5 &   12 & -  & \citep{1973NASSP.319..183C} & nca\\
 & 3/02/SIA &  0.859 &  0.218 &  134.3 &  309.1 &    5.3 &  332.9 &  -14.7  &  30.5 & 110.1 &  129.5 &  200.1 &   -3.3 &  353 & R & \citep{2008Icar..195..317B} &nca\\
\hline 
 & 127/00/MCA &  0.600 &  0.930 &  146.3 &  351.2 &   15.2 &    0.2 &   50.7  &  13.4 &  66.2  &  351.2 &   34.9 &   45.2 &    4 & P & \citep{1994PSS...42..151P}  & nca\\
  & 127/02/MCA &  0.636 &  0.934 &  148.8 &  349.7 &   14.3 &    4.7 &   48.1  &  13.6 &  62.7 &  349.7 &   37.8 &   41.4 &      & P & \citep{1989JIMO...17..242T} & nca\\
\hline 
 & 321/00/TCB &  0.166 &  0.924 &  124.9 & 296.5 &   77.0 &  232.3 &   35.8  &  38.7 & 127.1 &  296.5 &  279.2 &   52.4 & 1123 & R & \citep{2008Icar..195..317B} & nca\\
 & 321/01/TCB &  0.172 &  0.860 &  98.2 &  296.0 &   76.0 &  233.6 &   34.4  &  37.7 & 127.5 &  296.0 &  282.1 &   51.5 & 3560 & R & \citep{2010Icar..207...66B} &nca\\
\hline
 & 11/00/EVI & 0.912& 0.234 &308.0 &334.5 &  3.5 &173.6 &  4.9 &36.0 & 107.7 &334.5 &197.7 & 1.9 &4 & P & \citep{1971SCoA...12...14L} & nca\\
*&11/01/EVI  & 0.851& 0.382 &349.1 &280.5 &  3.5 &182.1 &  2.6 &29.2 & 169.9 &280.5 &260.4 & 3.2 &7 & -  & \citep{2006mspc.book.....J} & D/1766 G1 \\ 
 & 11/06/EVI  & 0.822& 0.443 &283.3 &357.2 &  5.5 &185.7 &  3.4 &27.3 &  96.7 &357.2 &186.7 & 5.4&158 &T & \citep{2022JIMO...50...38S} & nca\\
\hline
&107/00/DCH & 0.930& 0.950 &330.0 &144.7&  70.2& 254.5& -86.0& 42.6& 112.6 & 324.7&  303.0& -62.7& 4 & R & \citep{1975AuJPh..28..591G} & nca\\
&107/01/DCH & 0.440& 0.930 &340.0 &145.7&  61.9& 179.6& -83.1& 34.2& 113.1 & 325.7&  287.4& -65.7&47 & R & \citep{1975AuJPh..28..591G} & nca\\
&107/02/DCH & 0.600& 0.940 &338.0 &145.7&  75.0& 208.2& -78.1& 42.0& 119.8 & 325.7& 283.6& -59.27&33 & R &  \citep{1999md98.conf..307J} & nca\\
\hline
*&121/00/NHY & 0.714 &0.718&  70.8& 184.6 &  9.6 &158.6 &-11.1 &19.3 &  71.1 &  4.6 &160.0 &-18.6 & - & P & \citep{1989JIMO...17..242T} & nca\\
&121/02/NHY & 0.711 &0.618&  84.7& 162.8 &  0.5 &161.6 &  7.2 & 20.9 &  87.3 &343.0 &177.3 &-0.6 &29  & R & \citep{1976Icar...27..265S} & nca\\
&121/03/NHY & 0.910 &0.750&  62.0& 178.7 &  7.3 &159.6 & -5.1 & 20.1 &  74.9 &358.7 &164.5 &-12.7 &3  & R & \citep{1975AuJPh..28..591G} & nca\\
\hline
&152/00/NOC & 0.889 &0.108 & 25.6 & 47.8&  42.0&   2.2&  17.8& 33.0& 126.9 &  47.8& 321.5& 15.4& 16 & R & \citep{1976Icar...27..265S} & nca\\
&152/01/NOC & 0.930 &0.170 & 42.6 & 64.4&  10.2&  37.7&  19.9& 35.8& 112.6 &  64.4& 337.3&  4.8& 10 & R & \citep{1964AuJPh..17..205N} & nca\\
*&152/02/NOC & 0.926 &0.118 & 33.1 & 45.1&  34.2&   9.0& 17.3& 36.8& 119.6 &  45.5& 329.6& 12.3&1256 & R  & \citep{2008Icar..195..317B} & nca\\
*&152/04/NOC & 0.956 &0.096 & 31.4 & 53.6&  40.2&  17.1& 19.9& 40.3& 119.5 &  53.6& 329.8&  11.7& 12 & T & \citep{2022JIMO...50...38S} & nca\\
\hline
&183/00/PAU & 0.960 &0.170& 134.0 &305.7&  45.0& 340.7& -25.9& 40.5& 115.4 & 125.7& 206.6& -16.4& 32 & R & \citep{1967SCoA...11..183K} & nca\\
&183/01/PAU & 0.961 &0.122& 142.8 &306.2&  64.1& 347.9& -23.7& 44.1& 121.3 & 126.5& 212.9& -17.0& 91 & R & \citep{2008Icar..195..317B} & nca\\
&183/04/PAU & 0.980 &0.135& 139.1 &315.3&  53.1& 352.5& -20.5& 43.8& 117.7 & 136.0& 208.8& -15.8& 23 & T & \citep{2016Icar..266..331J}& nca\\
\hline
&188/00/XRI & 0.990 &0.080& 211.6& 301.2&  32.8&  93.7&  15.0& 44.0& 117.3 & 121.2& 332.4&  -8.4& 3 & R & \citep{1964AuJPh..17..205N} & nca\\
&188/01/XRI & 0.980 &0.180& 228.0& 297.7&  16.0&  98.7&  16.0& 38.4& 109.1 & 117.7& 340.8&  -7.2&23 & R & \citep{1967SCoA...11..183K} & nca\\
&188/02/XRI & 0.993 &0.048& 204.1& 311.3&  33.2& 102.9&  16.6& 45.4& 118.9 & 131.5& 330.9&  -6.3&204& R & \citep{2008Icar..195..317B} & nca\\
\hline
&219/00/SAR& 0.950 &0.180 &312.9 &183.3& 14.8& 19.3& 15.4& 36.7& 110.2 & 183.3& 200.3 & 6.8&   5 & R & \citep{1964AuJPh..17..205N} & nca\\
&219/01/SAR& 0.947 &0.068 &336.8 &179.5& 22.9& 28.2& 18.4& 36.3& 123.0 & 179.5& 213.2 & 6.4&  17 & R & \citep{1976Icar...27..265S} & nca\\
&219/02/SAR& 0.876 &0.146 &326.6 &180.2& 17.1& 23.8& 18.8& 31.2& 118.4 & 180.2& 208.7 & 8.3&  22 & R & \citep{1976Icar...27..265S} & nca\\
&219/03/SAR& 0.926 &0.091 &333.6 &178.3&  3.6& 27.2& 12.3& 33.7& 121.3 & 178.3& 211.3 & 1.1&  46 & R & \citep{1976Icar...27..265S} & nca\\
\hline
  & 0025/00/NOA &  0.950 &  0.220 &  307.0 &  201.7 &   12.0 &   34.7 &   20.1  &  36.3 & 107.3 &  201.7 &  197.4 &    5.9 &   15 & R & \citep{1967SCoA...11..183K} & nca\\
  & 0025/01/NOA &  0.857 &  0.274 &  305.4 &  206.1 &    3.6 &   38.6 &   17.8  &  30.1 & 106.8 &  205.0 &  196.8 &    2.5 &   53 & T & \citep{2016Icar..266..331J} & nca\\
\hline 
 * & 0032/00/DLM &  0.953 &  0.554 &  265.6 &  262.2 &  133.8 &  156.1 &   32.7  &  62.3 & 146.5 &  262.2 &  243.3 &   21.1 &    6 & T & \citep{2006mspc.book.....J} & nca\\
 * & 0032/01/DLM &  0.810 &  0.710 &  249.0 &  261.7 &  138.0 &  163.7 &   29.9  &  63.0 & 152.2 &  261.7 &  251.4 &   21.1 &    9 & R & \citep{1967SCoA...11..183K} & nca\\
\hline 
  & 0040/00/ZCY &  0.768 &  0.898 &  139.8 &   19.2 &   66.4 &  303.7 &   44.5  &  39.0 & 112.4 &   19.2 &  306.9 &   61.5 &   30 & R & \citep{1976Icar...27..265S} & nca\\
  & 0040/02/ZCY &  0.780 &  0.900 &  140.5 &   31.5 &   74.9 &  309.5 &   42.5  &  43.0 & 117.6 &   32.0 &  299.6 &   57.8 &   64 & T & \citep{2016Icar..266..331J} & nca\\
\hline 
  & 0076/00/KAQ &  0.655 &  0.884 &   45.2 &    0.5 &    1.0 &  334.7 &  -13.7  &  12.8 &  61.1 &  180.5 &  151.1 &   -3.0 &    3 & P & \citep{1994PSS...42..151P} & 11P, 103P\\
 * & 0076/01/KAQ &  0.705 &  0.867 &  229.2 &  186.6 &    2.1 &  341.6 &   -2.9  &  19.0 & 65.4 &  186.7 &  155.3 &    4.5 &    4 & P & \citep{1971SCoA...12....1L} & 11P\\
\hline 
  & 0088/00/ODR &  0.768 &  1.006 &  192.2 &  115.4 &   46.2 &  284.9 &   61.2  &  28.5 &  94.5 & 115.4 &  212.5 &   81.6 &   14 & R & \citep{1976Icar...27..265S} & nca\\
 * & 0088/01/ODR &  0.637 &  1.013 &  188.2 &  104.4 &   30.3 &  259.3 &   55.8  &  19.6 &  80.9 &  110.0 &  129.6 &   78.1 &   63 & T & \citep{2016Icar..266..355J} & nca\\
\hline
\hline
\end{tabular}
\end{center}
\normalsize
\end{table*}
\addtocounter{table}{-1}
\begin{table*}
\begin{center}
\scriptsize
\caption{ Continuation.}
\begin{tabular}{@{ }p{0.1cm}@{ } l@{$\,\,\,\,$} c@{$\,\,\,\,$} c@{$\,\,\,\,$} c@{$\,\,\,\,$} c@{$\,\,\,\,$} c@{$\,\,\,\,$} c@{$\,\,\,\,$} c@{$\,\,\,\,$} c@{$\,\,\,\,$} c@{$\,\,\,\,$} c@{$\,\,\,\,$} c@{$\,\,\,\,$} c@{$\,\,\,\,$} c@{$\,\,\,\,$} c@{$\,\,\,\,$} l | l}
\multicolumn{2}{c}{Shower code} & \multicolumn{1}{c}{e} &\multicolumn{1}{c}{q} & \multicolumn{1}{c}{$\omega$} &  \multicolumn{1}{c}{$\Omega$} &  \multicolumn{1}{c}{inc} & 
\multicolumn{1}{c}{RA} &  \multicolumn{1}{c}{DE} & \multicolumn{1}{c}{Vg}  & \multicolumn{1}{c}{$\theta$} & 
\multicolumn{1}{c}{$\lambda_S$} &  \multicolumn{1}{c}{$\lambda_{RSC}$} & \multicolumn{1}{c }{$\beta_R$}  &
\multicolumn{1}{c}{N} & To &   References & Parent comet\\
\hline
\hline 
  & 0093/00/VEL &  0.600 &  0.930 &   33.0 &  145.7 &   62.0 &  152.4 &  -65.1  &  35.3 & 111.1 &  325.7 &  240.3 &  -65.5 &    9 & R & \citep{1975AuJPh..28..591G} & nca\\
  & 0093/01/VEL &  0.930 &  0.860 &   43.0 &  143.7 &   49.9 &  133.4 &  -50.1  &  33.2 &  99.3 &  323.7 &  200.8 &  -62.8 &    3 & R & \citep{1975AuJPh..28..591G} & nca\\
\hline 
  & 0100/00/XSA &  0.780 &  0.383 &   66.6 &  296.1 &    4.3 &  284.8 &  -18.7  &  26.3 & 101.9 &  296.0 &  348.1 &    4.0 &   14 & R & \citep{1976Icar...27..265S} & nca\\
  & 0100/01/XSA &  0.736 &  0.285 &   46.9 &  305.9 &    1.1 &  284.4 &  -21.9  &  24.4 & 112.2 &  305.6 &  337.8 &    0.9 &   26 & R & \citep{1976Icar...27..265S} & nca\\
\hline 
  & 0105/00/OCN &  0.640 &  0.980 &    7.0 &  122.7 &   74.3 &  160.4 &  -63.1  &  41.5 & 118.4 &  302.7 &  264.2 &  -61.5 &    3 & R & \citep{1975AuJPh..28..591G} & nca\\
 * & 0105/01/OCN &  0.588 &  0.980 &    0.0 &  119.7 &   70.0 &  156.4 &  -65.3  &  38.4 & 113.3 &  323.4 &  245.1 &  -64.2 &    3 & R & \citep{1964PhDT........28N} & nca\\
\hline 
  & 0106/00/API &  0.890 &  0.980 &   12.0 &  145.7 &   47.2 &  110.1 &  -65.0  &  30.0 &  94.50 &  325.7 &  213.2 &  -81.7 &    3 & R & \citep{1975AuJPh..28..591G} & nca\\
  & 0106/01/API &  0.670 &  0.980 &  354.0 &  145.7 &   48.4 &   98.5 &  -76.0  &  28.9 &  99.1 &  325.7 &  292.2 &  -80.2 &    5 & R & \citep{1975AuJPh..28..591G} & nca\\
\hline 
  & 0108/00/BTU &  0.870 &  0.980 &  346.0 &  178.7 &   58.3 &   49.6 &  -77.9  &  36.3 & 103.9 &  358.7 &  300.5 &  -73.9 &   10 & R & \citep{1975AuJPh..28..591G} & nca\\
  & 0108/01/BTU &  0.530 &  0.980 &  346.0 &  178.7 &   55.3 &   50.3 &  -80.9  &  32.1 & 106.2 &  358.7 &  291.0 &  -72.6 &   11 & R & \citep{1975AuJPh..28..591G} & nca\\
\hline 
 * & 0113/00/SDL &  0.664 &  0.729 &   69.0 &  146.4 &    4.3 &  137.7 &   17.9  &  17.4 &  78.4 &  326.4 &  168.4 &    1.6 &  -  & P  & \citep{1989JIMO...17..242T} & 26P \\
  & 0113/01/SDL &  0.676 &  0.580 &   91.3 &  134.5 &    6.4 &  135.9  &    7.4  &  20.9 &  91.6 &  314.5 &  181.6 &   -9.0 &   37 & R & \citep{1976Icar...27..265S} & nca\\
\hline 
  & 0118/00/GNO &  0.620 &  0.980 &   13.0 &  178.7 &  121.6 &  263.1 &  -56.0  &  59.0 & 147.2 &  358.7 &  266.7 &  -32.7 &    6 & R & \citep{1975AuJPh..28..591G} & nca\\
 * & 0118/01/GNO &  0.430 &  0.660 &   97.0 &  179.7 & 137.4 &  250.9 &  -43.0  &  55.9 & 155.0 &  359.7 &  255.5 &  -20.6 &    3 & R & \citep{1975AuJPh..28..591G} & nca\\
  & 0124/00/SVI &  0.738 &  0.565 &   91.2 &  182.0 &    6.1 &  179.7 &   -8.3  &  22.9 &  91.1 &    2.0 &  181.1 &   -7.8 &   13 & R & \citep{1973Icar...18..253S} & nca\\
  & 0124/01/SVI &  0.670 &  0.640 &   83.0 &  175.7 &    0.1 &  172.6 &    2.9  &  20.1 &  86.4 &  355.7 &  176.4 &   -0.3 &    5 & R & \citep{1967SCoA...11..183K} & nca\\
\hline 
  & 0128/00/MKA &  0.890 &  0.180 &   42.0 &  359.7 &    1.8 &  338.6 &   -7.9  &  33.2 & 112.4 &  359.7 &  337.6 &    1.0 &    7 & R & \citep{1975AuJPh..28..591G} & nca\\
 * & 0128/01/MKA &  0.860 &  0.300 &   59.7 &  354.4 &   2.5 &  340.1 &   -7.5  &  29.8 & 105.5 &  354.4 &  344.5 &    0.8 &    3 & R & \citep{1964AuJPh..17..205N} & nca\\
\hline 
  & 0133/00/PUM &  0.562 &  1.002 &  174.6 &   27.1 &    6.8 &  117.0 &   52.9  &   8.4 &  32.2 &   27.1 &   81.5 &   31.2 &  -   & P & \citep{1989JIMO...17..242T} & 79P, 79P-B\\
  & 0133/01/PUM &  0.473 &  0.993 &  183.5 &   28.9 &    9.4 &  150.1 &   54.8  &   8.6 &  40.8 &   28.9 &  101.5 &   39.4 &   21 & R & \citep{1976Icar...27..265S} & nca\\
\hline
  & 0150/01/SOP &  0.792 &  0.331 &  120.8 &  244.4 &    4.3 &  257.9 &  -26.6  &  27.5 & 104.0 &   65.2 &  194.0 &   -3.6 &    7 & T & \citep{2010MNRAS.404..867J} & nca\\
  & 0150/08/SOP &  0.849 &  0.383 &  113.9 &  228.1 &    4.7 &  237.8 &  -24.2  &  29.0 &  99.5 &   51.3 &  189.5 &   -4.0 &  115 & T & \citep{2022JIMO...50...38S} & nca\\
\hline
  & 0151/00/EAU &  0.594 &  0.354 &  318.3 &   59.5 &   59.6 &  284.7 &   15.5  &  30.8 & 126.2 &   59.5 &  228.6 &   38.1 &   17 & R & \citep{1976Icar...27..265S} & nca\\
  & 0151/01/EAU &  0.513 &  0.405 &  322.8 &   62.5 &   64.6 &  294.1 &   20.4  &  31.5 & 129.4 &   63.0 &  237.6 &   41.3 &   11 & T & \citep{2016Icar..266..331J} & nca\\
\hline
   & 0154/00/DEA &  0.708 &  0.592 &   90.0 &   48.0 &    2.8 &   44.3 &   21.0  &  20.6 &  90.1 &   48.1 &  359.9 &    4.0 &   25  & R   & \citep{1976Icar...27..265S} & nca\\
   & 0154/01/DEA &  0.710 &  0.600 &   89.5 &   62.9 &    2.7 &   59.5 &   23.8  &  21.0 &  90.7 &   63.0 &  359.3 &    3.2 &    6  &  R &  \citep{1964AuJPh..17..205N} & nca\\
\hline 
  & 0167/00/NSS &  0.707 &  0.332 &  309.4 &   93.0 &    8.2 &  293.1 &  -14.1  &  23.2 & 109.4 &   93.0 &  199.6 &    7.6 &   45 & R & \citep{1976Icar...27..265S} & nca\\
  & 0167/01/NSS &  0.780 &  0.380 &  296.6 &   85.9 &    4.5 &  277.7 &  -20.0  &  26.5 & 101.3 &   85.9 &  191.3 &    3.2 &    4 & R & \citep{1964AuJPh..17..205N} & nca\\
\hline 
 * & 0170/00/JBO &  0.596 &  1.000 &  168.0 &   91.2 &   21.7 &  207.8 &   63.9  &  15.1 &  67.3 &   91.2 &   65.4 &   64.9 &   54 & R & \citep{1976Icar...27..265S} & 7P, 73P, 197P\\
  & 0170/04/JBO &  0.663 &  1.012 &  187.6 &   90.1 &   18.9 &  225.8 &   47.5  &  14.2 &  61.5 &   92.5 &  105.3 &   60.4 &    9 & T & \citep{2022JIMO...50...38S} & 7P, 73P\\
\hline
 & 0179/00/SCA &  0.792 &  0.272 &  311.2 &  110.1 &    4.5 &  311.3 &  -14.6  &  26.9 & 109.5 &  110.2 &  199.6 &    3.3 &   40 & R & \citep{1976Icar...27..265S} & nca\\
 * & 0179/01/SCA &  0.911 &  0.192 &  314.8 &  111.8 &    8.4 &  319.5 &  -10.9  &  34.1 & 110.4 &  118.0 &  200.5 &    4.6 &   20 & T & \citep{2016Icar..266..355J} & nca\\
 \hline 
  & 0186/00/EUM &  0.659 &  0.980 &  156.3 &  105.8 &   20.0 &  192.5 &   61.9  &  15.2 &  68.1 &  105.8 &   45.0 &   58.2 &   -  & P  & \citep{1989JIMO...17..242T} & 7P, 73P\\
  & 0186/01/EUM &  0.617 &  1.015 &  185.5 &   89.4 &   22.1 &  233.3 &   53.9  &  15.2 &  69.5 &   89.0 &  108.7 &   68.3 &   45 & T & \citep{2016Icar..266..355J} & 7P, 73P\\
\hline 
  & 0189/00/DMC &  0.742 &  0.443 &   71.0 &  128.6 &    2.1 &  122.4 &   22.3  &  24.3 &  98.9 &  128.7 &  351.1 &    2.2 &   43 & R & \citep{1976Icar...27..265S} & nca \\
  & 0189/01/DMC &  0.900 &  0.250 &   53.2 &  124.1 &    5.1 &  108.4 &   24.8  &  33.1 & 107.4 &  124.1 &  342.6 &    2.4 &    5 & R & \citep{1964AuJPh..17..205N} & nca \\
\hline 
  & 0197/00/AUD &  0.335 &  1.007 &  185.6 &  141.9 &   30.4 &  272.4 &   64.9  &  17.3 &  89.4 &  141.9 &  160.5 &   88.1 &   54 & R & \citep{1976Icar...27..265S} & nca \\
  & 0197/01/AUD &  0.644 &  1.008 &  188.7 &  142.6 &   33.8 &  271.7 &   58.9  &  21.1 &  88.4 &  143.0 &  133.6 &   82.3 &   17 & T & \citep{2016Icar..266..331J} & nca \\
\hline 
  & 0199/00/ADC &  0.753 &  0.597 &   87.3 &  327.8 &    2.8 &  329.4 &  -15.9  &  21.6 &  90.0 &  147.7 &  178.3 &   -3.2 &    6 & P & \citep{1994PSS...42..151P} & 45P \\
  & 0199/02/ADC &  0.811 &  0.551 &  270.9 &  143.8 &    1.8 &  325.3 &  -11.5  &  23.9 &  90.0 &  143.7 &  180.0 &    2.2 &  123 & T & \citep{2022eMetN...7..293R} & 45P \\
\hline 
 * & 0202/00/ZCA &  0.990 &  0.050 &  206.5 &  326.9 &   21.1 &  120.4 &   18.9  &  43.8 & 118.3 &  146.9 &  331.7 &   -1.5 &    3 & R & \citep{1964AuJPh..17..205N} & nca\\
  & 0202/01/ZCA &  0.981 &  0.088 &  212.6 &  340.0 &   16.6 &  136.1 &   11.7  &  42.1 &  114.8 &  160.0 &  335.1 &   -4.8 &  949 & R & \citep{2010Icar..207...66B} & nca\\
\hline 
  & 0220/00/NDR &  0.609 &  1.004 &  181.4 &  162.6 &   32.3 &  265.4 &   59.8  &  20.3 &  82.9 &  162.6 &   88.1 &   82.9 &   49 & R & \citep{1976Icar...27..265S} & nca\\
  & 0220/01/NDR &  0.654 &  1.004 &  182.0 &  171.7 &   28.8 &  267.7 &   54.0  &  18.9 &  77.4 &  172.0 &   91.8 &   77.4 &   51 & T & \citep{2016Icar..266..355J} & nca \\
\hline
  & 0233/00/OCC &  0.730 &  0.990 &  193.2 &  190.0 &    2.8 &  302.2 &   -8.6  &  10.0 &  25.0 &  190.0 &  112.5 &   11.3 &   -   & V & \citep{1988JIMO...16..191W} & 15P, D/1978 R1\\
  & 0233/01/OCC &  0.768 &  0.987 &  190.8 &  204.0 &    0.8 &  315.7 &  -13.9  &  10.4 &  20.5 &  203.8 &  110.3 &    2.8 &   -  & P &  \citep{1989JIMO...17..242T} & 11P, 85P, 103P\\
\hline 
  & 0253/00/CMI &  0.961 &  0.074 &  152.5 &   72.2 &   29.2 &  102.4 &   14.5  &  40.2 & 119.8 &  252.0 &  210.1 &   -8.4 &   32 & T & \citep{2016Icar..266..355J}& nca\\
  & 0253/03/CMI &  0.949 &  0.107 &  146.3 &   93.8 &   23.1 &  121.2 &   11.4  &  39.0 & 116.8 &  273.8 &  207.1 &   -8.8 &  183 & T & \citep{2022JIMO...50...38S}& nca\\
\hline 
 * & 0254/00/PHO &  0.670 &  0.990 &  359.0 &   74.7 &   13.0 &   15.6 &  -44.9  &  11.7 &  46.8 &  253.7 &   97.5 &  -46.3 &    -  & V & \citep{1973NASSP.319..183C} & 289P, 46P\\
   & 0254/01/PHO &  0.708 &  0.946 &   26.6 &   52.2 &    2.7 &    6.7 &   -7.7  &  10.9 &  41.8 &  232.2 &  130.9 &   -9.7 &    9 & T & \citep{2022JIMO...50...38S} & 289P, 46P, 104P\\
\hline 
  & 0324/00/EPR &  0.971 &  0.130 &   39.7 &   96.0 &   63.0 &   58.2 &   37.9  &  44.8 & 119.8 &   95.5 &  328.7 &   17.3 &  203 & R & \cite{2008Icar..195..317B} & nca\\
  & 0324/01/EPR &  0.982 &  0.167 &   46.8 &   87.9 &   53.0 &   53.8 &   37.8  &  43.8 & 115.9 &   88.0 &  332.6 &   18.0 &    4 & T & \citep{2016Icar..266..331J} & nca\\
\hline 
  & 0326/00/EPG &  0.771 &  0.173 &  334.9 &  105.2 &   55.4 &  326.3 &   14.7  &  29.9 & 132.1 &  105.5 &  228.6 &   26.5 &   62 & R & \citep{2008Icar..195..317B} & nca\\
  & 0326/01/EPG &  0.806 &  0.144 &  337.8 &  109.3 &   49.0 &  330.2 &   13.0  &  28.4 & 133.2 &  109.0 &  228.2 &   23.5 &   33 & T & \citep{2016Icar..266..331J} & nca\\
  \hline

\hline
\end{tabular}
\end{center}
\normalsize
\end{table*}

\addtocounter{table}{-1}
\begin{table*}
\begin{center}
\scriptsize
\caption{ Continuation.}
\begin{tabular}{@{ }p{0.1cm}@{ } l@{$\,\,\,\,$} c@{$\,\,\,\,$} c@{$\,\,\,\,$} c@{$\,\,\,\,$} c@{$\,\,\,\,$} c@{$\,\,\,\,$} c@{$\,\,\,\,$} c@{$\,\,\,\,$} c@{$\,\,\,\,$} c@{$\,\,\,\,$} c@{$\,\,\,\,$} c@{$\,\,\,\,$} c@{$\,\,\,\,$} c@{$\,\,\,\,$} c@{$\,\,\,\,$} l | l}
\multicolumn{2}{c}{Shower code} & \multicolumn{1}{c}{e} &\multicolumn{1}{c}{q} & \multicolumn{1}{c}{$\omega$} &  \multicolumn{1}{c}{$\Omega$} &  \multicolumn{1}{c}{inc} & 
\multicolumn{1}{c}{RA} &  \multicolumn{1}{c}{DE} & \multicolumn{1}{c}{Vg}  & \multicolumn{1}{c}{$\theta$} & 
\multicolumn{1}{c}{$\lambda_S$} &  \multicolumn{1}{c}{$\lambda_{RSC}$} & \multicolumn{1}{c }{$\beta_R$}  &
\multicolumn{1}{c}{N} & To &   References & Parent comet\\
\hline
\hline 
  & 0327/00/BEQ &  0.816 &  0.163 &  330.3 &  106.2 &   49.7 &  321.5 &    8.7  &  31.6 & 126.8 &  106.5 &  220.4 &   22.6 &   89 & R & \citep{2008Icar..195..317B}& nca\\
  & 0327/01/BEQ &  0.849 &  0.157 &  327.6 &   84.8 &   46.5 &  305.1 &    1.1  &  33.2 & 124.2 &   91.0 &  216.7 &   20.1 &   38 & T & \citep{2016Icar..266..331J} & nca\\
\hline 
  & 0334/01/DAD &  0.603 &  0.983 &  177.4 &  254.8 &   71.8 &  210.8 &   58.6  &  40.8 & 117.2 &  256.0 &  272.0 &   62.8 &   47 & T & \citep{2016Icar..266..331J} & nca\\
  & 0334/03/DAD &  0.639 &  0.975 &  185.8 &  250.9 &   72.9 &  203.0 &   62.3  &  41.2 & 117.7 &  250.9 &  265.7 &   62.2 &  487 & T & \citep{2022JIMO...50...38S} & nca\\
\hline 
  & 0347/00/BPG &  0.890 &  0.304 &   61.1 &   36.0 &   62.7 &  350.5 &   27.8  &  41.0 & 118.1 &   36.0 &  327.4 &   29.1 & 1105 & R & \citep{2010Icar..207...66B} & nca\\
  & 0347/01/BPG &  0.944 &  0.347 &   68.0 &   42.3 &   69.1 &  354.3 &   30.8  &  44.2 & 118.6 &   42.0 &  326.4 &   30.2 &   11 & T & \citep{2016Icar..266..355J} & nca\\
\hline 
  & 0372/00/PPS &  0.590 &  0.856 &  125.0 &  106.0 &  152.6 &   20.1 &   24.1  &  62.9 & 161.5 &  106.0 &  281.7 &   14.5 & 1395 & R & \citep{2010Icar..207...66B}& nca\\
  & 0372/02/PPS &  0.882 &  0.889 &  136.9 &  102.9 &  150.4 &   17.0 &   25.0  &  66.5 & 159.6 & 103.0 &  282.4 &   16.4 &  379 & T & \citep{2016Icar..266..331J}& nca\\
\hline 
  & 0386/00/OBC &  0.936 &  0.417 &  281.5 &  214.0 &   80.9 &   66.8 &   56.2  &  47.6 & 122.7 &  214.0 &  220.7 &   34.0 &  355 & R & \citep{2010Icar..207...66B} & nca\\
  & 0386/01/OBC &  0.932 &  0.387 &  286.3 &  205.2 &   68.6 &   45.8 &   52.3  &  43.6 & 118.0 &  205.0 &  214.3 &   33.5 &   28 & T & \citep{2016Icar..266..355J} & nca\\
\hline 
  & 0392/00/NID &  0.737 &  0.987 &  181.1 &  241.0 &   74.9 &  200.1 &   64.5  &  43.0 & 117.5 &  241.0 &  270.1 &   62.5 & 2059 &  R & \citep{2010Icar..207...66B} & nca\\
 *& 0392/01/NID &  0.734 &  0.973 &  194.7 &  254.4 &   72.9 &  196.5 &   68.0  &  41.9 & 116.4 &  242.0 &  260.9 &   63.2 &   65 & T & \citep{2016Icar..266..331J} & nca\\
\hline 
  & 0490/00/DGE &  0.818 &  0.716 &   66.8 &   69.3 &   23.5 &   69.5 &  -13.6  &  23.8 &  86.8 &  249.3 &  176.1 &  -35.3 &    7 & T & \citep{2014me13.conf..217R} & nca\\
  & 0490/01/DGE &  0.851 &  0.666 &   72.5 &   73.3 &   22.2 &   73.3 &   -7.6  &  24.7 &  87.2 &  254.0 &  176.8 &  -29.9 &   20 & T & \citep{2016Icar..266..371J} & nca\\
\hline 
  & 0507/00/UAN &  0.968 &  0.688 &  110.3 &   98.0 &  116.4 &   19.8 &   42.5  &  58.8 & 139.1 &   98.0 &  297.7 &   31.4 &   13 & T & \citep{2013JIMO...41...43H} & nca\\
  & 0507/01/UAN &  0.910 &  0.849 &  130.0 &  101.0 &  117.8 &    7.1 &   40.3  &  59.3 & 142.1 &   96.0 &  288.4 &   33.8 &   28 & T & \citep{2016Icar..266..384J} & nca\\
\hline 
  & 0512/00/RPU &  0.913 &  0.985 &    9.0 &   43.0 &  106.4 &  120.0 &  -24.0  &  57.5 & 136.4 &  223.0 &  266.0 &  -43.5 &   16 & T & \citep{2013JIMO...41...70S} & C/1879 M1 \\
   & 0512/01/RPU &  0.915 &  0.987 &  349.4 &   50.8 &  107.0 &  130.4 &  -26.3  &  57.8 & 137.3 &  231.0 &  271.2 &  -42.7 &   22 & T & \citep{2016Icar..266..355J} & nca\\
\hline 
  & 0555/00/OCP &  0.948 &  0.897 &  217.8 &  191.3 &   89.7 &   63.3 &   72.9  &  50.8 & 125.8 &  191.0 &  247.0 &   50.6 &   16 & T & \citep{2014JIMO...42...90A} & nca\\
  & 0555/01/OCP &  0.942 &  0.804 &  234.4 &  188.7 &   89.5 &   48.8 &   67.4  &  50.9 & 125.8 &  189.0 &  239.2 &   47.1 &   14 & T & \citep{2016Icar..266..355J}& nca\\
\hline 
  & 0574/00/GMA &  0.891 &  0.918 &  211.2 &  251.7 &   98.7 &  173.3 &   54.7  &  54.4 & 131.7 &  252.0 &  254.0 &   46.2 &   21 & T & \citep{2014JIMO...42..132G} & nca\\
  & 0574/01/GMA &  0.705 &  0.986 &  179.4 &  245.1 &   97.3 &  184.1 &   52.1  &  52.4 & 132.1 &  245.1 &  270.8 &   47.9 &   15 & T & \citep{2022JIMO...50...38S}& nca\\
\hline 
 * & 0644/00/JLL &  0.947 &  0.098 &  327.3 &  277.7 &   22.4 &  140.2 &   23.4  &  38.6 & 117.1 &  288.0 &  207.3 &    7.5 &   24 & T & \citep{2016Icar..266..384J}& nca\\
  & 0644/02/JLL &  0.957 &  0.081 &  331.4 &  288.1 &   21.8 &  141.3 &   21.2  &  39.7 & 118.7 &  288.1 &  208.9 &    5.7 &  172 & T & \citep{2022JIMO...50...38S}& nca\\
\hline 
  & 0709/00/LCM &  0.826 &  0.833 &   48.5 &  105.8 &   33.2 &   98.4 &  -31.5  &  25.4 &  87.9 &  286.0 &  176.4 &  -54.6 &   18 & T & \citep{2016Icar..266..384J} & nca\\
  & 0709/01/LCM &  0.829 &  0.779 &   56.6 &   93.3 &   29.2 &   88.8 &  -22.5  &  25.1 &  86.6 &  273.3 &  175.1 &  -45.9 &  192 & T & \citep{2022JIMO...50...38S}& nca\\
\hline 
  & 1048/00/JAS &  0.990 &  0.209 &   54.0 &  294.6 &  144.2 &  243.5 &  -10.2  &  60.2 & 139.9 &  294.6 &  308.8 &   10.8 &   10 & T & \citep{2021PSS..19505152H}& nca \\
  & 1048/01/JAS &  0.978 &  0.197 &   51.4 &  297.4 &  150.9 &  245.2 &  -12.4  &  60.0 & 141.0 &  297.4 &  308.1 &    8.9 &    9 & T & \citep{2021PSS..19505152H}& nca\\
\hline
\hline
\end{tabular}
\end{center}
\normalsize
\end{table*}

%% file: Shower-duplicity-PII.bbl
\begin{thebibliography}{10}

\bibitem{Vaubaillon_etal2019}
J{\'e}r{\'e}mie {Vaubaillon}, Lubo{\v{s}} {Neslu{\v{s}}an}, Aswin {Sekhar}, Regina {Rudawska}, and Galina~O. {Ryabova}.
\newblock {\em {From Parent Body to Meteor Shower: The Dynamics of Meteoroid Streams}}, page 161.
\newblock Cambridge University Press, 2019.

\bibitem{Egal_etal2023}
Auriane {Egal}, Paul~A. {Wiegert}, Peter~G. {Brown}, and Denis {Vida}.
\newblock {Modeling the 2022 {\ensuremath{\tau}}-Herculid Outburst}.
\newblock {\em The Astrophysical Journal}, 949(2):96, June 2023.

\bibitem{Moorhead_etal2017}
Althea~V. {Moorhead}, William~J. {Cooke}, and Margaret~D. {Campbell-Brown}.
\newblock {Meteor shower forecasting for spacecraft operations}.
\newblock In {\em 7th European Conference on Space Debris}, page~9, April 2017.

\bibitem{2025AdSpR..75.1145M}
Althea~V. {Moorhead}, William~J. {Cooke}, Peter~G. {Brown}, and Margaret~D. {Campbell-Brown}.
\newblock {The threshold at which a meteor shower becomes hazardous to spacecraft}.
\newblock {\em Advances in Space Research}, 75(1):1145--1162, January 2025.

\bibitem{2024A&A...682A.159J}
T.~J. {Jopek}, L.~{Neslu{\v{s}}an}, R.~{Rudawska}, and M.~{Hajdukov{\'a}}.
\newblock {Search for duplicates of showers in the IAU MDC database. Methods and general results}.
\newblock {\em Astronomy \& Astrophysics}, 682:A159, February 2024.

\bibitem{Odeh_etal2023}
Mohammad~Sh. {Odeh}, Mashhoor~A. {Al-Wardat}, and Peter {Jenniskens}.
\newblock {New showers identified among meteors observed in the UAE}.
\newblock {\em Experimental Astronomy}, 56(2-3):793--819, July 2023.

\bibitem{2023A&A...671A.155H}
M.~{Hajdukov{\'a}}, R.~{Rudawska}, T.~J. {Jopek}, M.~{Koseki}, G.~{Kokhirova}, and L.~{Neslu{\v{s}}an}.
\newblock {Modification of the Shower Database of the IAU Meteor Data Center}.
\newblock {\em Astronomy \& Astrophysics}, 671:A155, March 2023.

\bibitem{2024book...Jen}
Peter {Jenniskens}.
\newblock {\em {Atlas of Earth's Meteor Showers}}.
\newblock Elsevier, 2023.

\bibitem{2023NewAR..9601671J}
Tadeusz~J. {Jopek}, M{\'a}ria {Hajdukov{\'a}}, Regina {Rudawska}, Masahiro {Koseki}, Gulchehra {Kokhirova}, and Lubo{\v{s}} {Neslu{\v{s}}an}.
\newblock {New nomenclature rules for meteor showers adopted}.
\newblock {\em New Astronomy Reviews}, 96:101671, June 2023.

\bibitem{1993Icar..106..603J}
Tadeusz~J. {Jopek}.
\newblock {Remarks on the Meteor Orbital Similarity D-Criterion}.
\newblock {\em Icarus}, 106(2):603--607, December 1993.

\bibitem{1997A&A...320..631J}
T.~J. {Jopek} and C.~{Froeschle}.
\newblock {A stream search among 502 TV meteor orbits. an objective approach.}
\newblock {\em Astronomy \& Astrophysics}, 320:631--641, April 1997.

\bibitem{2017P&SS..143...43J}
Tadeusz~Jan {Jopek} and Ma{\l}gorzata {Bronikowska}.
\newblock {Probability of coincidental similarity among the orbits of small bodies - I. Pairing}.
\newblock {\em Planetary and Space Science}, 143:43--52, September 2017.

\bibitem{2024CoSka..54a..57N}
L.~{Neslu{\v{s}}an}, R.~{Rudawska}, M.~{Hajdukov{\'a}}, S.~{{\v{D}}uri{\v{s}}ov{\'a}}, and T.~J. {Jopek}.
\newblock {The computer programs to check the internal consistency of the meteor-shower data}.
\newblock {\em Contributions of the Astronomical Observatory Skalnate Pleso}, 54(1):57--71, April 2024.

\bibitem{2009JIMO...37...21K}
D.~{Koschny}, R.~{Arlt}, G.~{Barentsen}, P.~{Atreya}, J.~{Flohrer}, T.~{Jopek}, A.~{Kn{\"o}fel}, P.~{Koten}, H.~{L{\"u}then}, J.~{McAuliffe}, J.~{Oberst}, J.~{T{\'o}th}, J.~{Vaubaillon}, R.~{Weryk}, and M.~{Wisniewski}.
\newblock {Report from the ISSI team meeting ``A Virtual Observatory for meteoroids''}.
\newblock {\em WGN, Journal of the International Meteor Organization}, 37(1):21--27, February 2009.

\bibitem{1957SCoA....1..183W}
Fred~L. {Whipple} and Luigi~G. {Jacchia}.
\newblock {Reduction Methods for Photographic Meteor Trails}.
\newblock {\em Smithsonian Contributions to Astrophysics}, 1:183--206, January 1957.

\bibitem{1961deor.book.....D}
A.~D. {Dubiago}.
\newblock {\em {The determination of orbits.}}
\newblock New York, Macmillan, 1961, 1961.

\bibitem{1987BAICz..38..222C}
Zdenek {Ceplecha}.
\newblock {Geometric, Dynamic, Orbital and Photometric Data on Meteoroids From Photographic Fireball Networks}.
\newblock {\em Bulletin of the Astronomical Institutes of Czechoslovakia}, 38:222, July 1987.

\bibitem{1990BAICz..41..391B}
J.~{Borovicka}.
\newblock {The Comparison of Two Methods of Determining Meteor Trajectories from Photographs}.
\newblock {\em Bulletin of the Astronomical Institutes of Czechoslovakia}, 41:391, December 1990.

\bibitem{1995A&AS..112..173B}
J.~{Borovicka}, P.~{Spurny}, and J.~{Keclikova}.
\newblock {A new positional astrometric method for all-sky cameras.}
\newblock {\em Astronomy and Astrophysics Supplement}, 112:173, July 1995.

\bibitem{2010epsc.conf..888J}
T.~J. {Jopek} and G.~B. {Valsecchi}.
\newblock {From meteor observations to meteoroid orbits: propagation of uncertainties}.
\newblock In {\em European Planetary Science Congress 2010}, page 888, September 2010.

\bibitem{2015P&SS..117..223D}
Vasily {Dmitriev}, Valery {Lupovka}, and Maria {Gritsevich}.
\newblock {Orbit determination based on meteor observations using numerical integration of equations of motion}.
\newblock {\em Planetary and Space Science}, 117:223--235, November 2015.

\bibitem{2007MNRAS.375..595W}
I.~P. {Williams} and D.~C. {Jones}.
\newblock {How useful is the `mean stream' in discussing meteoroid stream evolution?}
\newblock {\em Monthly Notices of the Royal Astronomical Society}, 375(2):595--603, February 2007.

\bibitem{1999md98.conf..199V}
Yu. {Voloshchuk}.
\newblock {Mean orbits of meteor streams}.
\newblock In W.~J. {Baggaley} and V.~{Porubcan}, editors, {\em Meteoroids 1998, Proceedings of the International Conference held at Tatranska Lomnica, Slovakia, August 17-21, 1998}, pages 199--202. Astronomical Institute of the Slovak Academy of Sciences, January 1999.

\bibitem{1999SoSyR..33..302V}
Yu.~I. {Voloshchuk} and B.~L. {Kashcheev}.
\newblock {A Method for Calculating the Mean Orbits of Meteor Streams}.
\newblock {\em Solar System Research}, 33:302, January 1999.

\bibitem{2006MNRAS.371.1367J}
T.~J. {Jopek}, R.~{Rudawska}, and H.~{Pretka-Ziomek}.
\newblock {Calculation of the mean orbit of a meteoroid stream}.
\newblock {\em Monthly Notices of the Royal Astronomical Society}, 371(3):1367--1372, September 2006.

\bibitem{2010pim8.conf...91J}
Tadeusz~T. {Jopek}, Regina {Rudawska}, and Halina {Ziomek-Pretka}.
\newblock {Remarks on the mean orbits of the meteoroid stream}.
\newblock In {\em Proc. of the International Meteor Conference, 27th IMC, Sachticka, Slovakia, 2008}, pages 91--97, August 2010.

\bibitem{2020P&SS..18204821J}
Peter {Jenniskens}, Tadeusz~J. {Jopek}, Diego {Janches}, Maria {Hajdukov{\'a}}, Gulchehra~I. {Kokhirova}, and Regina {Rudawska}.
\newblock {On removing showers from the IAU Working List of Meteor Showers}.
\newblock {\em Planetary Space Science}, 182:104821, March 2020.

\bibitem{2024MNRAS.535.3661D}
S.~{{{D}}uri{\v{s}}ov{\'a}}, L.~{Neslu{\v{s}}an}, M.~{Hajdukov{\'a}}, R.~{Rudawska}, and T.~J. {Jopek}.
\newblock {Parent comets of IAU MDC meteoroid streams unaltered by dynamical evolution}.
\newblock {\em Monthly Notices of the Royal Astronomical Society}, 535(4):3661--3685, December 2024.

\bibitem{2006mspc.book.....J}
Peter {Jenniskens}.
\newblock {\em {Meteor Showers and their Parent Comets}}.
\newblock Cambridge University Press, 2006.

\bibitem{2011msss.conf....7J}
T.~J. {Jopek} and P.~M. {Jenniskens}.
\newblock {The Working Group on Meteor Showers Nomenclature: A History, Current Status and a Call for Contributions}.
\newblock In W.~J. {Cooke}, D.~E. {Moser}, B.~F. {Hardin}, and D.~{Janches}, editors, {\em Meteoroids: The Smallest Solar System Bodies}, pages 7--13, July 2011.

\bibitem{1971SCoA...12...14L}
B.~A. {Lindblad}.
\newblock {A computerized stream search among 2401 photographic meteor orbits.}
\newblock {\em Smithsonian Contributions to Astrophysics}, 12:14--24, January 1971.

\bibitem{2022JIMO...50...38S}
Yasuo {Shiba}.
\newblock {Jupiter Family Meteor Showers by SonotaCo Network Observations}.
\newblock {\em WGN, Journal of the International Meteor Organization}, 50(2):38--61, April 2022.

\bibitem{2016JIMO...44..151K}
M.~{Koseki}.
\newblock {Research on the IAU meteor shower database}.
\newblock {\em WGN, Journal of the International Meteor Organization}, 44(5):151--169, October 2016.

\bibitem{1975AuJPh..28..591G}
G.~{Gartrell} and W.~G. {Elford}.
\newblock {Southern hemisphere meteor stream determinations.}
\newblock {\em Australian Journal of Physics}, 28:591--620, October 1975.

\bibitem{1999md98.conf..307J}
T.~J. {Jopek}, G.~B. {Valsecchi}, and Cl. {Froeschle}.
\newblock {Meteor stream identification a new approach. Application to 3675 radio meteors}.
\newblock In W.~J. {Baggaley} and V.~{Porubcan}, editors, {\em Meteoroids 1998, Proceedings of the International Conference held at Tatranska Lomnica, Slovakia, August 17-21, 1998}. Astronomical Institute of the Slovak Academy of Sciences, January 1999.

\bibitem{1989JIMO...17..242T}
A.~K. {Terentjeva}.
\newblock {Fireball streams.}
\newblock {\em WGN, Journal of the International Meteor Organization}, 17(6):242--245, December 1989.

\bibitem{1976Icar...27..265S}
Z.~{Sekanina}.
\newblock {Statistical Model of Meteor Streams. IV. A Study of Radio Streams from the Synoptic Year}.
\newblock {\em Icarus}, 27(2):265--321, February 1976.

\bibitem{1964AuJPh..17..205N}
C.~S. {Nilsson}.
\newblock {A southern hemisphere radio survey of meteor streams}.
\newblock {\em Australian Journal of Physics}, 17:205, June 1964.

\bibitem{2008Icar..195..317B}
P.~{Brown}, R.~J. {Weryk}, D.~K. {Wong}, and J.~{Jones}.
\newblock {A meteoroid stream survey using the Canadian Meteor Orbit Radar. I. Methodology and radiant catalogue}.
\newblock {\em Icarus}, 195(1):317--339, May 2008.

\bibitem{1967SCoA...11..183K}
B.~L. {Kashcheyev} and V.~N. {Lebedinets}.
\newblock {Radar studies of meteors}.
\newblock {\em Smithsonian Contributions to Astrophysics}, 11:183, January 1967.

\bibitem{2016Icar..266..331J}
P.~{Jenniskens}, Q.~{N{\'e}non}, J.~{Albers}, P.~S. {Gural}, B.~{Haberman}, D.~{Holman}, R.~{Morales}, B.~J. {Grigsby}, D.~{Samuels}, and C.~{Johannink}.
\newblock {The established meteor showers as observed by CAMS}.
\newblock {\em Icarus}, 266:331--354, March 2016.

\bibitem{Shiba2023}
Yasuo {Shiba}.
\newblock {Halley Type and Long Period Meteor Shower Luminous Altitude Characteristics}.
\newblock {\em WGN, Journal of the International Meteor Organization}, 51(4):93--108, August 2023.

\bibitem{2020eMetN...5...93K}
Masahiro {Koseki}.
\newblock {Confusions in IAUMDC Meteor Shower Database (SD)}.
\newblock {\em eMeteorNews}, 5(2):93--111, March 2020.

\bibitem{1994PSS...42..151P}
V.~{Porub{\v{c}}an} and M.~{Gavajdov{\'{a}}}.
\newblock {A search for fireball streams among photographic meteors}.
\newblock {\em Planetary and Space Science}, 42(2):151--155, February 1994.

\bibitem{1971SCoA...12....1L}
B.~A. {Lindblad}.
\newblock {A stream search among 865 precise photographic meteor orbits.}
\newblock {\em Smithsonian Contributions to Astrophysics}, 12:1--13, January 1971.

\bibitem{2023P&SS..23505737N}
L.~{Neslu{\v{s}}an}, T.~J. {Jopek}, R.~{Rudawska}, M.~{Hajdukov{\'a}}, and G.~{Kokhirova}.
\newblock {Showers with both Northern and Southern solutions}.
\newblock {\em Planetary and Space Science}, 235:105737, October 2023.

\bibitem{1964PhDT........28N}
Carl~Sigurd {Nilsson}.
\newblock {\em {A radio survey of the orbits of meteors}}.
\newblock PhD thesis, University of Adelaide, Australia, January 1964.

\bibitem{1988msdc.book.....K}
Gary~W. {Kronk}.
\newblock {\em {Meteor showers. A descriptive catalog}}.
\newblock 1988.

\bibitem{2021Bruzzone}
Juan~Sebastian {Bruzzone}, Robert~J. {Weryk}, Diego {Janches}, Carsten {Baumann}, Gunter {Stober}, and Jose~Luis {Hormaechea}.
\newblock {Observation of the A Carinid Meteor Shower 2020 Unexpected Outburst}.
\newblock {\em The Planetary Science Journal}, 2(2):56, April 2021.

\bibitem{2021Jenniskens}
P.~{Jenniskens}.
\newblock {Beta Tucanids (BTU \#108) meteor outburst in 2021}.
\newblock {\em eMeteorNews}, 6(4):330--331, April 2021.

\bibitem{1994A&A...287..990J}
P.~{Jenniskens}.
\newblock {Meteor stream activity I. The annual streams}.
\newblock {\em Astronomy \& Astrophysics}, 287:990--1013, July 1994.

\bibitem{2014JIMO...42...68M}
Sirko {Molau} and Steve {Kerr}.
\newblock {Meteor showers of the southern hemisphere}.
\newblock {\em WGN, Journal of the International Meteor Organization}, 42(2):68--75, April 2014.

\bibitem{1973Icar...18..253S}
Z.~{Sekanina}.
\newblock {Statistical Model of Meteor Streams. III. Stream Search Among 19303 Radio Meteors}.
\newblock {\em Icarus}, 18(2):253--284, February 1973.

\bibitem{2010Icar..207...66B}
P.~{Brown}, D.~K. {Wong}, R.~J. {Weryk}, and P.~{Wiegert}.
\newblock {A meteoroid stream survey using the Canadian Meteor Orbit Radar. II: Identification of minor showers using a 3D wavelet transform}.
\newblock {\em Icarus}, 207(1):66--81, May 2010.

\bibitem{2010MNRAS.404..867J}
T.~J. {Jopek}, P.~{Koten}, and P.~{Pecina}.
\newblock {Meteoroid streams identification amongst 231 Southern hemisphere video meteors}.
\newblock {\em Monthly Notices of the Royal Astronomical Society}, 404(2):867--875, May 2010.

\bibitem{2016Icar..266..355J}
P.~{Jenniskens}, Q.~{N{\'e}non}, P.~S. {Gural}, J.~{Albers}, B.~{Haberman}, B.~{Johnson}, D.~{Holman}, R.~{Morales}, B.~J. {Grigsby}, D.~{Samuels}, and C.~{Johannink}.
\newblock {CAMS confirmation of previously reported meteor showers}.
\newblock {\em Icarus}, 266:355--370, March 2016.

\bibitem{2022eMetN...7..293R}
P.~{Roggemans}, D.~{{\v{S}}egon}, D.~{Vida}, J.~{Greaves}, T.~{Sekiguchi}, A.~{Angelsky}, and A.~{Davydov}.
\newblock {Near anti-helion meteor shower outburst recorded by Global Meteor Network}.
\newblock {\em eMeteorNews}, 7(5):293--301, September 2022.

\bibitem{1988JIMO...16..191W}
J.~{Wood}.
\newblock {The October Capricornid meteor stream.}
\newblock {\em WGN, Journal of the International Meteor Organization}, 16(6):191--194, January 1988.

\bibitem{2017A&A...607A...5D}
B.~A. {Dumitru}, M.~{Birlan}, M.~{Popescu}, and D.~A. {Nedelcu}.
\newblock {Association between meteor showers and asteroids using multivariate criteria}.
\newblock {\em Astronomy \& Astrophysics}, 607:A5, October 2017.

\bibitem{1973NASSP.319..183C}
A.~F. {Cook}.
\newblock {A Working List of Meteor Streams}.
\newblock In Curtis~L. {Hemenway}, Peter~M. {Millman}, and Allan~F. {Cook}, editors, {\em NASA Special Publication}, volume 319, page 183. NASA 1973, 1973.

\bibitem{1962R}
H.~B. {Ridley}.
\newblock {The Phoenicid Meteor Shower of 1956 December 5}.
\newblock {\em J. Brit. Astron. Assoc.}, 72:266--272, 1962.

\bibitem{1963MNSSA..22...42R}
H.~B. {Ridley}.
\newblock {The Phoenicid Meteor Shower of 1956 December 5}.
\newblock {\em Monthly Notes of the Astronomical Society of South Africa}, 22:42, January 1963.

\bibitem{2009JIMO...37...55S}
{SonotaCo}.
\newblock {A meteor shower catalog based on video observations in 2007-2008}.
\newblock {\em WGN, Journal of the International Meteor Organization}, 37:55--62, April 2009.

\bibitem{2014me13.conf..217R}
R.~{Rudawska} and P.~{Jenniskens}.
\newblock {New meteor showers identifed in the CAMS and SonotaCo meteoroid orbit surveys}.
\newblock In T.~J. {Jopek}, F.~J.~M. {Rietmeijer}, J.~{Watanabe}, and I.~P. {Williams}, editors, {\em Meteoroids 2013}, pages 217--224, July 2014.

\bibitem{2016Icar..266..371J}
Peter {Jenniskens} and Quentin {N{\'e}non}.
\newblock {CAMS verification of single-linked high-threshold D-criterion detected meteor showers}.
\newblock {\em Icarus}, 266:371--383, March 2016.

\bibitem{2013JIMO...41...43H}
David {Holman} and Peter {Jenniskens}.
\newblock {Discovery of the Upsilon Andromedids (UAN, IAU \#507)}.
\newblock {\em WGN, Journal of the International Meteor Organization}, 41(2):43--47, April 2013.

\bibitem{2016Icar..266..384J}
P.~{Jenniskens}, Q.~{N{\'e}non}, P.~S. {Gural}, J.~{Albers}, B.~{Haberman}, B.~{Johnson}, R.~{Morales}, B.~J. {Grigsby}, D.~{Samuels}, and C.~{Johannink}.
\newblock {CAMS newly detected meteor showers and the sporadic background}.
\newblock {\em Icarus}, 266:384--409, March 2016.

\bibitem{2013JIMO...41...70S}
Damir {{{S}}egon}, {\v{Z}}eljko {Andrei{\'c}}, Korado {Korlevi{\'c}}, Filip {Novoselnik}, Denis {Vida}, and Ivica {Skoki{\'c}}.
\newblock {8 new showers from Croatian Meteor Network data}.
\newblock {\em WGN, Journal of the International Meteor Organization}, 41(3):70--74, June 2013.

\bibitem{2014JIMO...42...90A}
{\v{Z}}eljko {Andrei{\'c}}, Peter {Gural}, Damir {{\v{S}}egon}, Ivica {Skoki{\'c}}, Korado {Korlevi{\'c}}, Denis {Vida}, Filip {Novoselnik}, and David {Gostinski}.
\newblock {Results of CMN 2013 search for new showers across CMN and SonotaCo databases I-0.5mm}.
\newblock {\em WGN, Journal of the International Meteor Organization}, 42(3):90--97, June 2014.

\bibitem{2014JIMO...42..132G}
Peter {Gural}, Damir {{\v{S}}egon}, {\v{Z}}eljko {Andrei{\'c}}, Ivica {Skoki{\'c}}, Korado {Korlevi{\'c}}, Denis {Vida}, Filip {Novoselnik}, and David {Gostinski}.
\newblock {Results of CMN 2013 search for new showers across CMN and SonotaCo databases II}.
\newblock {\em WGN, Journal of the International Meteor Organization}, 42(4):132--138, August 2014.

\bibitem{2021PSS..19505152H}
M.~{Hajdukov{\'a}} and L.~{Neslu{\v{s}}an}.
\newblock {Modeling the meteoroid streams of comets C/1894 G1 (Gale) and C/1936 O1 (Kaho-Kozik-Lis)}.
\newblock {\em Planetary and Space Science}, 195:105152, January 2021.

\bibitem{Gural2011}
P.~S. {Gural}.
\newblock {The California All-sky Meteor Surveillance (CAMS) System}.
\newblock In {\em Proceedings of the International Meteor Conference, 29th IMC, Armagh, Northern Ireland, 2010}, pages 28--31, January 2011.

\bibitem{Jenniskens_etal2011}
P.~{Jenniskens}, P.~S. {Gural}, L.~{Dynneson}, B.~J. {Grigsby}, K.~E. {Newman}, M.~{Borden}, M.~{Koop}, and D.~{Holman}.
\newblock {CAMS: Cameras for Allsky Meteor Surveillance to establish minor meteor showers}.
\newblock {\em Icarus}, 216:40--61, November 2011.

\bibitem{Jenniskens_etal2016a}
P.~{Jenniskens}, Q.~{N{\'e}non}, J.~{Albers}, P.~S. {Gural}, B.~{Haberman}, D.~{Holman}, R.~{Morales}, B.~J. {Grigsby}, D.~{Samuels}, and C.~{Johannink}.
\newblock {The established meteor showers as observed by CAMS}.
\newblock {\em Icarus}, 266:331--354, March 2016.

\bibitem{Jenniskens_etal2016b}
P.~{Jenniskens}, Q.~{N{\'e}non}, P.~S. {Gural}, J.~{Albers}, B.~{Haberman}, B.~{Johnson}, D.~{Holman}, R.~{Morales}, B.~J. {Grigsby}, D.~{Samuels}, and C.~{Johannink}.
\newblock {CAMS confirmation of previously reported meteor showers}.
\newblock {\em Icarus}, 266:355--370, March 2016.

\bibitem{Jenniskens_etal2016c}
P.~{Jenniskens}, Q.~{N{\'e}non}, P.~S. {Gural}, J.~{Albers}, B.~{Haberman}, B.~{Johnson}, R.~{Morales}, B.~J. {Grigsby}, D.~{Samuels}, and C.~{Johannink}.
\newblock {CAMS newly detected meteor showers and the sporadic background}.
\newblock {\em Icarus}, 266:384--409, March 2016.

\bibitem{Jenniskens_Nenon2016}
P.~{Jenniskens} and Q.~{N{\'e}non}.
\newblock {CAMS verification of single-linked high-threshold D-criterion detected meteor showers}.
\newblock {\em Icarus}, 266:371--383, March 2016.

\bibitem{2016JIMO...44...42S}
{SonotaCo}.
\newblock {Observation error propagation on video meteor orbit determination}.
\newblock {\em WGN, Journal of the International Meteor Organization}, 44:42--45, April 2016.

\bibitem{SonotaCo2017}
{SonotaCo}.
\newblock {Exhaustive error computation on 3 or more simultaneous meteor observations}.
\newblock {\em WGN, Journal of the International Meteor Organization}, 45(5):95--97, October 2017.

\bibitem{SonotaCo_etal2021}
{SonotaCo}, T.~{Masuzawa}, T.~{Sekiguchi}, T.~{Miyoshi}, Y.~{Fujiwara}, K.~{Maeda}, and S.~{Uehara}.
\newblock {Ongoing Meteor Work. SNMv3: A Meteor Data Set for Meteor Shower Analysis}.
\newblock {\em WGN, Journal of the International Meteor Organization}, 49:64--70, 2021.

\end{thebibliography}
